\documentclass[a4paper,11pt]{article}
\pdfoutput=1 

\usepackage{jcappub} 
\usepackage[T1]{fontenc} 
\usepackage{amsfonts,latexsym,amssymb,amsmath,mathrsfs}
\pdfoutput=1
\usepackage{graphicx}

\newcommand{\ma}[1]{\mbox{$\mathcal{#1}$}}
\newcommand{\mas}[1]{\mbox{$\mathscr{#1}$}}
\newcommand{\ti}{\tilde}
\newcommand{\we}{\wedge}
\newcommand{\tr}{{\rm Tr}}
\newcommand{\jh}{\,{\rm jh}}
\newcommand{\tjh}{\,{\rm tjh}}

\def\FigDir{.}

\def\defscript{\mathscr}

\def\H{{\defscript H}}

\def\LL{{\defscript L}}

\def\N{{\defscript N}}

\def\P{{\defscript P}}

\def\SS{{\defscript S}}
\def\T{{\defscript T}}
\def\U{{\defscript U}}

\def\mf#1{{\mathfrak{#1}}}

\def\ifempty#1{\def\tmpdata{#1}\ifx\tmpdata\empty }

\def\linebreak{\hfill\break}

\def\mpl{m_{\rm pl}}

\def\bra<#1|{\langle #1\rvert}
\def\ket|#1>{\lvert#1 \rangle}
\def\braket<#1|#2>{\langle #1|#2 \rangle}

\def\pfrac#1#2{\left(\frac{#1}{#2}\right)}

\def\otop#1{\hbox{$#1\kern-0.1em$\llap{\hbox{\raise1.7ex\hbox{$\scriptstyle\circ$}}}} }

\def\inpare#1{\left(#1\right)}
\def\bigpare(#1){\left(#1\right)}
\def\inrbra#1{\left\{ #1 \right\}}
\def\insbra#1{\left[ #1 \right]}
\def\inang#1{\left\langle {#1} \right\rangle}
\def\EXP#1{\inang{#1}}
\def\bigbra[#1]{\left[ #1 \right]}

\def\Slash#1{\hbox{{\setbox1=\hbox{$#1$}\rlap{\hbox to 1.4\wd1{\hfil $/$ \hfil}}\box1}}}

\def\tend{\rightarrow}

\def\therefore{\mbox{\setbox0=\hbox{X}\hbox{$\ldotp$}\raise0.7\ht0\hbox{$\ldotp$}\hbox{$\ldotp$}} \quad }
\def\because{\mbox{\setbox0=\hbox{X}\raise0.7\ht0\hbox{$\ldotp$}\hbox{$\ldotp$}\raise0.7\ht0\hbox{$\ldotp$}}\kern0pt }

\def\e#1{{10^{#1}}}
\def\bm#1{\boldsymbol{#1}}

\def\ZR{{{\mathbb Z}}}

\def\RF{{{\mathbb R}}}

\def\SL{{\rm SL}}

\def\SO{{\rm SO}}
\def\so{\mathfrak{so}}

\def\SU{{\rm SU}}

\def\Usp{{\rm Usp}}

\def\upin{\hbox{\setbox0=\hbox{$\cup$} \vrule width 0.05 \wd0 height \ht0 depth 0pt \kern - 0.5\wd0 \box0 }}
\def\Frac(#1/#2){\left(\frac{#1}{#2}\right)}
\def\lsim{\lesssim}
\def\gsim{\gtrsim}

\def\Mat#1{\begin{pmatrix} #1 \end{pmatrix}}

\def\Tr{{\rm Tr}}
\def\tr{{\rm tr}}
\def\Tp#1{\,{}^T \! #1}

\def\Im{{\rm Im\,}}
\def\Re{{\rm Re\,}}

\def\sdprod{\mathrel{{\setbox0=\hbox{$\displaystyle\times$}\lower0.3\wd0\hbox{$\stackrel{\box0}{\scriptstyle\sim}$}}}}

\def\w{\wedge}

\def\tosigma#1,{%
    \ifx\tmpindex\relax \def\tmpindex{#1} \let\next=\tosigma
    \else \ifnum\tmpindex=0 1 \else \sigma_\tmpindex \fi
          \ifx#1\relax  \let\next=\relax
          \else \otimes \let\next=\tosigma \def\tmpindex{#1} \fi
    \fi \next}
\def\tspb(#1){\let\tmpindex=\relax\tosigma#1,\relax,}
\def\pd{\partial}

\def\HyperG(#1,#2;#3;#4){F\inpare{\textstyle #1,#2;#3;#4}}

\def\Eq#1{\begin{equation} #1 \end{equation}}

\def\Eqr#1{\begin{eqnarray} #1 \end{eqnarray}}

\def\Eqrsub#1{\begin{subequations}\Eqr{#1}\end{subequations}}
\def\Eqrsubl#1#2{\begin{subequations}
  \expandafter\ifx\csname Rlabel\endcsname \relax \label{#1}
  \else \Rlabel{#1} \fi \Eqr{#2}\end{subequations}}
\def\Bitm{\begin{itemize}}
\def\Eitm{\end{itemize}}
\def\Blist#1#2{\begin{list}{#1}{\parsep=0pt \itemsep=0pt%
  \listparindent=0pt #2}}
\def\Elist{\end{list}}
\long\def\ignore#1#2{\def\ignoreflag{#1}\long\def\tmptext{#2}
  \ifnum\ignoreflag>1 #2 \fi}

\title{\boldmath Inflation in maximal gauged supergravities}

\author[a,b]{Hideo Kodama}
\author[c]{Masato Nozawa}

\affiliation[a]{Theory Center, KEK, Tsukuba 305-0801, Japan}
\affiliation[b]{Department of Particles and Nuclear Physics,
The Graduate University for Advanced Studies, Tsukuba 305-0801, Japan}
\affiliation[c]{Dipartimento di Fisica, Universit\`a di Milano, and INFN, Sezione di Milano,\\
Via Celoria 16, 20133 Milano, Italy}

\emailAdd{Hideo.Kodama@kek.jp}
\emailAdd{Masato.Nozawa@mi.infn.it}

\abstract{
We discuss the dynamics of multiple scalar fields and the possibility of realistic inflation in the maximal gauged supergravity. In this paper, we address this problem in the framework of recently discovered 1-parameter deformation of ${\rm SO}(4,4)$ and ${\rm SO}(5,3)$ dyonic gaugings, for which the base point of the scalar manifold corresponds to an unstable de Sitter critical point. In the gauge-field frame where the embedding tensor takes the value in the sum of the {\bf 36} and {\bf 36'} representations of ${\rm SL}(8)$, we present a scheme that allows us to derive an analytic expression for the scalar potential. With the help of this formalism, we derive the full potential and gauge coupling functions in analytic forms for the ${\rm SO}(3)\times {\rm SO}(3)$-invariant subsectors of ${\rm SO}(4,4)$ and ${\rm SO}(5,3)$ gaugings,  and argue that there exist no new critical points in addition to those discovered so far. For the ${\rm SO}(4,4)$ gauging, we also study the behavior of 6-dimensional scalar fields in this sector near the Dall'Agata-Inverso de Sitter critical point at which the negative eigenvalue of the scalar mass square with the largest modulus goes to zero as the deformation parameter $s$ approaches a critical value $s_{\rm c}$. We find that when the deformation parameter $s$ is taken sufficiently close to the critical value,  inflation lasts more than 60 e-folds even if the initial point of the inflaton allows an $O(0.1)$ deviation in Planck units from the Dall'Agata-Inverso critical point. It turns out that the spectral index $n_s$ of the curvature perturbation at the time of the 60 e-folding number is always about $0.96$ and within the $1\sigma$ range $n_s=0.9639\pm0.0047$ obtained by Planck, irrespective of the value of the $\eta$ parameter at the critical saddle point. The tensor-scalar ratio predicted by this model is around $10^{-3}$ and is close to the value in the Starobinsky model.
}


\begin{document}
\maketitle
\flushbottom










\section{Introduction}

The recent observational data for CMB anisotropy~\cite{Ade:2013zuv} have definitely confirmed that the early universe underwent an accelerated expanding phase, dubbed inflation~\cite{Starobinsky:1980te,Sato:1980yn,Guth:1980zm}.  Inflation is one of the most exciting topics in contemporary cosmology.  Still, we do not have yet a satisfactory description of  inflation from the viewpoint of fundamental theory, although a lot of endeavors have been dedicated since the proposal of the KKLT scenario~\cite{Kachru:2003aw,Kachru:2003sx}. The most difficult issue to this problem is how to drive the cosmic
acceleration while keeping all moduli stabilized.  The no-go theorem~\cite{nogo,Maldacena:2000mw} forbids the accelerating solution, provided we are discussing the problem within the framework of classical higher-dimensional supergravity
(see also \cite{Caviezel:2008tf,Hertzberg:2007wc} for the no-go theorem against stable de Sitter vacua in IIA theory). This theorem requires the necessity of introducing noncompact extra dimensions, higher-order corrections, non-perturbative effects and so on.

Currently, not all of the lower-dimensional supergravities have been derived by  dimensional reduction and dualities from the 11-dimensional supergravity. This fact motivates us to investigate inflationary models within 4-dimensional supergravity settings. Many people have focused attention to the $N=1$ supergravity, since chiral fermions are naturally incorporated in its theoretical framework. The $N=1$ supergravity has been able to provide  desired inflaton potentials by fine-tuning the (Hodge-)K\"ahler metric, the superpotential, and possibly the D-term (see e.g, \cite{Yamaguchi:2011kg,Baumann:2014nda}  for a recent review). It is, however, difficult in general to embed these phenomenological models into string theory. Moreover, the $N=1$ supergravity fails to have a strong predictive ability to explain the observational data, since the K\"ahler potential and superpotential are arbitrary functions. To the contrary,  the potential for $N \ge 2$ theories arises only from the gauging prescription so that the potential is highly restrictive, viz, we might be able to circumvent the landscape problem~\cite{Susskind:2003kw}. It deserves to comment that the  implementation of this program is fairly nontrivial and requires an independent study, since $N\ge 2$ supergravities do not in general fit into the framework of the $N=1$ supergravity~\cite{Andrianopoli:2001zh}. Extended gauged supergravities have attracted much attention recently also in the context of flux compactification and gauge/gravity correspondence. In particular, potentials arising from gaugings may have their origin in  non-perturbative quantum effects in higher-dimensional theories. This expectation is supported by the fact that the potential is identified in some cases as (minus) the internal curvature of generalized geometry~\cite{Aldazabal:2013mya}.  This prompts us to examine the possibility of inflation in extended supergravities. To this aim, the maximal $N=8$ gauged supergravity certainly provides an optimal arena for exploring the gravity sector of string dynamics, since the IIA/IIB string theory and M-theory are formulated with maximal supercharges corresponding to $N=8$.   Over the past 30 years, the $N=8$ gauged supergravity~\cite{DN} has been intensively studied from various points of view~\cite{Hull1,Hull2,Hull3,Hull4,Cordaro:1998tx,Hull:2002cv,Warner:1983vz,Warner:1983du,Hull:1984ea,
Ahn:2001by,Ahn:2002qga,Kallosh:2001gr,Bobev:2010ib,Bobev:2011rv,Fischbacher:2010ec,Fischbacher:2009cj,Fischbacher:2010ki,Fischbacher:2011jx}.\footnote{
See ref.~\cite{Meissner:2014joa} for a recent attempt to embed the standard model
fermions in $N=8$ supergravity. }
A standard lore that had emerged from these analyses is that the maximal gauged supergravity is unique.
This stereotyped idea was dispelled by the recent discovery of a new 1-parameter family of deformation of ${\rm SO}(8)$ gauged supergravity~\cite{Dall'Agata:2012bb}, based upon the embedding tensor formulation~\cite{deWit:2002vt,deWit:2007mt}.  Although the original ${\rm SO}(8)$ gauged supergravity has been obtained via gaugings and also by the $S^7$ reduction of the 11-dimensional supergravity~\cite{DN2}, the uplifting of the deformed theory into the 11-dimensional framework is
yet unknown (see~\cite{deWit:2013ija} for a work along this direction). A distinctive feature of the new deformation is that it introduces ``magnetic'' vector potentials, in addition to the standard electric ones. The introduction of magnetic potentials manifests itself in a variety of ways. An interesting consequence of this dyonic gaugings is that they possess a completely distinct structure of extrema, compared to the standard electric ones. Combined with the ``go to the origin'' approach developed in refs.~\cite{Dibitetto:2011gm,DI}, we can now obtain dozen of new critical points, new supersymmetry breaking patterns, the analytic mass spectrum for seventy  scalars and a black hole solution~\cite{DI,KN,Borghese:2011en,Dall'Agata:2012cp,Dall'Agata:2012sx,
Borghese:2012qm,Borghese:2012zs,Borghese:2013dja,Catino:2013ppa,Lu:2014fpa,Dall'Agata:2014ita,Gallerati:2014xra}.
In our previous paper~\cite{KN}, we revealed that the special ${\rm SL}(8)$-type critical points\footnote{We refer to a gauging with its gauge group contained in the standard $\SL(8)$ maximal subgroup of $E_{7(7)}$ as {\it a  $\SL(8)$-type gauging}. If a critical point of the potential in a $\SL(8)$-type gauging can be further transformed into the origin corresponding to the coset $\SU(8)$ in the scalar manifold $E_{7(7)}/\SU(8)$ by a duality transformation in $\SL(8)$, it is referred to as {\it a special $\SL(8)$-type critical point}.}
can be completely classified and their mass spectrum can be evaluated group-theoretically. It turned out that many of ${\rm SL}(8)$-type gaugings allow a deformation parameter as in the ${\rm SO}(8)$ theory. An intriguing facet of  special ${\rm SL}(8)$-type (anti-)de Sitter extrema is that their mass spectrum is only dependent on the residual gauge symmetry, insensitive to the original gaugings and the deformation parameter. Further, special de Sitter critical points appear only for ${\rm SO}(4,4)$ and ${\rm SO}(5,3)$ gaugings and are unstable in the directions that are singlet with respect to the residual symmetry. The parameter $\eta =V''/V$ characterizing the instability growth rate is $\eta=-2$, which implies that the instability grows in a time scale comparable to the cosmic expansion time in the cosmological context.  Remarkably, the adjustable deformation parameter can be absorbed in the overall cosmological constant and does not affect the $\eta$ parameter. This is a typical $\eta$-problem by which many of supergravity theories are plagued.  Accordingly, a de Sitter critical point around which a realistic inflation can be derived must be of the general type sitting off the origin of the scalar manifold in any ${\rm SL}(8)$-type gauge frame, if it exists.

In ref.~\cite{Dall'Agata:2012sx}, Dall'Agata and Inverso found a new de Sitter critical point of general type in the ${\rm SO}(4,4)$ gauging (we shall refer to this extremum as the DI critical point) which evades the $\eta$-problem  (see ref.~\cite{Baron:2014bya} for the 11-dimensional uplift). The modulus of the $\eta$ parameter at this critical point can be made as small as we wish by choosing the deformation parameter close to a critical value.  In the present paper, we will make deeper studies of the scalar field dynamics in this gauged supergravity by extending the analysis of~\cite{Dall'Agata:2012sx}. We present a simple method to compute the scalar potential away from the origin, based on an inventive diagonalization of the coset representative . Applying this prescription to the ${\rm SO}(3)\times {\rm SO}(3)$-invariant sector of ${\rm SO}(4,4)$ and ${\rm SO}(5,3)$ gaugings, we examine whether there exist additional critical points in these six-dimensional sectors or not.  As far as the authors know, the present study provides the highest-dimensional analytic search of critical points in $N=8$ gauged supergravity with noncompact gaugings. We further deeply investigate whether a realistic inflation is realized around the DI saddle point by solving the equations of motion in the six-dimensional ${\rm SO}(3)\times {\rm SO}(3)$-invariant subspace of the 70-dimensional scalar manifold in the $\SO(4,4)$ gauging. We also take into account the flux contributions and examine if the chromo-natural inflation~\cite{Adshead:2012kp,Dimastrogiovanni:2012st} and the anisotropic inflation~\cite{Watanabe:2009ct,Yokoyama:2008xw} occur.
We find that by choosing the deformation parameter close to the critical value, the universe can inflate for a sufficiently long time even if the initial offset of the inflaton from the DI saddle point is of order unity in the Planck units. We also find that the spectral index $n_s$ of the scalar curvature perturbation at the present horizon scale is quite insensitive to the initial condition and the existence of flux and consistent with the observations by Planck~\cite{Ade:2013zuv}.  We will also see that the tensor-scalar ratio predicted by our model is of order $10^{-3}$ and about one third of the value predicted in the Starobinsky model~\cite{Starobinsky:1980te}.

Because the potential of this gauged maximal supergravity has no lower bounds, we cannot discuss the reheating phase and subsequent big-bang stage.  In order to investigate these post inflation stages, we have to find a modification of the theory in such a way that the potential is non-negative and have to couple the system to a matter sector with a lower local supersymmetry. In the present paper we do not attempt to make such modifications, and examine the scalar dynamics by assuming that the inflationary stage is well described by the maximal gauged supergravity.

The present paper is organized as follows. In the next section, after fixing our notations, we develop a general formulation that can be used to derive an explicit expression for the potential in the
${\rm SL}(8)$-type gauge frame.
In sections~\ref{sec:SO44} and ~\ref{sec:SO53}, we apply this formulation to the $\SO(3)\times\SO(3)$-invariant sector of the $\SO(4,4)$ and $\SO(5,3)$ gaugings, respectively. We derive explicit expressions for the kinetic term and  potential for the scalar fields as well as the gauge coupling functions, and perform the search of new critical points in this sector. Next, we do cosmology in section~\ref{sec:inflation}. We first derive the effective action describing the cosmological evolution of the scalar-gravity-gauge system obtained in section~\ref{sec:SO44}. We clarify the definitions and meanings of slow-roll parameters in our multi-component system. In particular, we point out the existence of special attractor slow roll trajectories that play a crucial role in the inflation dynamics in our system. After analyzing the $\eta$-problem in our system in detail, we explore the possibility of a realistic inflation by solving the multi-dimensional dynamical evolution equations numerically. Section~\ref{sec:summary} summarizes the results of the present work.

\section{${\rm SL}(8)$ gauge-field frame}

We begin this section with a brief review of $N=8$ supergravity results in the existing
literatures~\cite{deWit:2007mt,DI,KN}, based on the embedding tensor formalism. Our notations basically follow refs.~\cite{deWit:2007mt,DI,KN}.

The embedding tensor  enables us to discuss the embedding of a gauge group $G$ into the duality group $E_{7(7)}$ in a group-theoretical fashion~\cite{deWit:2002vt,deWit:2007mt}.  The embedding tensor $\Theta_M{}^\alpha $ specifies the gauge generators $X_M$ in terms of the $E_{7(7)}$ generators $t_\alpha $ as $X_M=\Theta_M{}^\alpha t_\alpha $. Here $M,N,...=1,...,56$ and
$\alpha, \beta,...=1,...,133$ are the fundamental  and adjoint representation indices of $E_{7(7)}$.  Using the gauge generators $X_M$, the gauge coupling of scalar fields can be introduced in the coset representation by the replacement $\partial_\mu \to \partial_\mu -gA_\mu{}^M X_M$, where $g$ is a gauge coupling constant and $A_\mu{}^M$ is a gauge potential.

Consistency of a gauging amounts to requiring that the embedding tensor obeys a set of linear and  quadratic constraints. The linear relation renders $\Theta_M{}^\alpha $ to belong to the ${\bf 912}$ representation of $E_{7(7)}$, which branches into ${\bf 912}\to {\bf 36}+{\bf 36}'+{\bf 420}+{\bf 420}'$ under its maximal subgroup ${\rm SL}(8)$. In this section, we restrict our consideration to $\SL(8)$-type gaugings and work in the ${\rm SL}(8)$ gauge-field frame.  Then,  only the ${\bf 36}$ and ${\bf 36}'$ representations are turned on, and the gauge generators $X_M={}^{T\!}(X_{ab} ~ X^{ab})$ take the form,
\begin{align}
 X_{ab} =\left(
\begin{array}{cc}
 \delta_{[a}{}^{[e}\theta_{b][c}\delta_{d]}{}^{f]} &
0
\\
0 &
-\delta_{[a}{}^{[c}\theta_{b][e}{\delta_{f]}}^{d]}
\end{array}
\right) \,, \qquad
 X^{ab} =\left(
\begin{array}{cc}
-{\delta_{[c} }^{[a}\xi^{b][e}\delta_{d]}{}^{f]}&
0
\\
0 &
{\delta_{[e} }^{[a}\xi^{b][c}{\delta_{f]}
 }^{d]}
\end{array}
\right) \,.
\label{N8_CP_dyonic_Xtensors}
\end{align}
Here, $\theta _{ab}$ and $\xi^{ab}$ are $8\times 8$ symmetric matrices, and $a,b,...=1,...,8$ are  ${\rm SL}(8)$ indices. The
quadratic constraint requires the following relation~\cite{DI}
\begin{align}
\theta \cdot \xi =c \mathbb I_8 \,,
\end{align}
where $c\in \mathbb R$ is a constant and $\mathbb I_n$ is the $n$-dimensional unit matrix. The electric gaugings $\xi=0$ trivially satisfy the quadratic constraints.

In maximal supergravity, scalar fields are described by the symmetric coset space
$E_{7(7)}/{\rm SU}(8)$. In gauged versions,
these scalars acquire a potential to keep supersymmetries. In the embedding tensor formulation,
the potential is also expressed in an $E_{7(7)}$ covariant way and reads
\begin{align}
V =\frac{g^2}{672}\left({X_{MN}}^R{X_{PQ}}^SM^{MP}M^{NQ}M_{RS}+7{X_{MN}}^Q{X_{PQ}}^N
 M^{MP}\right) \,,
\label{N8_potential}
\end{align}
where $M_{MN}=(L\cdot {}^{T\!}L)_{MN}$ is the $56\times 56 $ scalar matrix with its inverse $(M^{-1})^{MN}=M^{MN}=\Omega^{MP}\Omega^{NQ}M_{PQ}$. Here, $\Omega^{MN}=\Omega_{MN}=(i\sigma_2 \otimes \mathbb I_{28})_{MN}$ is an ${\rm Sp}(56,\mathbb R)$-invariant metric, and $L=L(\phi)$ is a coset representative in the ${\rm Sp}(56,\mathbb R)$ representation described below in more detail.  The kinetic term of the scalar fields can be expressed as $e^{-1}\ma L=-\frac 1{12} \ma P_{\mu ijkl} \ma P^{\mu ijkl} $,
where $\ma P_{\mu ijkl}$ is the $\eta$-selfdual part of the $T$-tensor [see eq.~(\ref{Pmuijkl}) below for definition].
In the absence of the gauge field, this kinetic term can be regarded as the Lagrangian for the harmonic map from the spacetime to the target scalar manifold $E_{7(7)}/{\rm SU}(8)$ with the metric
\begin{align}
\label{kinetic}
d s_T^2\equiv \frac 1{12} \ma P_{ijkl} \otimes \ma P^{ijkl}
=-\frac 1{96}\tr (d M d M^{-1}) \,.
\end{align}
We call this metric {\it a kinetic-term metric} or {\it a scalar manifold metric} in the present paper.

In appearance, the potential~(\ref{N8_potential}) is a highly nonlinear function of 70 scalars, which is the main obstruction when one attempts to find critical points. A proposal put forward in~\cite{DI,Dibitetto:2011gm} is to move a critical point to the origin of the scalar manifold by an $E_{7(7)}$ transformation acting naturally on the scalar manifold.  This procedure transforms the embedding tensor as well. It follows that the critical point condition $\partial_\rho V=0$ gives rise to another quadratic condition on the embedding tensor~\cite{DI}. Here $\rho=1,...,70$ runs over only the noncompact directions of the coset space $E_{7(7)}/{\rm SU}(8)$. This additional relation must be solved together with the quadratic constraints of the embedding tensor, and these algebraic equations are indeed amendable to analytic study. In our previous paper~\cite{KN}, we demonstrated that all special $\SL(8)$-type critical points, which can be mapped to the origin by ${\rm SL}(8)$ transformations, are exhausted by known ones.\footnote{
In ref.~\cite{KN}, it has been argued that the mass spectrum of the scalars is dependent only on the residual gauge symmetry, rather than the original gaugings when the cosmological constant is nonvanishing. We have also confirmed that this property is shared also by other fields with spin.}
It turned out that only the ${\rm SO}(4,4)$ and ${\rm SO(5,3)}$ gaugings admit de Sitter critical points of the special type. Both of these critical points possess the tachyonic directions, irrespective of the deformation parameter, providing candidates of an inflaton. Unfortunately, the curvature of these critical points turned out to be too large to allow a sufficient duration of inflation.
In the present paper, we explore their ``off-center'' behavior, which was outside the scope of our
previous analysis~\cite{KN}.

\subsection{Coset representative}
\label{sec:coset}

To find critical points away from the origin, we need to compute explicitly the exponential map of the coset representative. In this section, we present an efficient formulation that facilitates to perform this program for $\SL(8)$-type gaugings.

The 70 scalar fields take values in the ${\bf 70}_R$ representation of ${\rm SU}(8)$ and are expressed as
%
\begin{align}
\phi_{ijkl}=\phi_{[ijkl]} =\pm\frac 1{4!} \epsilon_{ijklmnpq} \phi^{mnpq}\,.
\end{align}
Here and in what follows, the ${\rm SU}(8)$ indices $i,j,...=1,...,8$ are raised and lowered by complex conjugation, i.e., $\bar \phi_{ijkl} =\phi^{ijkl}$. In the ${\rm Usp}(56)$ representation~\cite{CJ},
we can adopt the unitary gauge in which the $E_{7(7)}$ coset representative $\mas S$ takes the form,
%
\begin{align}
\mas S =\exp \left(
\begin{array}{cc}
0 & \phi \\
\phi^\dagger & 0
\end{array}
\right)=\left(
\begin{array}{cc}
\cosh(\sqrt{\phi\phi^\dagger }) & \phi \jh (\sqrt{\phi^\dagger \phi}) \\
\phi^\dagger \jh (\sqrt{\phi\phi^\dagger }) & \cosh (\sqrt{\phi^\dagger \phi})
\end{array}
\right)\in {\rm Usp}(56)\,,
\label{cosetrep_USP}
\end{align}
where we have defined
\begin{align}
\jh (z)\equiv z^{-1} \sinh (z) \,.
\end{align}
In the double-index notation, $\phi_{ijkl}=\phi_{[ij][kl]}$ should be viewed as a $28\times 28$ symmetric matrix.
Noticing $\phi \jh(\sqrt{\phi^\dagger\phi})=\jh(\sqrt{\phi\phi^\dagger})\phi$, the following properties hold
%
\begin{align}
\mas S =\mas S^\dagger \,, \qquad \mas S^\dagger \ti \Omega \mas  S =\ti \Omega
\equiv -i\left(
\begin{array}{cc}
\mathbb I_{28} & \\
& -\mathbb I_{28}
\end{array}
\right) \,.
\end{align}

Our remaining task is to find how to implement the exponential map for the coset representative~(\ref{cosetrep_USP}).
This can be tackled most easily as follows. Since $\phi$ can be viewed as a $28\times 28$ complex symmetric matrix,
the Autonne-Takagi factorization~\cite{mat} enables us to find matrices $\Lambda$ and $U$ such that
%
\begin{align}
\phi =U\Lambda {}^{T\!} U\,, \qquad U\in {\rm U}(28) \,.
\label{phi_ATfactorization}
\end{align}
Here $\Lambda ={\rm diag}(\lambda_1,...,\lambda_{28})$ is the diagonal matrix
with real entities.
Equation (\ref{phi_ATfactorization}) is tantamount to the following eigenvalue problem,
\begin{align}
\phi u= \lambda \bar u \,,\qquad U_{ij}=(\bar u_j)_i \,.
\label{AT2}
\end{align}
If this system is solved,
the square-root matrix can be easily computed to give
$\sqrt{\phi\phi^\dagger }=U\Lambda U^\dagger $, thereby
the coset representative~(\ref{cosetrep_USP}) now takes the simple form
\begin{align}
\mas S = \left(
\begin{array}{cc}
U\cosh \Lambda U^\dagger  &  U\sinh \Lambda {}^{T\!}U\\
\bar U \sinh \Lambda U^\dagger  & \bar U \cosh \Lambda {}^{T\!}U
\end{array}
\right)\,.
\label{coset_ATfactorization}
\end{align}
The computation of the scalar matrix $M_{MN}$ asks for
the change of the basis from the ${\rm Usp}(56)$ basis to the ${\rm Sp}(56,\mathbb R)$
basis. This can be done through the Majorana-Weyl ${\rm SO}(8)$ gamma matrices~\cite{CJ},
\begin{align}
\mas S_{\underline{M}}{}^{\underline{N}}
=\ma S_{\underline{M}}{}^PL_P{}^Q
(\ma S^{-1})_Q{}^{\underline{N}}\,, \qquad
\ma S \equiv \frac i{4\sqrt 2}\left(
\begin{array}{cc}
\Gamma_{ij}{}^{ab} & i \Gamma_{ijab} \\
\Gamma^{ijab} & -i \Gamma^{ij}{}_{ab}
\end{array}
\right)=(\ma S^\dagger )^{-1} \,.
\end{align}
Here $L_M{}^N$ is the coset representative in the ${\rm Sp}(56,\mathbb R)$
representation and the underlined indices refer to the ${\rm SU}(8)$.
In the  ${\rm Sp}(56,\mathbb R)$ basis,
we have
\Eq{
\frac{1}{16}(\Gamma\phi\Gamma )_{ab}{}^{cd}=2 S_{[a}{}^{[c}\delta_{b]}{}^{d]}+i U_{abcd} \,, \qquad
S_{ab}=S_{(ab)}\,, \qquad
U_{abcd}=U_{[abcd]}=\star U_{abcd}\,.
\label{phibyS&U}
}
Each of real tensors ($S_{ab}, U_{abcd}$) has 35 components.
 Since there is no need to distinguish the upper and lower indices of gamma matrices,
 we can identify from the scratch,
 \begin{align}
 \frac 1{16}\Gamma \phi \Gamma \to \phi \,, \qquad
 \ma S \to  \T_0 \otimes \mathbb I_{28}\,,
 \end{align}
where
\Eq{
\T_0\equiv \frac {i}{\sqrt 2}  \Mat{  1 & i \\  1 & -1 } \,.
}
With this identification, the matrix $M_{MN}$ reads
 \begin{align}
 M_{MN}= (\ma S^\dagger \mas S^{2} \ma S )_{MN}
 =\left(
\begin{array}{cc}
M_{abcd} &M_{ab}{}^{cd} \\
M^{ab}{}_{cd} & M^{abcd}
\end{array}
\right)\,,
\label{rule}
 \end{align}
where
\begin{subequations}
\begin{align}
M_{abcd} &=\frac 14 \left[(U+\bar U)e^{2\Lambda}\,{}^{T\!}(U+\bar U) -(U-\bar U)e^{-2\Lambda}\,{}^{T\!}(U-\bar U) \right]_{abcd} \,,\\
M_{ab}{}^{cd} &=\frac i4 \left[(U-\bar U)e^{-2\Lambda}\,{}^{T\!}(U+\bar U) -(U+\bar U)e^{2\Lambda}\,{}^{T\!}(U-\bar U) \right]_{ab}{}^{cd}\,,\\
M^{ab}{}_{cd} &=\frac i4 \left[(U+\bar U)e^{-2\Lambda}\,{}^{T\!}(U-\bar U) -(U-\bar U)e^{2\Lambda}\,{}^{T\!}(U+\bar U) \right]^{ab}{}_{cd}\,,\\
M^{abcd} &=\frac 14 \left[(U+\bar U)e^{-2\Lambda}\,{}^{T\!}(U+\bar U) -(U-\bar U)e^{2\Lambda}\,{}^{T\!}(U-\bar U) \right]^{abcd}\,.
\end{align}
\end{subequations}
Then, if we can find the Autonne-Takagi factorization~(\ref{phi_ATfactorization}), the
matrix $M_{MN}$ can be obtained in the above manner. Some specific examples
are provided in the next sections.

\subsection{Potential}

Having illustrated the way to find out the scalar matrix $M_{MN}$, we shall next provide an explicit expression of the potential in the ${\rm SL}(8)$ gauge-field frame. Due to the property $M^{MN}=(\Omega M {}^{T\!}\Omega )^{MN}$, we have
\begin{align}
M^{MN}=\left(
\begin{array}{cc}
(M^{-1})^{abcd} & (M^{-1})^{ab}{}_{cd} \\
(M^{-1})_{ab}{}^{cd} & (M^{-1})_{abcd}
\end{array}
\right)=\left(
\begin{array}{cc}
M^{abcd} & -M^{ab}{}_{cd} \\
-M_{ab}{}^{cd} & M_{abcd}
\end{array}
\right)
\end{align}
Using this relation, some elementary calculations reveal that the potential (\ref{N8_potential}) is simplified in the ${\rm SL}(8)$ gauge-field frame to
\begin{align}
V= &\frac{g^2}{672}
\left[
M^{abcd}(\theta_{be}\theta_{dg}W^{eg}{}_{ac}+21\theta_{ad }\theta
 _{bc}) +
M_{abcd}(\xi^{be}\xi^{dg}W^{ac}{}_{eg}+21\xi^{ad }\xi^
 {bc}) \right. \nonumber \\ &\left.
 \quad
 +M^{ab}{}_{cd}\left(\theta_{be}\xi^{dg}W^{ec}{}_{ag}+21{\delta_a}^c
 (\theta \cdot \xi )_b{}^d\right)
+M_{ab}{}^{cd}
 \left(\theta_{dg}\xi^{be}W^{ag}{}_{ec}+21{\delta_c}^a (\xi \cdot \theta
 )^b{}_d\right)
\right]\,,
\label{V_SL8}
\end{align}
where
\begin{align}
W^{ab}{}_{cd}\equiv 2 (M^{aebf}M_{cedf}+M^{ae}{}_{df}M_{ce}{}^{bf}) \,.
\end{align}
At the origin of the scalar manifold, we have
$M_{abcd}=M^{abcd}=\delta_{[a}{}^c\delta_{b]}{}^d$ and
$M^{ab}{}_{cd}=M_{ab}{}^{cd}=0$, recovering the result~\cite{DI,KN}
\begin{align}
V_0=\frac{g^2}{64} \left[2 \tr (\theta^2+\xi^2)-\tr(\theta)^2
-\tr(\xi)^2\right]\,.
\end{align}
It follows that the potential can be computed straightforwardly with the formula (\ref{V_SL8}),
provided we can implement the diagonalization of $\mas S$ (\ref{phi_ATfactorization}).

\subsection{Kinetic term}

In terms of the mixed coset representative $\ma V=\ma S^{-1} \mas S$,
the $\eta$-selfdual part of the $T$-tensor $\ma P_{\mu ijkl}$ can be
defined by~\cite{deWit:2007mt}
\begin{align}
\label{Pmuijkl}
\ma P_{\mu ijkl}= [\ma V^{-1}(\partial_\mu -g A_\mu {}^M X_M)\ma V]_{ijkl} \,.
\end{align}
This quantity defines the scalar kinetic term through (\ref{kinetic}).
The Autonne-Takagi  factorization (\ref{phi_ATfactorization}) leads to
\begin{align}
\label{}
\ma P_{\mu ijkl}=- (UK_\mu {}^{T\!}U)_{ijkl } \,,
\end{align}
where
\begin{align}
\label{}
K_\mu =&\partial_\mu \Lambda -\sinh \Lambda{}^{T\!}UA^s{}_\mu \sinh\Lambda
+\cosh\Lambda U^\dagger A^s{}_\mu \bar U \cosh \Lambda  \nonumber \\
&-\cosh \Lambda U^\dagger (A^a{}_\mu-\partial_\mu U U^\dagger )U \sinh \Lambda
+\sinh \Lambda  {}^{T\!}U(A^a{}_\mu -\partial_\mu \bar U {}^{T\!} U)\bar U \cosh \Lambda  \,,
\end{align}
with
\begin{align}
\label{}
A^s{}_\mu =\frac 12 A^M{}_\mu (\mathbb X_M+{}^{T\! }\mathbb X_M )\,, \qquad
A^a{}_\mu =\frac 12 A^M{}_\mu (\mathbb X_M-{}^{T\! }\mathbb X_M )\,.
\end{align}
Here,
$\mathbb X_M$ is a $28\times 28 $ matrix $\mathbb X_{Mcd}{}^{ef}=(X_{abcd}{}^{ef}, X^{abcd}{}_{ef})$.
Using these expressions,  one can obtain the scalar kinetic term by (\ref{kinetic}) without any difficulty.

\subsection{Gauge kinetic functions}
\label{sec:gaugekin}

Next, we proceed to find expressions for the gauge kinetic functions in terms of matrices
$\Lambda$ and $U$. Let us first focus on the case of the electric gauging, for which the Lagrangian for the vector fields is given by~\cite{DN,deWit:2002vt,deWit:2007mt}
\begin{align}
 e^{-1}\ma L_F=-\frac{1}{4}{\rm Re }({i}N_{\Lambda\Sigma })F^\Lambda{} _{\mu\nu
 }F^\Sigma{}^{\mu\nu }
-\frac 14 {\rm Im} ({ i } N_{\Lambda\Sigma }) F^\Lambda{} _{\mu\nu
 }*  F^{\Sigma }{}^{\mu\nu}\,,
\end{align}
where $(* F^\Lambda )_{\mu\nu }=(1/2)\epsilon_{\mu\nu\rho\sigma}F^{\Lambda\rho\sigma }$ and $\Lambda, \Sigma=1,...,28$ denotes the ${\rm SL}(8)$ antisymmetric index pairs $[ab]$. In terms of the mixed coset representative
\begin{align}
 \ma V_M{}^{\underline N}=(\ma S^{-1})_M{}^{\underline P}
\mas S _{\underline P}{}^{\underline N} \,,
\end{align}
the gauge kinetic function $N_{\Lambda\Sigma }=N_{(\Lambda \Sigma)}$ is defined by
\begin{align}
 \ma V^{\Sigma ij}N_{\Lambda\Sigma }=-\ma V_\Lambda{}^{ij}\,.
\end{align}
In the unitary gauge~(\ref{cosetrep_USP}), we immediately obtain
\begin{align}
{\rm i}N_{abcd}= \frac{1}{16} \tr \left[
\Gamma_{ab}\left(1+ \phi^\dagger \,{\rm tjh}(\sqrt{\phi\phi^\dagger})\right)
\left(1-\phi ^\dagger\,{\rm tjh}(\sqrt{\phi \phi^\dagger})\right)^{-1}\Gamma_{cd}
\right]\,,
\end{align}
where
\begin{align}
 {\rm tjh}(z)\equiv z^{-1} {\tanh (z)} \,.
\end{align}
Equation~(\ref{phi_ATfactorization}) yields
\begin{align}
{\rm i}N=
(1+\bar U \tanh \Lambda U^\dagger )
(1-\bar U \tanh \Lambda U^\dagger )^{-1}\,,
\end{align}

Since the hermitian matrix $\phi\phi^\dagger$ is positive and $\phi^\dagger (\phi\phi^\dagger)^{-1/2}$ is unitary, we have
\begin{align}
\label{}
| \tanh (\sqrt{\phi\phi^\dagger})|<1 \,.
\end{align}
Suppose that ${\rm det}[1-\phi^\dagger \tjh (\sqrt{\phi\phi^\dagger})]=0$ is satisfied, then there exists an eigenvector $v$ satisfying
\begin{align}
\label{}
\phi^\dagger \tjh (\sqrt{\phi\phi^\dagger}) v =v \,.
\end{align}
This  leads to the contradiction since
\begin{align}
v^\dagger  v=v^\dagger \tanh^2 (\sqrt{\phi\phi^\dagger}) v <v^\dagger v \,.
\end{align}
It follows that the gauge kinetic function $N_{\Lambda\Sigma}$ is always bounded locally.

Let us next discuss the dyonic gaugings. In this case,
we need an extra
Stueckelberg tensorial field $B_{\alpha\mu\nu }$ to compensate the
degrees of freedom for the magnetic gauge potentials~\cite{deWit:2007mt}.
This field can be eliminated by going back to the electric frame by the
symplectic rotation~\cite{deWit:2005ub}.
Let
\begin{align}
L_M{}^N=
\left(\begin{array}{cc}L_{\Lambda}{}^{\Sigma}&L_{\Lambda\Sigma} \\L^{\Lambda\Sigma}&L^{\Lambda}{}_{\Sigma}\end{array}\right)
\end{align}
denote the  desired ${\rm Sp}(56,\mathbb R)$ matrix for which
${\Theta'}_M{}^\alpha =L_M{}^N\Theta_{N}{}^\alpha ={}^{T\!}(\Theta'_{\Lambda}{}^\alpha , 0)$.
Under the symplectic rotation,
the gauge kinetic function $N_{\Lambda\Sigma}$ transforms into
\begin{align}
\label{gaugekin_tr}
N'_{\Lambda\Sigma}=(L_{\Lambda}{}^\Xi N_{\Xi\Gamma}-L_{\Lambda\Gamma})
(L^{\Sigma}{}_{\Gamma}-L^{\Sigma\Pi}N_{\Pi \Gamma})^{-1}\,.
\end{align}
In the ${\rm SL}(8)$-frame~(\ref{N8_CP_dyonic_Xtensors}), the no
magnetic-charge condition leads to
\begin{align}
 L^{abd[g}\theta_{d[e}\delta_{f]}{}^{h]}-
L^{ab}{}_{d[e}\xi^{d[g}\delta_{f]}{}^{h]} =0\,.
\end{align}
Assuming $\theta \xi =c \ne 0$, this equation can be solved to give
\begin{align}
 L^{ab}{}_{cd}=-\frac{1}{c} L^{abef}\theta_{ec}\theta_{df} \,.
\end{align}
Combined with the symplectic condition $L\Omega \,{}^{T\!}L=\Omega$,
we have
\begin{subequations}
\begin{align}
L&=\left(
\begin{array}{cc}
S\cdot \omega  & ~-c^{-1} \left\{({}^{T\!}\omega)^{-1}+S\cdot \omega \cdot (\theta \we \theta )\right\}\\
c\omega & -\omega \cdot (\theta \we \theta)
\end{array}
\right)  \,, \\
L^{-1} &=\left(
\begin{array}{cc}
 -(\theta \we \theta )\cdot {}^{T\!} \omega & ~c^{-1} \{(\omega )^{-1}+(\theta\we \theta )\cdot
{}^{T\!}\omega\cdot S\} \\
-c{}^{T\!}\omega & {}^{T\!}\omega \cdot S
\end{array}
\right)\,,
\end{align}
\end{subequations}
where $\omega^{\Lambda\Sigma}$ and $S_{\Lambda\Sigma}=S_{(\Lambda\Sigma)}$ are arbitrary $28\times 28$ matrices.  Here and in the following, we will use the notation $(S_{(1)}\we S_{(2)})_{abcd}=S_{(1)[a}{}^{[c}S_{(2)b]}{}^{d]}$ for the $8\times 8 $ tensors $S_{(1,2)}$. If we make the choices,
\begin{align}
\omega ={}^{T\!}\omega = -(\theta \we \theta )^{-1}\,, \qquad
 S=-({}^{T\!}\omega )^{-1}\cdot (\theta\we \theta )^{-1}\cdot \omega ^{-1} =\omega ^{-1}\,,
\end{align}
the gauge fields in the electric frame is
\begin{align}
\mas A' =A^{\prime ab}X'_{ab} = \left(
\begin{array}{cc}
(A'\cdot \theta) \we \mathbb I_8      &    \\
      &   -(A'\cdot \theta) \we \mathbb I_8
\end{array}
\right)\,.
\end{align}
The gauge kinetic function now reduces to
\begin{align}
\label{gaugekin_tr2}
N' =N [\mathbb I_{28} +c(\theta \we \theta)^{-1} N]^{-1} \,.
\end{align}
We will use this equation later when computing gauge kinetic functions in the dyonic gaugings.

\section{${\rm SO}(3)\times {\rm SO}(3)$ invariant sector of ${\rm SO}(4,4)$ gaugings}
\label{sec:SO44}

Although the prescription developed in the previous section can be used  in principle to obtain the potential for all seventy scalars with any types of consistent gaugings, this is not so practical as it stands. In particular, the most difficult and laborious task in this program is how to obtain the matrix $U$ in the Autonne-Takagi factorization (\ref{phi_ATfactorization}), since it admits a large extent of non-uniqueness.  In this section, we will restrict the consideration to a subsector of $E_{7(7)}/{\rm SU}(8)$ coset space in such a way that one can circumvent this complexity.  We will illustrate this by the ${\rm SO}(3)\times {\rm SO}(3)$-invariant sector of ${\rm SO}(4,4)$ gaugings in this section. Still, this system is quite rich and  yields physically interesting dynamics of scalar fields.

In the $\SL(8)$ gauge-field frame of the $\SO(4,4)$ gauging, the embedding tensor takes the form
\begin{align}
 \theta =\mathbb I_4\oplus (-\mathbb I_4) \,, \qquad
 \xi =c[\mathbb I_4\oplus (-\mathbb I_4)  ] \,,
\end{align}
Here $c$ is the deformation parameter. According to the analysis given in~\cite{Dall'Agata:2012bb,Dall'Agata:2012sx}, $c$ can be put into the range $c\in [0,1]$ due to the automorphisms of the theory.

Among the 70 generators acting translationally on the scalar manifold,  the following six exhaust the ${\rm SO}(3)\times {\rm SO}(3)$ invariant generators:
\Eqrsub{
({\bf 1},{\bf 1}) &:&  g_1 = S_1 \we \mathbb I_8 ; \qquad S_1 =(\mathbb I_4, -\mathbb I_4)\,,\\
 ({\bf 1},{\bf 1})&:& g_2  = \frac i2 (e_{1234}+e_{5678}) \,, \\
({\bf 9},{\bf 1}) &:& g _3 = S_3 \we \mathbb I_8 ; \qquad S_3=(\mathbb I_3,-3, \mathbb O_4) \,, \\
({\bf 1},{\bf 9}) &:& g _4 = S_4 \we \mathbb I_8 ; \qquad S_4=(\mathbb O_4, \mathbb I_3,-3) \,, \\
({\bf 4},{\bf 4}) &:& g_5 = S_5 \we \mathbb I_8 ; \qquad S_5=e_{4}\circ e_8+e_8\circ e_4  \,, \\
({\bf 4},{\bf 4}) &:& g_6 = \frac i2 (e_{1238}+e_{4567}) \,,
\label{SO44_SO3SO3generators}
}
where the boldface letters were intended to denote representations of residual gauge symmetry ${\rm SO}(4)\times{\rm SO}(4)$, and we have used the abbreviation $e_{abcd}=e_a\we e_b\we e_c \we e_d$.  Here, it is understood that each $g_i$ is used as $\phi_{ijkl}$ in \eqref{cosetrep_USP} through \eqref{phibyS&U} to define a generator in the Lie algebra. Our generators are related to those of Dall'Agata-Inverso~\cite{Dall'Agata:2012sx}
by
\begin{align}
& g_1^{\rm DI} =g_2\,,\quad
g_1^{\rm DI} =g_2\,, \quad
g_3^{\rm DI}=\frac 12 (g_1+g_3-g_4) \,, \quad
g_4 ^{\rm DI}=g_5 \,, \quad
 g_5 ^{\rm DI}=g_3-g_4 \,, \quad g_6 ^{\rm DI}= g_6 \,.
\end{align}
In the next two subsections, we shall limit ourselves to further subsectors to illustrate the diagonalization method.
We derive an analytic expression of the potential for each case and discuss the structure of critical points in detail. The full analysis will be done subsequently.

\subsection{${\rm SO}(4)\times {\rm SO}(4)$ invariant subsector}
\label{sec:SO4SO4}

Let us start with the simplest example, the ${\rm SO}(4)\times {\rm SO}(4)$ invariant subsector, which can be treated without using the method explained above and provides an example to be used to check the result for a larger sector obtained by the above method. In this sector, only the generators $g_1$ and $g_2$ survive. Let us denote the scalar fields in this subsector as
\begin{align}
 \phi \simeq x_1 g_1 + y_1 g_2 \,,
\end{align}
where $x_1$ and $ y_1$ are the ${\rm SO}(4)\times {\rm SO}(4)$ singlet scalar and pseudoscalar, respectively.

Since the matrix $\phi$ can be regarded as a map from $\mathbb R^8 \we \mathbb R^8 \simeq \mathbb R^{28}$ onto itself,
it is convenient here to decompose the indices as
\begin{align}
\mathbb R^8 &\simeq V_{1234}\oplus V_{5678}\,, \nonumber \\
\mathbb R^8 \we \mathbb R^8 &\simeq [V_{1234}\we V_{1234}]
\oplus [V_{5678}\we V_{5678}]\oplus [V_{1234}\we V_{5678}]
\simeq \mathbb R^6\oplus \mathbb R^6 \oplus \mathbb R^{16} \,,
\end{align}
where $V_{1234}$ is the vector space spanned by $1-4$ indices and similarly for $V_{5678}$.
It follows that the generators $g_{1,2}$ can be decomposed into each block as
\begin{align}
g_1 \simeq \mathbb I_6  \oplus (-\mathbb I_6)\oplus
\mathbb O_{16} \,, \qquad
g_2 \simeq \star_4 \oplus \star_4 \oplus \mathbb O_{16} \,,
\end{align}
where $\star_4$ is the map $[\star_4 X]_{ab}=\frac 12\epsilon_{abcd}X_{cd}$ onto $X_{ab}=X_{[ab]}\in \mathbb R^6$, thence
\begin{align}
 \phi \simeq (x_1+i y_1\star_4) \mathbb I_6 \oplus
(-x_1+i y_1 \star _4) \mathbb I_6 \oplus \mathbb O_{16} \,, \qquad
\phi \phi ^\dagger \simeq r_1^2 [\mathbb I_6\oplus \mathbb
I_6 \oplus \mathbb O_{16}] \,.
\end{align}
Here we have defined $x_1+i y_1=r_1e^{i\theta_1}$.
Instead of computing the Autonne-Takagi factorization~(\ref{AT2}),  it is easier in the present case to directly calculate the matrix $M_{MN}$ by using the above matrix representation for $\phi$. First, the coset representative $\mas S$ decomposes as $\mas S \simeq\mas S_1\otimes\mathbb I_6\oplus\mas S_2\otimes\mathbb I_6\oplus \mathbb I_{32}$, where
\begin{align}
 \mas S_1 =\left(
\begin{array}{cc}
 \cosh r_1 & \hat z{\jh} r_1 \\
\hat z^* {\jh}r_1 & \cosh r_1
\end{array}
\right)\,, \qquad
\mas S_2 =\mas S_1({x_1\to -x_1}) \,.
\end{align}
with
\begin{align}
 \hat z_{abcd} \equiv x_1
{\delta_{[a}}^c{\delta_{b]}}^d +\frac 12 {i}y_1 \epsilon_{abcd} \,.
\end{align}
This yields $M\simeq M_1\otimes\mathbb I_6\oplus M_2\otimes\mathbb I_6\oplus \mathbb I_{32}$, where
\begin{align}
 M_1 &=\left(
\begin{array}{cc}
 \cosh (2r_1)+2x_1{\jh}(2r_1)& 2y_1{\jh}(2r_1)\star_4  \\
2y_1 {\jh}(2 r_1) \star_4 & \cosh (2r_1) -2 x_1 {\jh}(2r_1)
\end{array}
\right) \,,  \qquad
 M_2 =M_1 (x_1\to -x_1)\,.
\end{align}
Putting all together,
the potential evaluates to
\begin{align}
 V=\frac{g^2}{4}(1+c^2) [2-\cosh(2r_1)] \,.
\end{align}
The kinetic term is independent of the deformation parameter and
described by the ${\rm SL}(2,\mathbb R)/{\rm SO}(2)$ coset space,
\begin{align}
 d s_T^2 =-\frac{1}{96} \tr (d M d M^{-1})
= d r_1^2 +\frac{1}{4}\sinh^2(2 r_1) d \theta_1^2\,.
\end{align}
It follows that the potential is constant in the angular direction, i.e, $\theta_1$
is a modulus, and there is no critical point other than the origin.
Both of the ${\rm SO}(4)\times {\rm SO}(4)$ invariant scalars $(x_1, y_1)$ have the mass spectrum $-2$
(in a unit of cosmological constant) at the origin, and immediately fall into the bottomless valley.
This behavior is an obvious difficulty in attempting to realize inflation around the origin.

Let us next compute the gauge coupling function.
We first focus upon the case for the electric gaugings ($c=0$), and
subsequently upon the dyonic gaugings by the symplectic rotation~\cite{deWit:2005ub}.
By virtue of  $\phi \simeq (x_1+i y_1 \star_4) \mathbb I_6\oplus (-x_1+{i}y_1 \star_4) \mathbb I_6\oplus \mathbb O_{16}$,
we have
\begin{align}
1\pm \phi^\dagger \,{\rm tjh}(\sqrt{\phi\phi^\dagger})& =[1\pm (x-{\rm i}y \star_4 )\,{\rm
 tjh}r_1]\otimes \mathbb I_6 \oplus [1\mp (x_1+{i}y_1\star_4) \,{\rm tjh}r_1]\otimes \mathbb I_6
 \oplus \mathbb I_{16}
 \,, \nonumber \\
\left[1+ (\mp x_1+ iy_1 \star_4)\,{\rm tjh}r_1\right]^{-1} &=\frac{\cosh^2
 r_1}{\cosh(2r_1)\mp (x_1/r_1)\sinh (2r_1)}\left[1+ (\mp x_1 -{i}y_1 \star_4)\,{\rm tjh}r_1\right]\,.
\end{align}
Hence
\begin{align}
 &\left[1+ \phi^\dagger \,{\rm tjh}(\sqrt{\phi\phi^\dagger})\right]
\left[1-\phi^\dagger \,{\rm tjh}(\sqrt{\phi \phi^\dagger})\right]^{-1}
\nonumber \\
& \qquad \simeq \frac{1-{i}(y_1/r_1) \sinh(2r_1)\star_4 }{f_-}\otimes\mathbb I_6 \oplus \frac{1-{\rm i}
(y_1/r_1) \sinh(2r_1)\star_4 }{f_+}\otimes\mathbb I_6
\oplus \mathbb I_{16} \,,
\end{align}
where
\begin{align}
  f_\pm \equiv &
\cosh(2r_1) \pm \frac{x_1}{r_1} \sinh (2r_1)\,.
\end{align}
It follows that
\begin{align}
e^{-1}\ma L_F =& -\frac{1}{4} \tr (F_{12} \cdot F_{12})
-\frac{1}{4}\left[\frac{1}{f_-}\tr (F_1 \cdot F_1)+\frac{1}{f_+}\tr (F_2
 \cdot F_2)\right]\nonumber \\ &+\frac{y_1}{4r_1}\sinh(2r_1)
 \left[\frac{1}{f_-} \tr (F_1 *\ti F_1)
+\frac{1}{f_+} \tr(F_2 * \ti F_2)
\right]\,,
\end{align}
where $\ti F_A{}^{ab}\equiv (1/2)\epsilon_{abcd}F_A{}^{cd}$  ($A=1,2$) and the alternate tensor $\epsilon_{abcd}$ in
$\mathbb R^6\simeq \mathbb R^4 \we \mathbb R^4 \mapsto \mathbb R^6$
should not be confused with the spacetime volume weight, for which the ordinary Hodge dual of the 2-form $F_{\mu\nu}$ is denoted by $*F_{\mu\nu}=\frac 12 \epsilon_{\mu\nu\rho\sigma}F^{\rho\sigma}$.

The term $F_{12}$ corresponds to the broken gauge symmetry, which will be massive by absorbing the Nambu-Goldstone bosons. Hence, we shall focus on $F_1$ and $F_2$, which are ${\rm SO}(4)$ gauge field strengths. Splitting these into the chiral components via the decomposition ${\rm SO}(4)\simeq {\rm SU}(2)\times {\rm SU}(2)$ as
\begin{align}
\label{}
F_{A(1)}{}^\pm =F_A{}^{12}\pm F_A{}^{34} \,, \qquad
F_{A(2)}{}^\pm =F_A{}^{13}\pm F_A{}^{42} \,, \qquad
F_{A(3)}{}^\pm =F_A{}^{14}\pm F_A{}^{23} \,,
\end{align}
we find that gauge fields with different chirality decouple as
\begin{align}
F_A{}^{ab} \cdot F_A{}^{ab}&=\sum_{i=1}^3 (F_{A(i)}{}^+ \cdot F_{A(i)}{}^++
F_{A(i)}{}^- \cdot F_{A(i)}{}^-) \,, \nonumber \\
F_A{}^{ab} \cdot *\ti F_A{}^{ab}&=\sum_{i=1}^3 (F_{A(i)}{}^+ \cdot * F_{A(i)}{}^+-
F_{A(i)}{}^- \cdot * F_{A(i)}{}^-) \,.
\end{align}

Let us move next to the case of dyonic gaugings. Following the prescription discussed in section~\ref{sec:gaugekin},
we can eliminate the tensorial subsidiary field by going to the electric frame by a symplectic rotation.
In this electric frame, we have
\begin{align}
 N'=\frac{N_-}{1-icN_-}\otimes \mathbb I_6 \oplus
\frac{N_+}{1-icN_+}\otimes \mathbb I_6 \oplus
\left(\frac{-i}{1+ic }\right)\otimes \mathbb I_{16}\,,
\end{align}
where
\begin{align}
 i N_\pm =\frac{1-i \sin\theta_1 \sinh (2r_1) \star _4}{f_\pm } \,.
\end{align}

\subsection{$\mathbb Z_2$ truncation}
\label{sec:SO3SO3Z2}

Let us next discuss the $\mathbb Z_2$ invariant subsector studied in~\cite{Dall'Agata:2012sx}, which we denote by $\Sigma_2$.
In this case, the generators except for $g_3-g_4$ and $g_6$ are odd under the reflection, hence the surviving scalars are described by
\begin{align}
 \phi \simeq x_2(g_3-g_4)+  y_2 g_6 \,.
\end{align}
It is therefore advantageous to decompose the indices as
\begin{align}
 \mathbb R^8 \simeq & V_{123}\oplus V_4 \oplus
V_{567} \oplus V_8 \,, \nonumber \\
\mathbb  R^8 \we \mathbb  R^8 \simeq & [(V_{123} \we
 V_{123})\oplus
(V_{123}\we V_8)]\oplus [(V_{567}\we
 V_{567})\oplus (V_{567} \we V_4)]\nonumber \\
&\oplus
[(V_{123}\we V_{567})\oplus (V_{123} \we V_4)]
\oplus
[(V_{567}\we V_{8})\oplus (V_{4} \we V_8)]\,.
\label{N8_tr_SO42_SO32sec_2form_dec}
\end{align}
The scalar field $\phi$ is then expressed as
\begin{align}
 \phi \simeq  (x_2\sigma_3 +{\rm i}y_2 \sigma_1 )\otimes \mathbb I_3
\oplus (x_2\sigma_3 -{\rm i}y_2  \sigma_1) \otimes \mathbb I_3 \oplus [x_2\mathbb I_9, -x_2 \mathbb
 I_6, -3 x_2 ] \,.
\end{align}
Since the point at issue is now reduced to the $2\times 2$ eigenvalue problem,
the Autonne-Takagi factorization~(\ref{AT2}) can be done easily,
\begin{align}
\left(
\begin{array}{cc}
x _2     & i y_2    \\
iy_2      & -x_2
\end{array}
\right) = U \left(
\begin{array}{cc}
   x_2-y_2   &    \\
      & x_2+y _2
\end{array}
\right) \,^{T\!} U \,,\qquad
U=\frac 1{\sqrt 2} \left(
\begin{array}{cc}
1      & -1   \\
-i      & -i
\end{array}
\right) \,.
\end{align}
Plugging these results into equations presented in section~\ref{sec:coset},
the potential is then given by
\begin{align}
 V= \frac{g^2}{64}e^{6x_2+2y_2}
&\left[(1+c^2e^{-12x_2})(1+e^{-2y_2})^2+3e^{-4x_2}\{
2(1-e^{-2y_2})^2(1+c^2 e^{-4x_2})\right. \nonumber \\
& \left. -(c^2+e^{-4x_2})
(1-6e^{-2y_2}+e^{-4y_2})\}\right]\,.
\label{DIpot}
\end{align}
This recovers the potential in~\cite{Dall'Agata:2012sx}
after the substitution
$x_2 \to -\frac 14 \ln x$,
$y_2 \to -\frac{1}{2}\ln \tau $ with $g \to 2\sqrt 2$.
The target space metric~(\ref{kinetic}) is $\mathbb R^2$ and reads
\begin{align}
 d s_T^2 =3 d x_2^2 + d y_2^2 \,.
\end{align}
As discussed in~\cite{Dall'Agata:2012sx},
the potential (\ref{DIpot}) has an off-center critical point for which
the mass spectrum hinges on the deformation parameter $c$, and can be
small and negative.

The gauge kinetic function can be evaluated following the argument given in section~\ref{sec:gaugekin}.
In the electric gaugings, we have
\begin{align}
N \simeq N_+\otimes \mathbb I_3 \oplus  N_-\otimes \mathbb I_3 \oplus
[e^{2x_2}\mathbb I_9,e^{-2x_2}\mathbb I_6,e^{-6x_2}] \,,
\end{align}
where
\begin{align}
N_\pm =\left(
\begin{array}{cc }
-i e^{2x_2}{\rm sech}(2y_2)   & \mp\tanh (2y_2)    \\
  \mp \tanh(2y_2)    &    -i e^{-2x_2}{\rm sech}(2y_2)
\end{array}
\right) \,.
\end{align}
In the dyonic gaugings, the symplectic rotation to the electric frame~(\ref{gaugekin_tr}) allows us to recast the result into
\begin{align}
N'\simeq N_+'\otimes \mathbb I_3 \oplus  N_-'\otimes \mathbb I_3 \oplus
[\ti f_+(2x_2)\mathbb I_9,\ti f_-(-2x_2)\mathbb I_6, \ti f_+(-6x_2)] \,,
\end{align}
where $\ti f_\pm(x)=-i e^{x}/(1\pm c i e^x)$ and
\begin{align}
N_\pm ' &=\frac{1}{1+c^2-2i c \sinh (2x_2){\rm sech}(2y_2)} \left(
\begin{array}{cc}
c\mp i e^{2x_2}{\rm sech} (2y_2)   &  \mp \tanh (2y_2)   \\
\mp \tanh (2y_2)       &   -c\mp i e^{-2x_2} {\rm sech}(2y_2)
\end{array}
\right)\,.
\end{align}
The gauge kinetic function is bounded in either case.

\subsection{Full analysis}

After having described the two subsectors in detail, we now come to the analysis of the full
${\rm SO}(3)\times {\rm SO}(3)$-invariant sector. The decomposition of indices is as follows:
\begin{align}
 \mathbb R^8 \simeq &  V_{123}\oplus   V_4 \oplus
  V_{567} \oplus   V_8 \,, \nonumber \\
\mathbb  R^8 \we \mathbb  R^8 \simeq & [(V_{123} \we
 V_{123})\oplus
(V_{123}\we V_4)
\oplus (V_{123}\we V_8)
] \nonumber \\
&
\oplus [(V_{567}\we
V _{567})\oplus (V_{567} \we V_8)\oplus {(V_{4} \we V_{567})}]
\oplus
[(V_{4} \we V_8)\oplus (V_{123}\we V_{567})]\,.
\end{align}
Then the generators $g_{1-6}$ are reduced to
\begin{subequations}
\begin{align}
g_1= & \Mat{ \mathbb I_2 & \\ & 0 } \otimes \mathbb I_3
     \oplus
       \Mat{ -\mathbb I_2 & \\ & 0} \otimes \mathbb I_3
     \oplus
      \mathbb O_{10}\,, \\
g_2 = &  \Mat{ i\sigma_1 & \\ & 0 } \otimes \mathbb I_3
      \oplus
       \Mat{ i\sigma _1 & \\ &  0 } \otimes \mathbb I_3
      \oplus
      \mathbb O_{10}\,, \\
g_3+g_4 =&  \left(
\begin{array}{cc}
1 & \\
& -\mathbb I_2
\end{array}
\right)\otimes \mathbb I_3\oplus
 \left(
\begin{array}{cc}
1 & \\
& -\mathbb I_2
\end{array}
\right)\otimes \mathbb I_3 \oplus
\left(
\begin{array}{cc}
 -3     &    \\
      &   \mathbb I_9
\end{array}
\right)  \,, \\
g_3-g_4=&  \left(
\begin{array}{cc}
\sigma_3 & \\
&  2
\end{array}
\right)\otimes \mathbb I_3 \oplus  \left(
\begin{array}{cc}
-\sigma_3 & \\
&  -2
\end{array}
\right)\otimes \mathbb I_3 \oplus \mathbb O_{10} \,,\\
g_5= & \left(
\begin{array}{cc}
0 & \\
&\frac 12 \sigma_1
\end{array}
\right)\otimes \mathbb I_3 \oplus  \left(
\begin{array}{cc}
0 & \\
& \frac 12 \sigma_1
\end{array}
\right)\otimes \mathbb I_3  \oplus {\mathbb O_{10}} \,, \\
g_6 = &  \Mat{ & & i\\ &0 & \\ i & & }\otimes \mathbb I_3
   \oplus
      \Mat{  & &-i\\ &0 & \\ -i & &} \otimes \mathbb I_3
   \oplus
   \mathbb O_{10} \,.
\end{align}
\end{subequations}
Let us parametrize the scalars as
\begin{align}
 \phi =& x_1 g_1  + y_1 g_2 +x_2 (g_3+g_4)+ y_2 g_6 +w_1 (g_3-g_4)+2
 w_2 g_5 \,,\label{so44_scalar_full}
 \end{align}
which can be decomposed into
\begin{align}
 \phi = \Phi_1 \otimes \mathbb I_3 \oplus \Phi_2 \otimes \mathbb I_3 \oplus (-3 x_2) \oplus
 (x_2 \mathbb I_9 ) \,,
\end{align}
where
\begin{subequations}
\begin{align}
 \Phi_1& = \left(
\begin{array}{ccc}
x_1+x_2+w_1 & i y_1 & i y_2  \\
i y_1 & x_1-x_2-w_1 & w_2 \\
i y_2 & w_2 & -x_2 +2 w_1
\end{array}
\right) \,, \\
 \Phi_2& = \left(
\begin{array}{ccc}
-x_1+x_2-w_1 &i y_1 & i y_2 \\
i y_1 & -x_1-x_2 +w_1 & -w_2 \\\
i y_2 & -w_2 & -x_2 -2 w_1
\end{array}
\right) \,.
\end{align}
\end{subequations}
The kinetic-term metric around the origin $O$ is given by
\begin{align}
 d s_T^2 |_{O}=\frac{1}{12}d \phi_{ijkl}d \phi^{ijkl}
=3(d w_1^2+d x_2^2)+d w_2^2+d x_1^2+d y_1^2+d y_2^2\,.
\label{SO44_kin_orig}
\end{align}

All we need to do is to perform the Autonne-Takagi factorization of $\Phi_{1,2}$.
This is yet actually redundant, since these $3\times 3$ matrices are related in a simple way as follows
\begin{align}
 \Phi_1 =x_2 I_{1,2}+\Phi_0 \,, \qquad
\Phi_2 =x_2 I_{1,2}-\bar \Phi_0 \,, \qquad
\Phi_0= \left(
\begin{array}{ccc}
x_1+w_1 &i y_1 & i y_2  \\
i y_1 & x_1-w_1 & w_2 \\
i y_2 & w_2 &  2 w_1
\end{array}
\right) \,,
\label{Phi0_SO44}
\end{align}
with $I_{1,2}={\rm diag}(1,-1,-1)$. It therefore suffices to compute
\begin{align}
 \Phi_0  =U\Delta\,^{T\!}U \,, \qquad U\in {\rm U}(3) \,, \qquad
 \Delta ={\rm diag}(\mu_i, \mu_2, \mu_3) \,.
 \label{SO44_Phi0}
\end{align}
In Appendix \ref{sec:ATF}, we describe how to implement the Autonne-Takagi factorization of the
matrix $\Phi_0$ of the form (\ref{Phi0_SO44}).
The eigenvalues $\mu_i$ can be obtained by the procedure laid out in appendix~\ref{sec:ATF}
and turn out to be the roots of the cubic equation $Q(\mu_i)=0$ given in (\ref{Qmueq}). In the present case, we have
\begin{align}
 Q(\mu )\equiv &  \mu^3-(x_1^2+y_1^2+y_2^2+3w_1^2+w_2^2)\mu \nonumber\\&
+2w_1^3+w_1(-2x_1^2-2y_1^2+y_2^2+w_2^2)+x_1(w_2^2-y_2^2)+2w_2y_1y_2 \,.
\end{align}
One sees that the one of the roots is not independent on account of the constraint $\sum_i \mu_i=0$.
The eigenvector $\vec v_i$~(\ref{N8_SO44_AT_vec}) takes the form
\begin{subequations}
\begin{align}
\vec v_i =\left(
\begin{array}{c}
\mu_i^2+(x_1+w_1)\mu_i +2 w_1(x_1-w_1)-w_2^2 \\
-(\mu_i+2w_1)y_1+w_2 y_2  \\
-(\mu_i+x_1-w_1)y_2+w_2 y_1
\end{array}
\right)\,.
\end{align}
\end{subequations}
As demonstrated in detail in appendix~\ref{sec:ATF}, the unitary matrix $U$
can be parametrized by ${\rm SU}(2)$ Euler angles $\theta_i$. Hence
we can view ($x_2, \mu_i, \theta_i$) under  a constraint $\sum_i\mu_i=0$
as new variables describing six scalar degrees of freedom.
We give in Eq.~(\ref{CartesianByPolar}) the expressions of the original coordinates
($x_1, x_2, y_1, y_2, w_1, w_2$) in terms of
the new variables ($x_2, \mu_i, \theta_i$).

After some tedious computations, we obtain the expression for the potential,
\begin{align}
g^{-2}V=& \left(e^{6x_2}+c^2 e^{-6x_2}\right)\bigl[
2C_{321}\cosh(2\mu_1-2\mu_2) +2C_{312}\cosh(2\mu_3-2\mu_1)
\nonumber \\ & \qquad \qquad
\qquad \qquad \quad +2C_{31-1}\cosh(2\mu_2-2\mu_3)
+C_{300}
\bigr]
\nonumber \\
&+ \left( e^{-2x_2}+c^2 e^{2x_2}\right)\bigl[
2 C_{111} \cosh(2\mu_1)+2 C_{110}\cosh(2\mu_2) +2 C_{101} \cosh(2\mu_3)+C_{100}
\bigr]
\nonumber \\
&+2 \left(e^{2x_2}+c^2e^{-2x_2}\right)
\bigl[C_{111}'\cosh(2\mu_1)+C_{110}'\cosh(2\mu_2)+C_{101}'\cosh(2\mu_3)\bigr]\,,
\label{SO44_fullpot}
\end{align}
where
\begin{subequations}
\begin{align}
C_{300}&=\frac{1}{64}\left[3+p^4-q^4\cos(4\theta_2 )\right]
-\frac{1}{512} \left[(1-p)^2 \cos(2\theta_2-2\theta_3
 )-(1+p)^2\cos(2\theta_2+2\theta_3 )\right]^2\,,
\\
C_{31-1}&=\frac{1}{1024} \left[(1-p)^2 \cos(2\theta_2-2\theta_3
 )-(1+p)^2\cos(2\theta_2+2\theta_3 )\right]^2  \,, \\
C_{321}&= \frac{1}{64}q^2 (p\sin2\theta_2 \cos\theta_3+\cos2\theta_2 \sin\theta_3
 )^2 \,, \\
C_{312}&= C_{321}(2\theta_3\to 2 \theta_3+\pi )\,, \\
C_{100}&=\frac{9}{32} \,, \\
C_{111}&=-\frac{3}{64} \cos^2\theta_1 \,, \\
C_{110}&=-\frac{3}{64}\sin^2\theta_1 \cos^2\theta_3 \,, \\
C_{101}&= -\frac 3{64} \sin^2\theta_1 \sin^2\theta_3 \,, \\
C_{111}'& = -\frac{3}{32} q^2\cos(2\theta_2 )\,, \\
C_{110}'&= \frac{3}{128} \left[2q^2 \cos(2\theta_2 )-(1-p)^2
 \cos(2\theta_2-2\theta_3 )-(1+p)^2 \cos(2\theta_2+2\theta_3 )\right]\,, \\
C_{101}'&=  C_{110}'(2\theta_3 \to 2\theta_3 +\pi) \,,
\end{align}
\end{subequations}
with $p=\cos\theta_1 $ and $q=\sin\theta_1 $.  Note that we have used the condition $\sum_i\mu_i=0$.
At the origin of the scalar manifold ($\mu_i=x_2=0$), one can get the value $V_0=(g^2/4)(1+c^2)$, as expected.

In order to recover the potential discussed in the previous subsections,
we need to be careful since some of $C_i^{-1}$ vanishes in those truncations.
Consequently we need to keep track
of higher-order terms in the computation of $V$ and truncate
irrelevant scalars at the final step. With this careful procedure
we are able to reproduce the correct results for the potential value and the mass spectrum
at the origin~\cite{KN}.

Under the constraint $\sum_i \mu_i=0$, the scalar manifold metric (\ref{kinetic}) is also expressed in terms of Euler angles as\footnote{
The six-dimensional scalar manifold $\ma M_T$ (\ref{SO44target}) has a product structure $\ma M_T \simeq \mathbb R\times \ma M_5$, where $\ma M_5$ is an Einstein space of negative curvature.
Note that $\ma M_T$ itself is a symmetric space where the Riemann tensor is covariantly constant.
}
\begin{align}
d s_T^2 =&3d x_2^2+\frac 12 \sum_{i=1}^3 d \mu_i^2
+\sinh^2(\mu_2-\mu_3)\chi_1^2
+\sinh^2(\mu_3-\mu_1)\chi_2^2
+\sinh^2(\mu_1-\mu_2)\chi_3^2\,,
\label{SO44target}
\end{align}
where $\chi_i$ is an ${\rm SU}(2)$ invariant form
\begin{align}
\label{}
\chi_1&=d \theta_3+\cos\theta_1 d \theta_2 \,, \nonumber \\
\chi_2&=-\sin\theta_3 d \theta_1 +\cos\theta_3\sin\theta_1 d \theta_2 \,, \\
\chi_3&= \cos\theta_3 d \theta_1 +\sin\theta_3 \sin \theta_1 d \theta_2 \,.\nonumber
\end{align}
Expanding at the origin, one recovers~(\ref{SO44_kin_orig}) as desired.
In the above coordinate system, the target space~(\ref{SO44target})
has a manifest invariance $\mathbb R\times{\rm SU}(2)$.

\subsection{Truncation of the gauge sector}
\label{subsec:gauge_sector}

If a generic component of the $\SO(4,4)$ gauge field does not vanish, the dynamics of the scalar fields
fail to be closed inside the $\SO(3)\times\SO(3)$-invariant subsector $\exp(\H_6)$, where $\H_6=\EXP{g_1,\cdots,g_6}$ is the six-dimensional $\SO(3)\times\SO(3)$-invariant subspace of the tangent space of the scalar manifold at the origin. Let us find the restriction on the gauge field configuration for the dynamics to be closed within $\exp(\H_6)$.

Let $\Lambda$ denote an element of $\mathfrak{e}_{7(7)}$ and $\ti \Lambda$
be the corresponding Usp representation, i.e.,
\Eq{
e^\Lambda \in E_{7(7)}\,,  \qquad  \ti \Lambda=\T_0 \Lambda \T_0^\dagger \in \rho_\Usp (\mf{e}_{7(7)}) \,.
}
For a generic transformation generated by $\Lambda$,
the mixed coset representative $\ma V$ transforms as
\Eq{
\ma V= \T_0^\dagger \SS \mapsto e^\Lambda \T_0^\dagger \SS = \T_0^\dagger e^{\ti \Lambda} \SS =\T_0^\dagger \SS' \U,
}
where $\U\in \SU(8)$, $\SS, \SS' \in H(64)\cap \Usp(56)$ with $H(n)$ being the set of hermitian matrices of degree $n$.  Let us express $\SS$ in terms of the exponential mapping as
\Eq{
\SS=e^X = 1 + X + \frac12 X^2 + \cdots\,, \quad X \in \LL(H(64)\cap \Usp(56)) \,.
}
It follows that the symmetric part of the matrix $e^{\ti \Lambda}\SS$ can be expanded as
\Eq{
(e^{\ti \Lambda}\SS)_{\rm sym} = 1+ X + \ti \Lambda_+ + \frac12 [\ti \Lambda_-,X] + \{X, \ti \Lambda_+\} + \frac 12 \ti \Lambda_-^2 +\cdots \,,
}
where $\ti \Lambda_\pm \equiv(\ti \Lambda\pm \ti \Lambda^\dagger)/2$.
Keeping the terms that are leading order with respect to $X$ and linear in $\ti\Lambda_\pm$, the tangent vector $X$ at the origin transforms under the infinitesimal $E_{7(7)}$  transformation $\ti \Lambda$ as
\Eqr{
 \delta X= \ti \Lambda_+ + \frac12 [\ti \Lambda_-,X] \,.
}
In particular, for $e^\Lambda\in \SL(8)$, we have
\Eqr{
&&\ti \Lambda=\Mat{A\w 1 & S\w 1 \\ S\w 1 & A\w 1}\in M(56)\,, \quad
A=-\Tp{A} , \quad S=\Tp{S} \in M(8,\RF) ,\notag\\
&& \ti \Lambda_+ = \Mat{0 & S\w1 \\ S\w 1 & 0},\quad
   \ti \Lambda_- =\Mat {A\w 1 & 0 \\ 0 & A\w 1}.
}
For $\Lambda\in \so(4,4)$, the $8\times 8$ matrices $A$ and $B$ take the
following $4\times 4$ forms
\Eq{
A=\Mat{A_1 &0 \\ 0 & A_2},\quad  S=\Mat{0 & B \\ \Tp{B} & 0}\,,
}
where $A_{1,2}$ are anti-symmetric matrices and $B$ is an arbitrary $4\times 4$ matrix.

Specializing this transformation formula to the $X=0$ case, we find that $\ti \Lambda_+ \in \H_6$ is required for $e^\Lambda$ to preserve $\H_6$, hence $\ti\Lambda_+ \propto g_5$. Further, $\H_6$ exhausts all the $\SO(3)\times\SO(3)$-invariant vectors in the tangent space at the origin of the scalar manifold, and $\SO(3)\times\SO(3)$ is a maximal proper subgroup of $\SO(4)\times\SO(4)$ containing $\ti\Lambda_-$. Hence, the set of all possible $\ti \Lambda_-$ is identical to $\so(3)\oplus\so(3)$. Thus, we have found that the gauge fields should be restricted to the sum of the two $\SO(3)$ gauge fields and the Abelian gauge field $A^{[4,8]}$ corresponding the generator $g_5$.

Among these gauge fields, the $\SO(3)$ gauge fields do not couple to the scalar fields in the $\H_6$ sector. Hence, the consistency is trivial for them. We do not consider the Abelian gauge field $A^{[4,8]}$ corresponding to the Goldstone direction because it has to vanish in the spatially homogeneous and isotropic universe.

Thus, we only have to add the gauge kinetic terms to the Lagrangian for the two $\SO(3)$ gauge fields
\Eqrsub{
&& A^{(+)I}=\frac12\epsilon_{IJK}A^{[JK]},\\
&& A^{(-)I}=\frac12\epsilon_{IJK}A^{[(4+J)(4+K)]},
}
where $I$, $J$ and $K$ run over $1$, $2$ and $3$. The corresponding field strengths can be written
\Eq{
F^{(\pm)I}= dA^{(\pm)I} \pm \frac{g}{2}\epsilon_{IJK} A^{(\pm)J}\w A^{(\pm)K}.
}
The gauge kinetic term is expressed in terms of these fields as
\Eq{
\kappa^2 e^{-1}\LL_1= \frac14\sum_{\epsilon=\pm}
\insbra{\Im(N_{\epsilon}) F^{(\epsilon)}\cdot F^{(\epsilon)}
 - \Re(N_{\epsilon}) F^{(\epsilon)}\cdot \ti F^{(\epsilon)}}.
\label{SO44:SO3xSO3:GaugeKineticTerm}
 }
From the general formulae in section \ref{sec:gaugekin}, we can find explicit expressions for the gauge coupling functions $N_{\epsilon}$ as given in  Appendix \ref{sec:GaugeCouplingFunctions}.

\subsection{Critical points of the potential}
\label{subsec:SO44:extremal}

Using the explicit expression of the potential (\ref{SO44_fullpot}) in terms of the Euler angles, we now move on to finding the critical points of the potential.  To this end, the ``go to the origin'' approach by the ${\rm SL}(8)$ transformation \cite{DI,Dibitetto:2011gm} does not work by the following reason. The orbit of ${\rm SL}(8)$ is 35 dimensional, which only covers the half of the coset space $E_{7(7)}/{\rm SU}(8)$. The overlapping orbit of ${\rm SL}(8)$ and the ${\rm SO}(3)\times{\rm SO}(3)$ invariant sector consists of the four dimensional space corresponding to $y_1=y_2=0$. It follows that the critical point with $(y_1, y_2)\ne (0,0)$ cannot be mapped to the origin by any ${\rm SL}(8)$ transformations. On account of this fact, we are forced to scan the critical points numerically. Unfortunately, the exact critical points we found are only the origin and the one found by Dall'Agata-Inverso~\cite{Dall'Agata:2012sx}.

It is enlightening to derive the mass spectrum at the DI critical point within our formalism,
although this has been already addressed in~\cite{Dall'Agata:2012sx}.  For this purpose, we define
\Eq{
y\equiv  \frac 12 (\mu_2-\mu_3) \,, \qquad
z\equiv \frac 12 (\mu_2+\mu_3)\,.
}
It then follows that the DI critical point~\cite{Dall'Agata:2012sx} corresponds to
\Eq{
\theta_1 =\frac \pi 2\,, \quad \theta_2 =0 \,, \quad \theta_3=\frac 34\pi \,, \quad
x_2=x_* \,, \quad y=y_* \,, \quad z=0\,.
}
where $X_*=e^{-2x_*}$ and $Y_*=e^{2 y_*}$ satisfy~\cite{Dall'Agata:2012sx}
\Eq{
c= \sqrt{\frac{1+6X_*^2-3X_*^4}{X_*^2(3-6X_*^2-X_*^4)}} \,, \qquad
Y_* =\frac{(-3\mp 2\sqrt 2)X_*^4 +2X_*^2-3\pm 2\sqrt 2}{X_*^4-6X_*^2+1}\,.
}

\begin{figure}
\centerline{
\includegraphics[width=8cm]{\FigDir/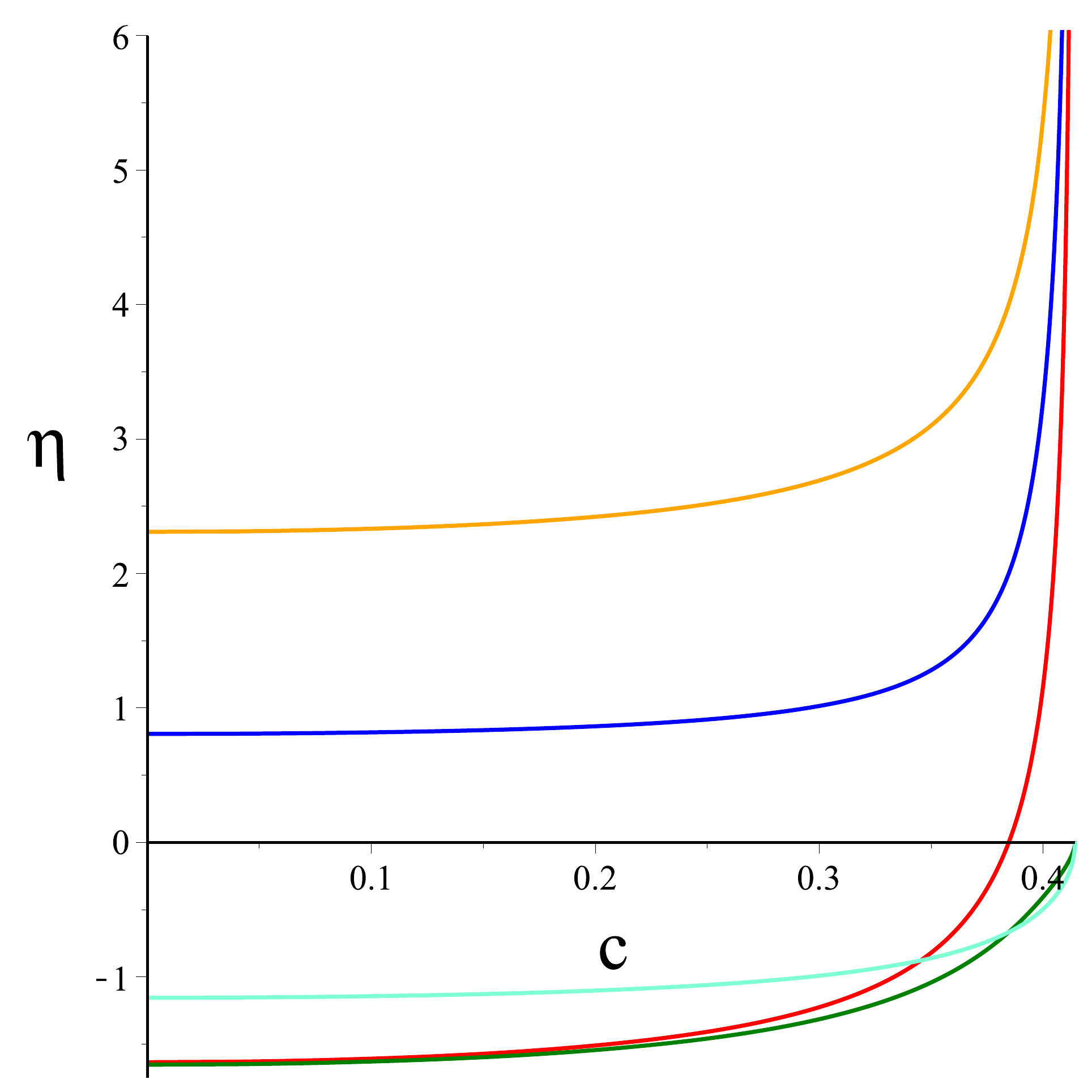}
\hspace{0.5cm}
\includegraphics[width=8cm]{\FigDir/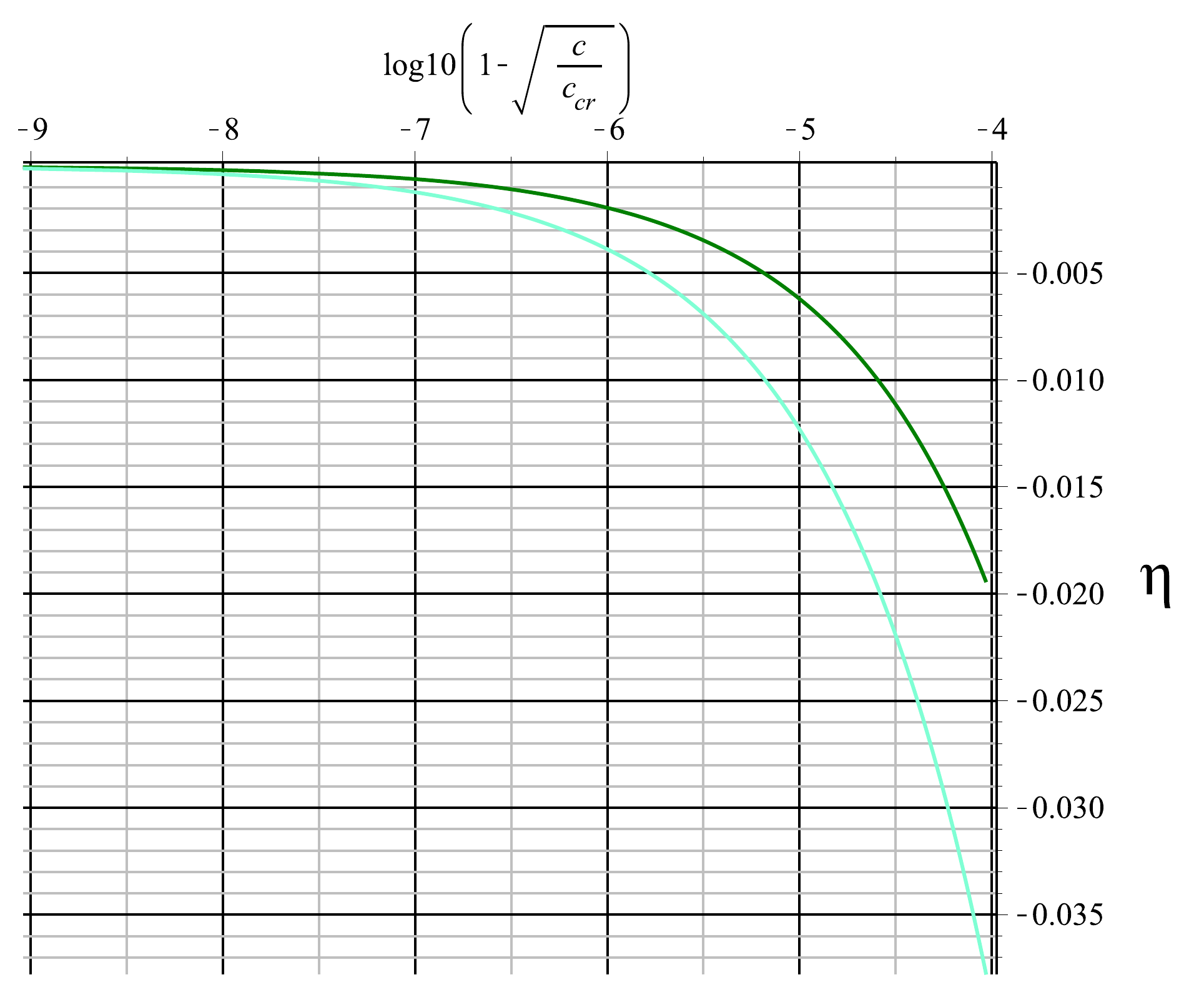}
}
\caption{The dependence of the $\eta$-parameters on the deformation parameter $c$.
In the critical limi $c\to (\sqrt 2-1)_-$, $\eta_1$ and $\eta_5$
are vanishing, whereas others tend to positively diverge.}
\label{fig:eta-s}
\end{figure}

On the $\ZR_2$-invariant plane $\Sigma_2$, we introduce the following basis in the tangent space of the scalar manifold $\Sigma_6=\exp(\H_6)$:
\Eqrsub{
&&  e_1=\frac1{\sqrt6}\frac{\pd}{\pd x_2}\,,\quad
    e_2 =\frac1{\sqrt2}\frac{\pd}{\pd y} \,, \quad
    e_3 = \frac1{\sqrt6}\frac{\pd}{\pd z} \,, \nonumber \\
&&  e_4 =\frac1{\sqrt2 \sinh(2 y)}\frac{\pd}{\pd \theta_3}\,\quad
    e_5 =\frac1{\sqrt2 \sinh y}\frac{\pd}{\pd \theta_2} \,, \quad
    e_6=\frac1{\sqrt2 \sinh y}\frac{\pd}{\pd \theta_1} \,.
\label{KineticMetric:ONbasis}
}
This basis is orthonormal with respect to the metric $K_{\alpha\beta}$ of the scalar manifold such that the scalar kinetic term reads
\Eq{
ds_{\rm T}^2 =-\kappa^2 e^{-1}\LL_K= \frac12 K_{\alpha\beta}d\phi^\alpha d\phi^\beta\,,
\label{ScalarManifold:metric}
}
where $(\phi^\alpha)=(x_2,y,z,\theta_3,\theta_2,\theta_1)$.
In terms of this basis, we define the $6\times 6$ $\eta$ matrix by
\Eq{
 (\eta _{ab})
  \equiv \inpare{ \frac{1}{V} e_a{}^\alpha e_b{}^\beta D_\alpha D_\beta V },
\label{eta-matrix:def}
}
where $D_\alpha$ is the covariant derivative in the scalar field manifold with respect to the metric $K_{\alpha\beta}$. Note that the projection via vielbein is necessary to obtain the mass spectrum for the canonically normalized fields.
Evaluating at the DI critical point, we find that $(\eta_{ab})$ has the following block-diagonal structure:
\Eq{
(\eta_{ab})_*
  = \Mat{ -\frac{2 (3X_*^4-2X_*^2+3)}{3(X_*^4-6 X_*^2+1)}& -2\sqrt{\frac{2}{3}} \\
         -2\sqrt{\frac{2}{3}} & 0 }
   \oplus
     \Mat{ -\frac{4(X_*^2-3)(3X_*^2-1)}{3(X_*^4-6X_*^2+1)} & -\frac{2}{\sqrt 3}\\
        -\frac{2}{\sqrt 3} & 0}
   \oplus
     \Mat{ -\frac{4(X_*^4-4X_*^2+1)}{X_*^4-6X_*^2+1} & 0 \\
            0    & 0 }
 \,.
}
A simple diagonalization gives the eigenvalues
\Eqrsub{
 \eta_{1,2}&=&  \frac{-3X_*^4+2X_*^2-3\pm\sqrt{33X_*^8-300X_*^6+934X_*^4-300X_*^2+33}}{3(X_*^4-6X_*^2+1)} \,, \\
 \eta_{3,4} &=& -\frac{2}{3}\frac{3X_*^4-10X_*^2+3\pm 2 \sqrt{3X_*^8-24X_*^6+58X_*^4-24X_*^2+3}}{X_*^4-6X_*^2+1} \,, \\
\eta_5  &=& -\frac{4(X_*^4-4X_*^2+1)}{X_*^4-6X_*^2+1}\,, \\
\eta_6 &=& 0 \,.
}
With the replacement $X_*^2\tend x_*$, this recovers the result in~\cite{Dall'Agata:2012sx} up to the two different signs in front of the square-root.
In the critical limit $c\to (\sqrt 2-1)^-$, $\eta_1$ and $\eta_4$ go to zero from below, while $\eta_2$ and $\eta_3$ positively diverge as shown in Fig. \ref{fig:eta-s}.

\section{${\rm SO}(3)\times {\rm SO}(3)$ invariant sector of
${\rm SO}(5,3)$ gaugings }
\label{sec:SO53}

Following the parallel argument as in the ${\rm SO}(4,4)$ case, let us next consider
the ${\rm SO}(3)\times {\rm SO}(3)$-invariant
sector of the ${\rm SO}(5,3)$ gaugings. The embedding tensor is given by~\cite{KN}
\begin{align}
  \theta =\lambda \mathbb I_5\oplus (-1/\lambda ) \mathbb I_3 \,, \qquad \xi=c\theta^{-1}
 \,, \qquad \lambda \equiv \sqrt{\frac{3c^2+1}{c^2+3}} \,.
\label{SO53_ET}
\end{align}
The electric gaugings  are specified by  $c=0$, whereas $c>0$ for the dyonic case. Note that
we employed a notation different from that in \cite{KN}. The above parametrization~(\ref{SO53_ET}) ensures that the origin of the scalar manifold becomes a critical point of the potential, which
corresponds to the unstable de Sitter critical point.

In the dyonic case, ref.~\cite{Dall'Agata:2012bb}
proposed that the inequivalent gaugings are characterized by the eigenvalues of the
tensor classifiers
\begin{align}
 B_{\alpha\beta }{}^{\gamma\delta }
=\Theta_M{}^\gamma \Theta_N{}^\delta
\Theta_P{}^\epsilon\Theta_Q{}^\zeta
q^{MNPQ}\eta_{\alpha \epsilon }\eta_{\beta\zeta } \,,
\end{align}
where
$\eta _{\alpha \beta }=\tr (t_\alpha t_\beta )$ and
$q_{MNPQ}=\eta^{\alpha\beta}(t_\alpha)_{(MN}(t_\beta)_{PQ)}$
are the Cartan-Killing metric and  the quartic invariant of $E_{7(7)}$,
respectively.  We follow the convention of the
$E_{7(7)}$ generators $(t_\alpha)_M{}^N=(t_\alpha)_{MP}\Omega^{PN}$
as in ref.~\cite{Godazgar:2013oba}.
Since the exact expression for the eigenvalues of $B_{\alpha\beta }{}^{\gamma \delta }$
are not so illuminating, we only comment  the results.
We found 11 inequivalent eigenvalues, all of which are invariant under the inversion
$c\to 1/c$.  This renders us to focus on the range $c\in [0, 1]$.

The ${\rm SO}(3)\times{\rm SO}(3)$ invariant  generators are given by
\begin{subequations}
\begin{align}
({\bf 1}, {\bf 1}): \qquad
g_1 &=S_1\wedge   \mathbb I_8 \,, \qquad  S_1 =3 \sum_{a=1}^5 (e^a)^2 -5 \sum_{a=6}^8 (e^a)^2 \,, \\
({\bf 14}, {\bf 1}): \qquad
g_2 &= S_2\wedge \mathbb I_8 \,, \qquad  S_2 = 2 \sum_{a=1}^3 (e^a)^2 -3 \sum_{a=4}^5
 (e^a)^2\,, \\
({\bf 14}, {\bf 1}): \qquad
g_3&=S_3 \wedge  \mathbb I_8 \,, \qquad  S_3 =e^4\otimes e^4-e^5\otimes e^5 \,, \\
({\bf 14}, {\bf 1}): \qquad
g_4&=S_4 \wedge \mathbb I_8 \,, \qquad  S_4 =e^4\otimes e^5+e^5\otimes e^4 \,,\\
({\bf 5}, {\bf 1}): \qquad
g_5&=\frac{i }2 (e^{1234}+e^{5678}) \,, \\
({\bf 5}, {\bf 1}): \qquad
g_6&=\frac{i}{2} (e^{1235}-e^{4678}) \,,
\end{align}
\end{subequations}
where the left end of each row specifies the representation of ${\rm SO}(5)\times{\rm SO}(3)$ to which each generator belongs in the full sector. In terms of these generators, the scalar fields in the invariant sector can be expressed as
\begin{align}
 \phi=\sum_{i=1}^4 x_i g_i+\sum_{j=1}^2 y_j g_{4+j } \,.
\end{align}

\subsection{$\mathbb Z_2$ truncation}

Before going into the full analysis, let us first discuss a simpler case for which some of the scalar fields are vanishing.
Consider the following index permutations
\begin{align}
 (1,2,3,4,5,6,7,8) \to (1,2,3,4,-5,-6,-7,-8) \,,
\end{align}
under which the embedding tensor is invariant and the generators $g_4$ and $g_6$ are odd. Then, setting $x_4=y_2=0$ is the consistent truncation. Decomposing the indices as
\begin{align}
 \mathbb R^8\we  \mathbb R ^8 \simeq  &
[V_{123}\we V_{123}\oplus V_{123}\we V_4]\oplus
[V_{678}\we V_{678}\oplus V_5 \we V_{678}]\nonumber \\ & \oplus[
V_{123}\we V_5\oplus V_4 \we V_{678}\oplus
V_{4}\we V_5\oplus V_{123}\we V_{678}]\,,
\end{align}
we have
\begin{align}
 \phi &\simeq \ti \Phi_1 \otimes  \mathbb I_3 \oplus
\ti \Phi_2 \otimes  \mathbb I_3 \oplus \ti \Phi_3 \,,
\end{align}
where
\begin{subequations}
\begin{align}
\ti \Phi_1 &= \left(
\begin{array}{cc}
3 x_1 +2 x_2 &i y_1 \\
i y_1 & 3 x_1 -\frac 12 x_2 +\frac 12 x_3
\end{array}
\right) \,, \\
\ti \Phi_2 &= \left(
\begin{array}{cc}
-5 x_1 & {i} y_1\\
{i } y_1 & -x_1 -\frac 32 x_2 -\frac 12 x_3
\end{array}
\right) \,, \\
\ti \Phi_3 &=[(3 x_1-\tfrac 12 x_2 -\tfrac 12 x_3) \mathbb I_3 \,,
(-x_1-\tfrac 32 x_2+\tfrac 12 x_3 ) \mathbb I_3 \,,
3(x_1-x_2)\,, -(x_1-x_2)  \mathbb I_9] \,.
\end{align}
\end{subequations}
In terms of new variable defined by
\begin{align}
&X_1= \frac 14 (12 x_1+3x_2+x_3) \,, \qquad
X_2 =  \frac 14 (4x_1+x_2-x_3)
 \,, \nonumber \\&
 x_-=x_2-x_1\,,  \qquad  re^{i \theta }
 =X_1 +i y_1  \,,
\end{align}
we can see that
\begin{align}
&\ti \Phi_1=x_- \sigma_3+\ti\Phi_0 \,, \qquad
\ti\Phi_2 =x_-\sigma_3 -\ti \Phi_0^*\,, \nonumber \\
&\ti \Phi_3 = [(2X_2-x_-)\mathbb I_3,(-2X_2-x_-) \mathbb I_3,-3x_-,x_-\mathbb  I_9]\,,
\end{align}
with
\begin{subequations}
\begin{align}
\ti \Phi_0&=\left(
\begin{array}{ cc}
 X_1+X_2   & i y_1    \\
i y_1      & X_1-X_2
\end{array}
\right)  =U\left(
\begin{array}{cc }
r+X_2      &    \\
      &   -r+X_2
\end{array}
\right){}^{T\!}U \,, \\
U&=\frac{1}{\sqrt{2 r}} \left(
\begin{array}{cc }
\sqrt{r+X_1}      & -\sqrt{r-X_1}   \\
i \sqrt{r-X_1}      & i \sqrt{r+X_1}
\end{array}
\right) \,, \qquad U {}^{T\!}U=\sigma_3 \,.
\end{align}
\end{subequations}
The kinetic-term metric is given by $\mathbb H^2 \times \mathbb R^2 $:
\begin{align}
d s^2_T &=d r^2 + \frac 14 \sinh^2 (2 r)d \theta ^2
+3 (d X_2^2+d x_-^2) \,, \nonumber \\
&\simeq 15 d x_1^2+\frac {15}4d x_2^2+\frac 14 d x_3^2+d y_1^2\,,
\end{align}
where at the second line we have Taylor-expanded the metric around the origin.
The potential is given by
\Eq{
V = g^2 (V_{(1)}+c^2V_{(2)})\,,  \qquad
V_{(2)}=V_{(1)}(c\to 1/c, X_i\to -X_i,x_-\to -x_-) \,,
}
where
\begin{align}
V_{(1)}=& \frac{3e^{2x_-}}{32}( e^{4X_2}-  \lambda^2 e^{-4X_2}- \lambda^2e^{2X_2} f_- + e^{-2X_2} f_+)
\nonumber \\ &
+\frac{3e^{-2x_-}}{64}\left(6  -\lambda^2 e^{-2X_2} f_--\frac{1}{\lambda^2}e^{2X_2}f_+\right)
+\frac{\lambda^2 e^{6x_-}}{64} ( e^{6X_2}f_-+e^{-6X_2}f_+-2 ) \,,
\end{align}
and
\begin{align}
f_\pm =\cosh (2r)\pm \cos\theta \sinh (2r) \,.
\end{align}
The potential for the electric gaugings is therefore given by $V=g^2 V_{(1)}$. Expanding the potential around the origin, we can get the
correct mass spectrum derived in~\cite{KN}
\begin{align}
V\simeq V_0 \left(-30x_1^2+5x_2^2+\frac 13 x_3^2-\frac 23 y_1^2\right) \,, \qquad
V_0= \frac{3g^2 (c^2+1)^3}{4(3c^2+1)(c^2+3)} \,.
\end{align}
The existence of critical points which lie outside the ${\rm SL}(8)$-type critical points
is $y_1\ne 0$, i.e, $\theta \ne n \pi $ ($n\in \mathbb N$).
In both of electric and dyonic cases, our numerical scan could not find any critical points other than the origin.

\subsection{Full analysis}

We next turn to evaluate the full ${\rm SO}(3)\times{\rm SO}(3)$ invariant sector. Let us decompose the indices as
\Eqr{
\mathbb R^8 \wedge \mathbb R^8
  &\simeq&
     [V_{123}\wedge V_{123}\oplus V_{123} \wedge V_4\oplus V_{123}\wedge V_5 ]
     \nonumber \\
  && \oplus [V_{678}\wedge V_{678}\oplus V_5\wedge V_{678}\oplus V_{678} \wedge V_{4}]
    \oplus [V_4\wedge V_5\oplus V_{123}\wedge V_{678}]\,,
}
leading to
\begin{subequations}
\begin{align}
g_1 & =\left(
\begin{array}{ccc}
3 & & \\
&3&\\
&& 3
\end{array}
\right)\otimes \mathbb I_3 \oplus
\left(
\begin{array}{ccc}
-5 &  & \\
 &-1  &\\
 &  &-1
\end{array}
\right)\otimes \mathbb I_3 \oplus
[3, -\mathbb I_9] \,, \\
g_2&=
\left(
\begin{array}{ccc}
2 &  & \\
 &-1/2  &\\
 &  & -1/2
\end{array}
\right)\otimes \mathbb I_3 \oplus
\left(
\begin{array}{ccc}
0 &  & \\
 & -3/2 &\\
 &  & -3/2
\end{array}
\right) \otimes\mathbb  I_3 \oplus [ -3, \mathbb I_9] \,, \\
g_3&= \left(
\begin{array}{ccc}
0 &  & \\
 & 1/2 &\\
 &  & -1/2
\end{array}
\right) \otimes \mathbb I_3 \oplus
\left(
\begin{array}{ccc}
 0&  & \\
 & 1/2 &\\
 &  & -1/2
\end{array}
\right)\otimes \mathbb I_3 \oplus \mathbb O_{10} \,, \\
g_4&=
\left(
\begin{array}{ccc}
0 &  & \\
 & 0 &1/2\\
 & 1/2 & 0
\end{array}
\right)\otimes \mathbb I_3 \oplus
\left(
\begin{array}{ccc}
0 &  & \\
 & 0 &1/2\\
 & 1/2  & 0
\end{array}
\right) \otimes \mathbb I_3 \oplus \mathbb O_{10}\,, \\
g_5 &= \left(
\begin{array}{ccc}
0 & i & \\
i &0  &\\
 &  &0
\end{array}
\right)\otimes \mathbb I_3 \oplus
\left(
\begin{array}{ccc}
0 &  i & \\
i  &0  &\\
&  & 0
\end{array}
\right)\otimes\mathbb  I_3 \oplus \mathbb O_{10} \,, \\
g_6 &= \left(
\begin{array}{ccc}
0 &  & i\\
 &0  &\\
i &  & 0
\end{array}
\right)\otimes \mathbb I_3
\oplus \left(
\begin{array}{ccc}
0 &  & i \\
 &0  &\\
i &  &0
\end{array}
\right)\otimes\mathbb  I_3 \oplus \mathbb O_{10} \,.
\end{align}
\end{subequations}
Hence, the coset representative in the invariant sector can be expressed in terms of the six scalar fields $x_i$ ($i=1,\cdots,4)$ and $y_j$ ($j=1,2$) as
\begin{align}
 \phi=\sum_{i=1}^4 x_i g_i+\sum_{j=1}^2
y_j g_{4+j } \simeq \Phi_1 \otimes \mathbb I_3 \oplus \Phi_2 \otimes
\mathbb I_3 \oplus
 \Phi_3 \,,
\end{align}
where
\begin{align}
\Phi_1 & =\left(
\begin{array}{ccc}
3 x_1+2 x_2 &i y_1 & i y_2 \\
i y_1 & 3 x_1 -\frac 12 x_2 +\frac 12 x_3 & \frac 12 x_4 \\
i y_2 & \frac 12 x_4 & 3 x_1 -\frac 12 x_2 -\frac 12 x_3
\end{array}
\right) \,, \\
\Phi_2 &= \left(
\begin{array}{ccc}
-5 x_1 & i y_1 & {i} y_2 \\
i y_1 &-x_1 -\frac 32 x_2 -\frac 12 x_3 &
-\frac 12 x_4 \\
{i}  y_2 & -\frac 12 x_4 & -x_1 -\frac 32 x_2 +\frac 12 x_3
\end{array}
\right) \,, \\
\Phi_3 &= (x_1-x_2)\times [3, -\mathbb I_9 ] \,.
\end{align}
The kinetic term at the origin $O$ reads
\begin{align}
d s_T^2|_O
&= \frac{1}{4}\left[
12 d x_-^2 +3 d x_+^2 +
d x_3^2+d x_4^2 +4(d y_1^2+d y_2^2)
\right]\,,
\end{align}
where we have defined
\begin{align}
 x_+\equiv 4x_1+x_2\,, \qquad x_-\equiv x_2-x_1 \,.
\end{align}

Following the same argument as in the ${\rm SO}(4,4)$ case,
the $3\times 3$ matrices
$\Phi_1$ and $\Phi_2$ are related in a simple manner
\begin{align}
 \Phi_1 =x_- I_{1,2}+\Phi_0 \,, \qquad \Phi_2 =x_- I_{1,2}
-\bar \Phi_0 \,,
\end{align}
with
\begin{align}
 I_{1,2}={\rm diag}(1,-1,-1) \,, \qquad
\Phi_0 \equiv \left(
\begin{array}{ccc}
x_+ & i y_1 & i y_2 \\
i y_1 & \tfrac 12 (x_++x_3) & \frac 12 x_4 \\
i y_2 & \frac 12 x_4 & \frac 12 (x_+-x_3)
\end{array}
\right) \,.
\end{align}
Since the matrix $\Phi_0$ falls into the family (\ref{Phi_univ}),
we get
\begin{align}
 \Phi_0 = U\Delta {}^{T\!}U \,, \qquad U\in {\rm U}(3) \,,
\end{align}
where the unitary matrix $U$  has exactly the same structure as~(\ref{N8_SO44_AT_Umat})
and the eigenvector takes the form
\begin{subequations}
\begin{align}
\vec v_i =\left(
\begin{array}{c}
\mu_i^2+x_+\mu_i+\frac 14 (x_+^2-x_3^2-x_4^2)       \\
-\{\mu_i +\frac 12 (x_+-x_3)\}y_1+\frac 12 x_4 y_2 \\
-\{\mu_i +\frac 12 (x_++x_3)\}y_2+\frac 12 x_4 y_1 \\
\end{array}
\right) \,.
\end{align}
\end{subequations}
The eigenvalues $\mu_i$ are the roots of the cubic equation $Q(\mu_i )=0$, where
\begin{align}
 Q(\mu )\equiv &\mu^3 -\frac 14 \left[3 x_+^2+x_3^2+x_4^2
 +4(y_1^2+y_2^2)\right] \mu  \nonumber \\ & +\frac{1}{4}\left[
-x_+^3+4x_4 y_1y_2 +2x_3 (y_1^2-y_2^2)+x_+\{x_3^2+x_4^2-2 (y_1^2+y_2^2)\}
\right] \,.
\end{align}
Hence we see $\sum_i \mu_i=0$.
Note that the variable $x_-$ decouples from the matrix $U$, hence also from $Q(\mu)$.
By replacing $x_2$ by $x_-$ in the ${\rm SO}(4,4) $ kinetic term (\ref{SO44target}), one obtains the full kinetic term for ${\rm SO}(5,3) $ gaugings, i.e.,
\begin{align}
d s_T^2= 3 d x_-^2+\frac 12 \sum_i d \mu_i^2
+\sinh^2(\mu_2-\mu_3)\chi_1^2
+\sinh^2(\mu_3-\mu_1)\chi_2^2
+\sinh^2(\mu_1-\mu_2)\chi_3^2\,.
\label{SO53_target}
\end{align}
The potential reads
\begin{align}
V=g^2 (V_{(1)}+c^2 V_{(2)}) \,,
\label{SO53_SO3SO3_fullpot}
\end{align}
where
\begin{align}
V_{(1)}= &  \frac{\lambda^2 e^{6x_-}}{32}
 \Bigl[ -1+\cos^2 \theta_1 \cosh (2(\mu_2-\mu_3))\nonumber \\
       &  +\sin^2\theta_1 \{\cos^2 \theta_3 \cosh (2(\mu_3-\mu_1))+\sin^2\theta_3 \cosh(2(\mu_1-\mu_2))\}
  \Bigr]
\nonumber \\
&
  +\frac{3e^{-2x_-}}{64}\Bigl[
       6 -\frac{1}{\lambda^2} \left\{e^{2\mu_1}\cos^2\theta_1 +\sin^2\theta_1
         (e^{2\mu_2}\cos^2\theta_3 +e^{2\mu_3}\sin^2\theta_3 )
       \right\}
\nonumber \\
&
     +\lambda^2 \left\{e^{-2\mu_1}\cos^2\theta_1 +\sin^2\theta_1
     (e^{-2\mu_2}\cos^2\theta_3 +e^{-2\mu_3}\sin^2\theta_3 ) \right\}
  \Bigr]
\nonumber \\
& +\frac{3e^{2x_-}}{32} \Bigl[
     \left\{e^{-2\mu_1}\sin^2\theta_1 +e^{-2\mu_2}(\sin^2\theta_3 +\cos^2\theta_1 \cos^2\theta_3 )
           +e^{-2\mu_3} (\cos^2\theta_3 +\cos^2\theta_1 \sin^2\theta_3 )\right\}
\nonumber \\
&  - \lambda^2 \left\{e^{2\mu_1}\sin^2\theta_1 +e^{2\mu_2}(\sin^2\theta_3 +\cos^2\theta_1 \cos^2\theta_3 )
           +e^{2\mu_3} (\cos^2\theta_3 +\cos^2\theta_1 \sin^2\theta_3 )\right\}
\Bigr]\,,
\label{SO53full_U1}
\end{align}
with
\begin{align}
V_{(2)}=V_{(1)} (x_-\to-x_-, \mu_i\to -\mu_i, c \to 1/c) \,.
\end{align}
A careful trace of higher-order terms at the origin ($\mu_i=x_-=0$), we can recover the desired result~\cite{KN}
\begin{align}
V\simeq V_0 \left[
-30 x_1^2+5 x_2^2 +\frac 13 (x_3^2+x_4^2) -\frac 23 (y_1^2+y_2^2)
\right]\,, \qquad
 V_0= \frac{3g^2(c^2+1)^3}{4(3c^2+1)(c^2+3)} \,.
\end{align}
It comes as a surprise to observe that the potential (\ref{SO53_SO3SO3_fullpot}) is  independent of the Euler angle $\theta_2 $. This means that the potential possesses the following flat direction
\begin{align}
\frac{\partial V}{\partial \theta_2}=
 \left[
2\left(x_3 \frac{\partial}{\partial x_4}-x_4 \frac{\partial }{\partial x_3}\right)
+\left(y_1\frac{\partial }{\partial y_2}-y_2\frac{\partial }{\partial y_1}\right)
\right] V=0 \,.
\end{align}
Note that $\partial/\partial \theta_2$ is a Killing vector of the scalar manifold (\ref{SO53_target}).
It would be intriguing to understand if there is a profound reason for this unexpected flat direction of the potential.

We do not attempt to write down the exact form of the coupling of the
gauge field to the kinetic term of the scalars,  since their expressions are quite long.
But this task is straightforward following the arguments of the ${\rm SO}(4,4)$ case.

As far as we scanned numerically, there appear no critical points other than origin.

\section{Inflation in the ${\rm SO}(4,4)$ gaugings}
\label{sec:inflation}

Let us now undertake the investigation whether a realistic inflationary solution can be realized in a  maximal gauged supergravity. As is clear from the expression for the scalar field potential in terms of the eigenvalues $\Lambda=(\lambda_j)$ and the unitary matrix $U$ in section 2, the potential at large fields is always dominated by the term $\exp(\sum_j 2n_j \lambda_j)$ with integer $n_j$ and bounded coefficients. The integer vector $\bm{n}=(n_j)$ runs over all possible combination $\bm{a}+\bm{b}+\bm{c}$ where $\bm{a}$, $\bm{b}$ and $\bm{c}$ are vectors that have only one non-vanishing component with value $\pm1$. Then, for any one-dimensional sector  $\lambda_j=c_j \phi$ with a canonically normalized kinetic term $-d\phi^2/2$, the exponent of the maximally growing exponential factor is given by $2\bm{n}\cdot\bm{\Lambda}=6 c_m |\phi|\ge (108/35)^{1/2}|\phi|$ from $1=\sum c_i^2/6 \le (70/6)c_m^2$, where $c_m$ is the maximum value of $|c_j|$.  Thus, the maximal coefficient in the exponents is too large for a power-law inflation to take place \cite{Lucchin:1984yf} unless all the coefficients of the exponential factors with such large exponents vanish in some special direction specified by the unitary matrix $U$ in \eqref{phi_ATfactorization}, and at the same time the potential is stationary in all directions orthogonal to that special direction. This implies that the chaotic and the power-law inflation do not occur in the maximal gauged supergravity except for accidental cases. Only an intermediate or small scale inflation is likely to occur. This includes the hill-top-type inflation \cite{Boubekeur:2005zm}   near a maximum point and the racetrack-type inflation \cite{BlancoPillado:2004ns} near a saddle point.
If an additional condition that all coefficients of the growing exponential factor vanish in some direction
is fulfilled, the Starobinsky-type inflation \cite{Starobinsky:1980te} might also be possible along an asymptotically de Sitter flat direction. We will see later that such an unexpected situation really happens in our system.

Another obstruction against inflation in the maximal gauged supergravity is the fact that the modulus of the $\eta$ parameter representing the curvature scale of the scalar potential is generally larger than the order unity, which spoils inflation.  In order for the inflation near the critical point to last for a sufficiently long time, the inflaton field must start from a point very close to the critical point. Since quantum fluctuations of order of the cosmic expansion rate $H$ would push the inflaton away from the critical point, such a classical fine tuning becomes unphysical if $|\eta|\gsim1$. Hence the $\eta$-problem
continues to reside in the maximal gauged supergravity. This should not be confused with the conventional $\eta$-problem in the $N=1$ supergravity~\cite{Baumann:2014nda}, since their origins are quite different.

As we see in more detail below, the theory obtained by the $\SO(4,4)$ gauging suffers from this $\eta$ problem for a generic value of the deformation parameter. In particular, the saddle point of the potential with the highest symmetry at the origin cannot provide an inflation since the negative eigenvalues of the mass square at the origin are independent of the deformation parameter and of the order unity. However, the $\SO(3)^2$-invariant saddle points may alleviate the $\eta$ problem if we fine-tune the deformation parameter to be close to the critical values as pointed out by Dall'Agata and Inverso~\cite{Dall'Agata:2012sx}.  We have found such a saddle point only for the $\SO(4,4)$ gauging so far. We therefore restrict our consideration to the $\SO(4,4)$ gauging. We also study the influence of the gauge flux to see the possibility of the chromo-natural type inflation~\cite{Adshead:2012kp,Dimastrogiovanni:2012st}.

In this section, we adopt the absolute units such that $\mpl=1/\kappa=1/\sqrt{8\pi G}=1$ and parametrize the deformation in terms of $s$ instead of $c$, following ref.~\cite{KN}. These are simply related by
\begin{align}
\label{}
g\to s g \,, \qquad c=1/s^2 \,,
\end{align}
hence $\theta =\xi^{-1}=s (\mathbb I_4, -\mathbb I_4)$.  In this case, the critical value of the deformation parameter $c=\sqrt 2-1$ corresponds to $s=s_c=\sqrt{\sqrt 2+1}$.
We can concentrate on the range $s\ge s_c$ due to the reflection invariance $s\to 1/s$ with a parity transformation of the scalar manifold.

\subsection{Description of the cosmological system without gauge flux}

The description of the cosmological system for Einstein's gravity sourced by scalar fields can be specified by the spatially homogeneous configuration. We assume that the metric is spatially flat and homogeneous. Namely,
we shall consider the Friedmann-Lema\^itre-Robertson-Walker (FLRW) universe,
\Eq{
d s^2=-\N^2 (t)d t^2+a^2(t)d \vec x^2 \,.
}
where the lapse function $\N=\N(t)$ will be set to unity at the level of equations of motion. Hence, the metric is specified by the cosmic scale factor $a(t)$ or the e-folding variable $\alpha(t)$ defined by $a=e^\alpha$. The cosmic expansion rate $H(t)$ is expressed as
\Eq{
H = \frac{\dot a}{\N a}=\frac1{\N}\dot \alpha.
}
Here and in the following, a dot denotes differentiation with respect to the proper cosmic time $t$.

The six scalar fields are also dependent only on time.  In the most part of our analysis,
we shall use the coordinate system
\Eqr{
\text{System-1} &:&  \bm{\phi}_{(1)}(t)=(x_2(t), \mu_2(t), \mu_3(t), \theta_1(t), \theta_2(t) , \theta_3(t))\,.
\label{CoordSys1}
}
In this parametrization, the system is governed by the following effective action
\begin{align}
S_{0+1}=\int dt\,e^{3\alpha}\left[ -\frac3{\N}e^{-2\alpha} \dot\alpha{}^2 - \N V +\frac{3}{\N} \Bigl(
3\dot x_2^2+\frac 12\sum_i \dot \mu_i^2 +\frac 1{2}\sum _{i,j,k}|\epsilon_{ijk}|\sinh^2(\mu_i-\mu_j)
\dot \chi_k^2
\Bigr)\right] \,,
\label{effaction}
\end{align}
where $V$ is given by (\ref{SO44_fullpot}) and the constraint $\sum_i\mu_i=0$ is understood. Since the above effective action is invariant under the transformation $\mu_2\tend -\mu_2, \mu_3\tend -\mu_3$, we can restrict the coordinate region to $\mu_2-\mu_3\ge0$.

The variation of (\ref{effaction}) with respect to $\N$ and $\phi^\alpha_{(1)}$ yields the Friedmann equation and the equations of motion for the scalars, respectively. We do not write them explicitly, but can be easily deduced. Although the final expressions are messy, equations of motion appear to take the simplest form in the coordinate system $\bm{\phi}_{(1)}$. However, this coordinate system is an analogue of the polar coordinates and is singular at the axes where two of $\mu_i$'s are pairwise equal, as discussed in Appendix \ref{sec:CoordTrf} in detail. This coordinate singularity prevents us to solve the equations of motion numerically around these axes. This problem can be circumvented by transforming to the original tangent space parametrization
\Eq{
\text{System-0} : \bm{\phi}_{(0)}=(x_1,y_1,x_2,y_2,w_1,w_2) \,.
}
Unfortunately, we fail to have an explicit expression of the Lagrangian in terms of this coordinate system. We find that the coordinate system $\bm{\phi}_{(2)}=(u,v,z,\theta_1,\theta_2)$ defined in
eq.~(\ref{phi2}) partially overcomes both of these two shortcomings. See Appendix \ref{sec:CoordTrf} for this technical problem.

\subsection{Slow roll parameters}

As we mentioned at the beginning of this section, the maximal gauged supergravity with $\SO(4,4)$ gauging suffers from the $\eta$-problem for a generic value of the deformation parameter $s$. Before discussing this problem, we first have to address the definitions of slow roll parameters for a multi-component system. For the system under consideration, the curvature of the potential is represented by the matrix $\bm{\eta}=(\eta_{ab})$ of rank six defined by \eqref{eta-matrix:def}. At an extremal point of the potential, this $\eta$ matrix is identical to the mass-square matrix of the scalar fields normalized by the value of the potential there. To the contrary, its physical meaning is less obvious at a generic point. Nevertheless, we will show that it plays an important role in the investigation of inflation. In particular, when the inflaton is a single component scalar field, $\eta=\mpl^2 V''/V$ together with the $\epsilon$ parameter defined by $\epsilon\equiv \mpl^2 (V'/V)^2/2$ determines the duration of inflation and various observational features of the primordial cosmological perturbations produced by quantum fluctuations during inflation. In the case of a slow-roll inflation for which $\epsilon, |\eta|\ll1$ at the time $t_{\rm O}$ when the comoving scale corresponding to the present horizon scale comes out of the Hubble horizon during inflation, the observed spectral index $n_s$ of curvature perturbations around the present horizon scale is determined by these parameters as
\Eq{
n_s = 1+2\eta-6\epsilon \,,
\label{nsbySRparam}
}
where it is understood that the right-hand side is evaluated at $t=t_{\rm O}$.

For a multiple-component inflation model it is far from obvious how to quantify
the ``slow roll" as in the single inflaton case.
We propose below the alternatives corresponding to $\epsilon$ and $\eta$ in the
single filed case. To this aim,
let us first define a sequence of parameters
\Eq{
\epsilon_n \equiv -\frac1{H^{n+1}} \frac{d^n H}{dt^n}  \,.
}
The equations of motion arising from the Lagrangian
$\ma L=(R-2V)*1-K_{\alpha \beta} d \phi ^\alpha\we *d \phi^\beta$
boil down to
\begin{align}
\ddot \phi^\alpha + \Gamma^\alpha{}_{\beta\gamma}\dot\phi^\beta\dot\phi^\gamma + 3H \dot\phi^\beta + D^\alpha V=0 \,,
\label{EOM_multi}
\end{align}
where $\Gamma^{\alpha}{}_{\beta\gamma}$ is the Christoffel symbol for the scalar manifold with the metric and $D^\alpha$ is the associate covariant derivative, i.e.,
$D^\alpha V=K^{\alpha\beta} \pd_\beta V$.  Differentiating the Friedmann equation
\Eq{
H^2 = \frac13\inpare{\frac12 K_{\alpha\beta} \dot\phi^\alpha \dot\phi^\beta+ V} \,,
}
with respect to $t$ and eliminating $\ddot\phi^\alpha$ with the help of the equations of motion,
we obtain
Raychaudhuri's equation
\Eq{
\dot H = -\frac12 K_{\alpha\beta}\dot\phi^\alpha \dot\phi^\beta \,.
 \label{dH}
}

Let us now introduce the slow roll approximation for which the equations of motion
(\ref{EOM_multi}) can be approximated by
\Eq{
3H \dot\phi^\alpha+D^\alpha V  \simeq 0 \,.
\label{SlowRollEOM}
}
Equation \eqref{dH} yields
\Eq{
\epsilon_H \equiv \epsilon_1 \equiv -\frac{\dot H}{H^2} \simeq \frac1{2V^2} K^{\alpha\beta}\pd_\alpha V \pd_\beta V\,.
}
Since the right-hand side of this equation coincides with the $\epsilon$ parameter in the single inflaton case, we define its counterpart in a multi-component system by
\Eq{
\epsilon_V \equiv \frac1{2V^2} K^{\alpha\beta}\pd_\alpha V \pd_\beta V\,.
}

Next, a differentiation of the slow roll approximation of \eqref{dH} with respect to $t$,
we get
\Eq{
H^2\dot H \simeq -\frac1{18} K^{\alpha\beta}\pd_\alpha V \pd_\beta V\,.
}
Eliminating $\dot\phi^\alpha$ by \eqref{SlowRollEOM}, we obtain
\Eq{
\epsilon_2 -2 \epsilon_1^2 \simeq - 2\epsilon_1 \eta_{\alpha\beta} \hat v^\alpha \hat v^\beta,
}
where $\hat v^\alpha$ is the gradient vector of the potential normalized to a unit vector with respect to the metric $K_{\alpha\beta}$:
\Eq{
\hat v^\alpha = - D^\alpha V / \inpare{K_{\alpha\beta}D^\alpha V D^\beta V}^{1/2}.
\label{UnitDV}
}
 Hence,
\Eq{
\eta_V\equiv \eta_{\alpha\beta} \hat v^\alpha \hat v^\beta \simeq \epsilon _1 - \frac{\epsilon_2}{2\epsilon_1} \equiv \eta_H.
}
This suggests that  $\eta_V$ defined in this equation provides a counterpart in the multi-component case to the $\eta$ parameter in the single component case.

Observe that it is more appropriate to work with $\epsilon_H$ and $\eta_H$ rather than $\epsilon_V$ and $\eta_V$ when we estimate the spectral index in \eqref{nsbySRparam}, because the $\eta$-dependence of $n_s$ comes from the time dependence of the $\epsilon$ parameter. So, in the present paper, we use $\epsilon_V$ and $\eta_V$ to get a physical understanding of the behavior of inflationary trajectories, while we use $\epsilon_H$ and $\eta_H$ for the terms determined only by the time-dependence of $H$ when we estimate $n_s$ and $r$. Of course, they should be equal approximately. In fact, in all numerical calculations of the inflation for which more than 60 e-folding was realized, we verified that the difference in these values are less than 2\% (see Table \ref{tbl:ns:nf} for some numerical examples).

\begin{figure}[t]
\centerline{
\includegraphics[width=12cm]{\FigDir/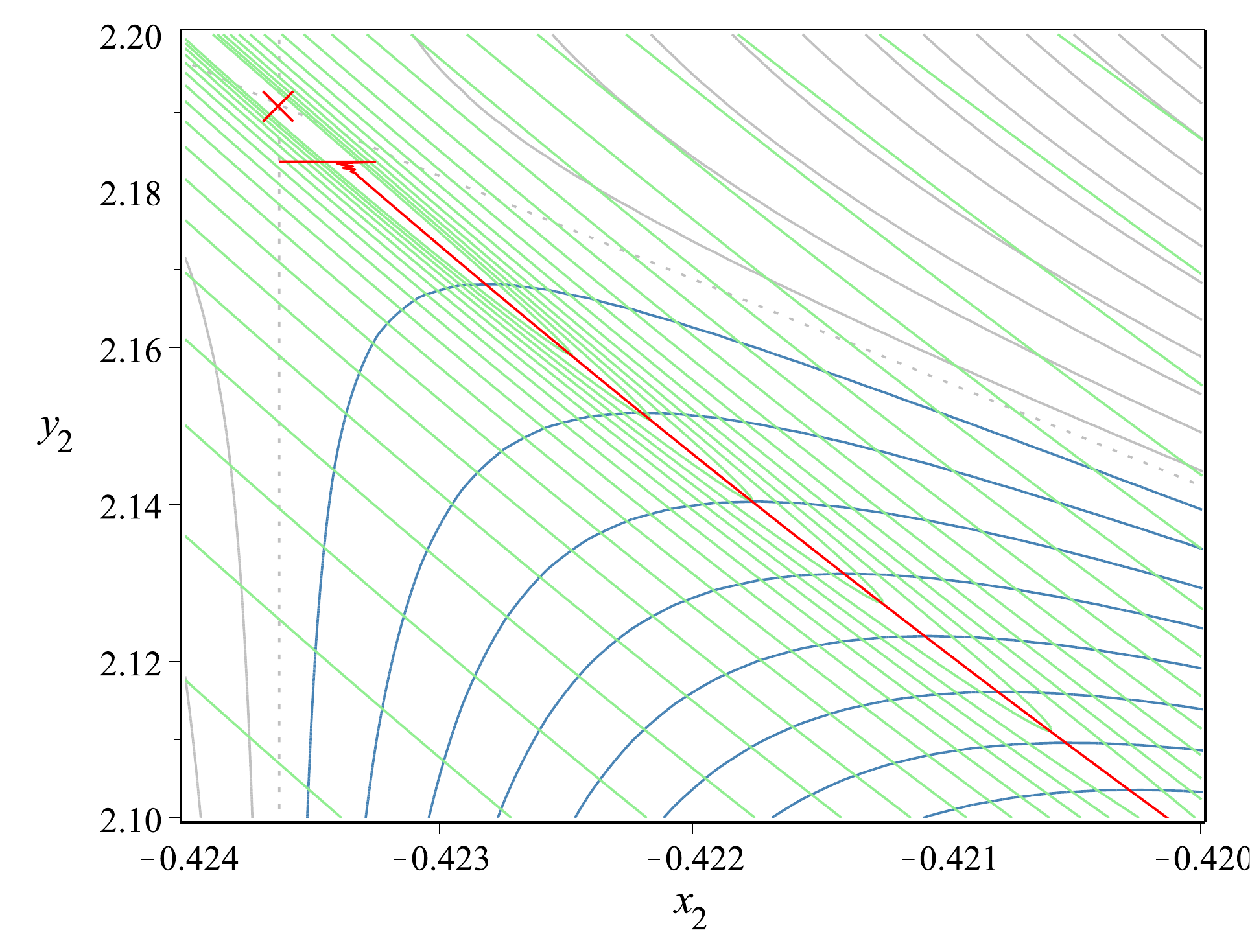}
}
\caption{An example of an inflationary trajectory on $\Sigma_2$ for $s=s_c+ 10^{-3}$. The green curves represent the contours of $\epsilon_V$, and the gray and the steel blue curves represent the contours of $V$ with $V>V_*$ and $V<V_*$, respectively. The red cross shows the DI saddle point, and the dotted curves represent the $V=V_*$ contour. The red curve represents the trajectory obtained numerically. We see that it first oscillates but soon settles down to a path on which the gradient vectors of $\epsilon_V$ and $V$ are parallel.}
\label{fig:SRtraj}
\end{figure}

Another subtlety of a multi-component system is the non-uniqueness of a slow roll trajectory: there are five-dimensional family of slow roll trajectories in the present case.  This non-uniqueness may make the observational predictions highly sensitive to the initial condition. As we will see in the next subsection,
this problem can be partially cured as far as slow roll trajectories near the $\ZR_2$-invariant 2D plane
$\Sigma_2$ are concerned, since there appear two attractor trajectories to which the projection of all sufficiently inflationary trajectories on this 2D plane converge.  In order to understand this behavior more clearly, let us calculate the divergence of the unit tangent vector field $\hat v^\alpha$ for the slow roll trajectories  \eqref{UnitDV}:
\Eq{
D_\alpha \hat v^\alpha = -\frac{1}{\sqrt{2\epsilon_V}} \inpare{\eta^\alpha{}_\alpha -\hat v^\alpha \hat v^\beta \eta_{\alpha\beta}}.
}
This equation implies that $D_\alpha\hat v^\alpha$ becomes minimum in the direction that minimizes $\hat v^\alpha \hat v^\beta \eta_{\alpha\beta}$, provided that the $\eta$ tensor is constant. This minimum direction is characterized by the eigenvector of the $\eta$ tensor corresponding to the minimum eigenvalue. Hence, if the right-hand side of the above equation is negative (this is actually the case in our system),  one expects that each slow roll trajectory converges to the one determined by the condition
\Eq{
 \eta_{\alpha\beta} \hat v^\beta = \eta_V K_{\alpha \beta} \hat v^\beta \,.
}
This condition amounts to the following relation
\Eq{
D_\alpha \epsilon _V= \frac{\eta_V-2\epsilon_V}{V} D_\alpha V \,.
}
This equation determines the unique trajectory at least locally. We have confirmed that the attractor trajectories satisfy this equation with good accuracy for our numerical solutions (see Fig.~\ref{fig:SRtraj}).

\begin{figure}
\centerline{
\includegraphics[width=8cm]{\FigDir/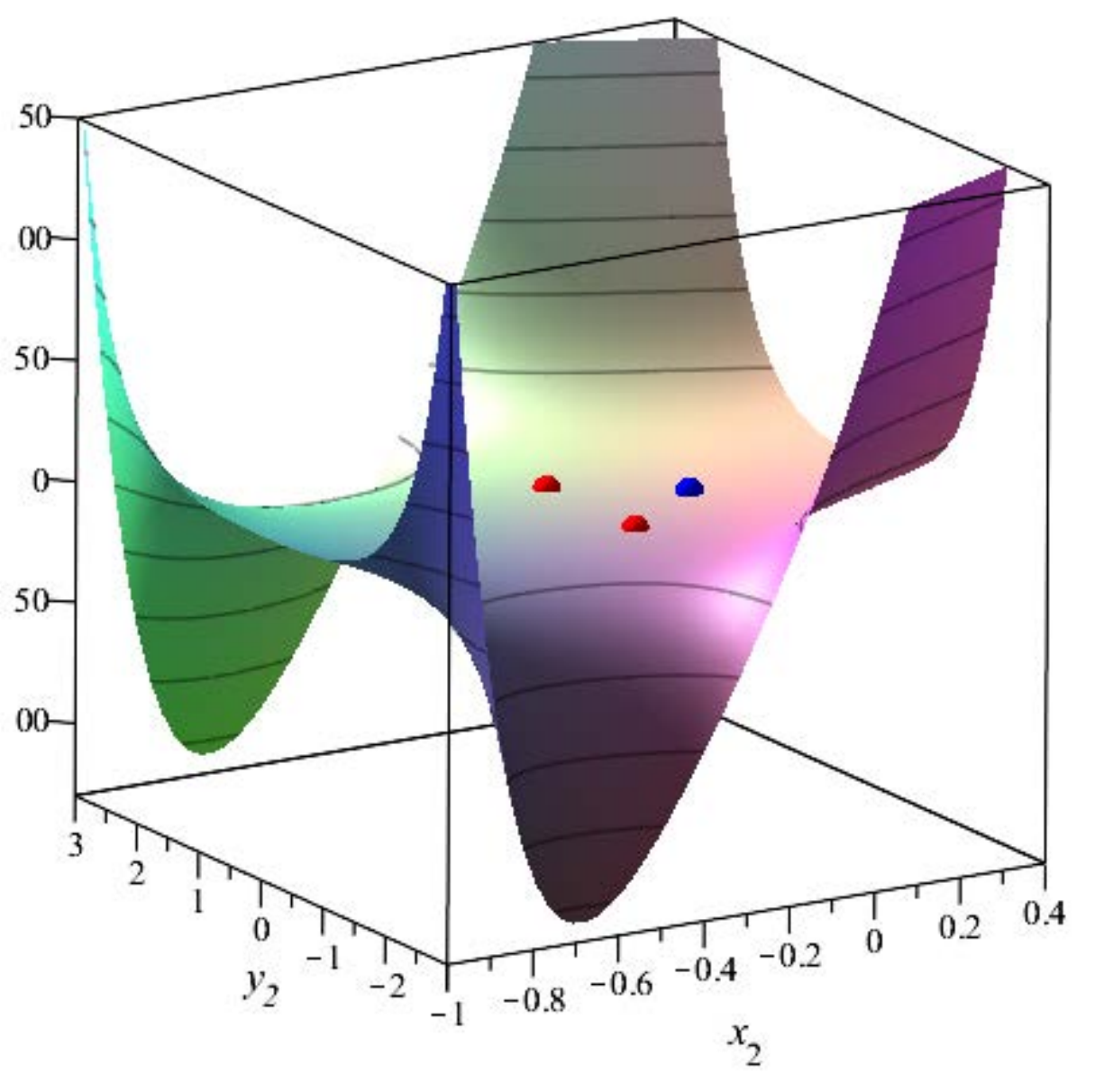}
\includegraphics[width=7cm]{\FigDir/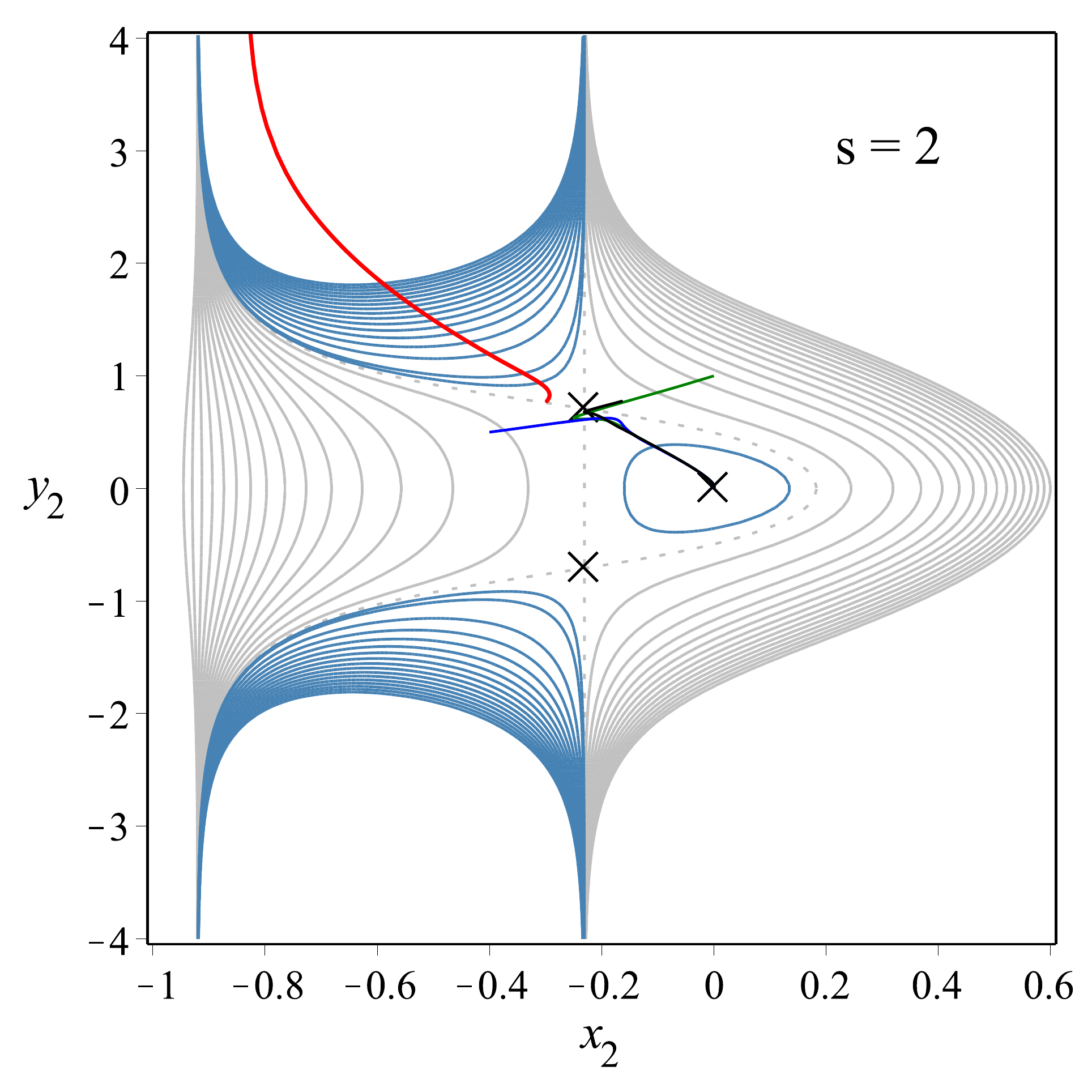}
}
\caption{The left panel is a 3D plot of the potential on $\Sigma_2$ for $s=2$, and the right panel is its contour plot. In the left panel, the blue dot shows the location of the $\SO(4)\times\SO(4)$-invariant extremum, and the red dots show the location of the DI saddle points. These critical points are shown by crosses in the right panel with some examples of trajectories.}
\label{fig:Potential:Z2:s=2}
\end{figure}

\begin{figure}
\centerline{%
\includegraphics[width=8cm]{\FigDir/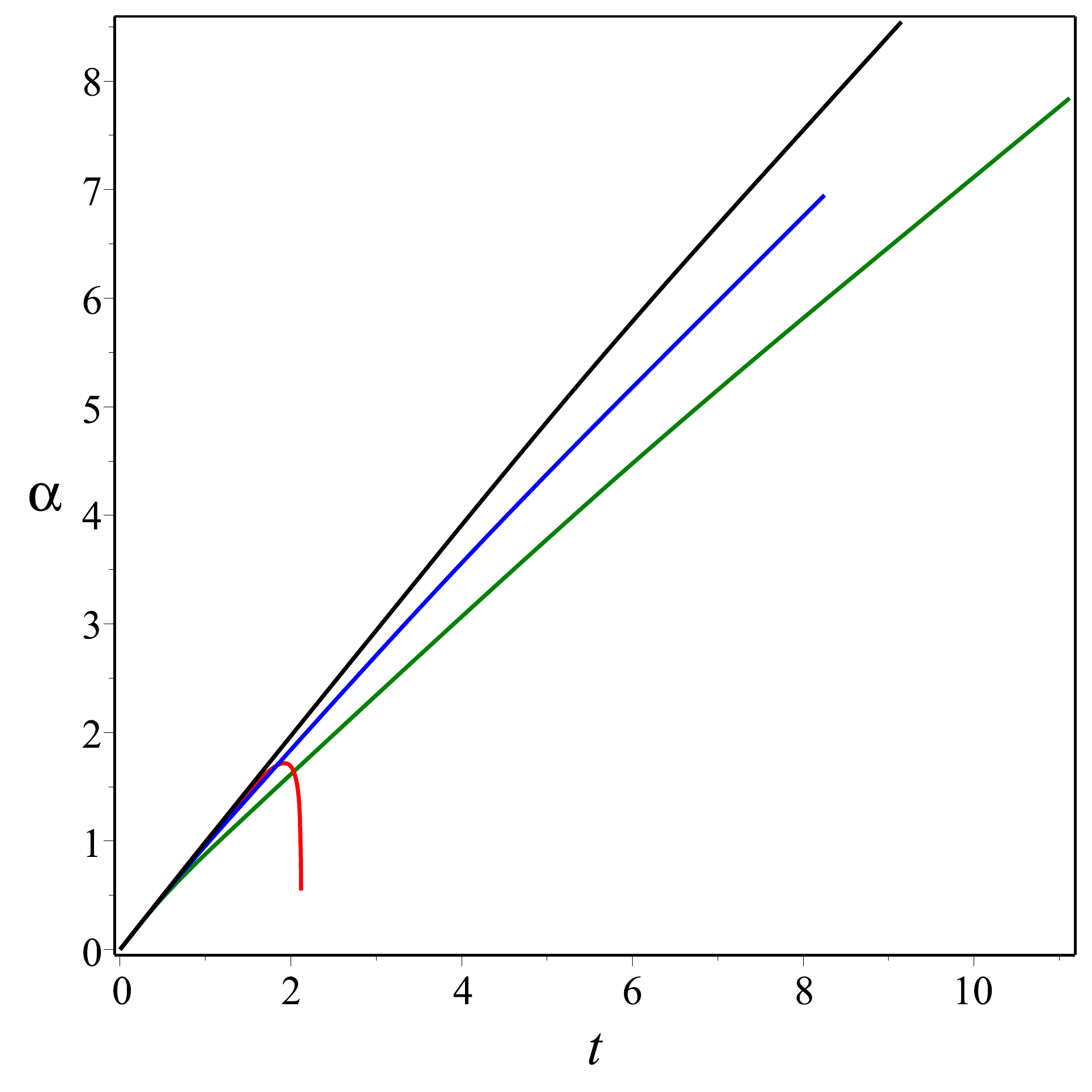}%
}
\caption{The time evolution of $\alpha(t)$ for the trajectories shown in Fig. \ref{fig:Potential:Z2:s=2}.  For trajectories arriving at the extremum at the origin, $\alpha(t)$ is drawn up to that point. The time variable $t$ is appropriately scaled for each trajectory to make the figure clearer.}
\label{fig:alpha:Z2:s=2}
\end{figure}

\subsection{$\eta$ problem}

Since the necessary ingredients and equipments are at hand, let us now discuss the $\eta$ problem in our system. As is clear from the arguments in the previous subsection, the $\eta$ parameter relevant to our multi-component system is the negative eigenvalue of the $\eta$ matrix with the maximal modulus. It coincides with $\eta_V=\hat v^\alpha \hat v^\beta \eta_{\alpha\beta}$ on a slow-roll attractor trajectory. We will henceforth denote it by $\eta_V$ at a generic point of the scalar manifold.

With respect to the orthonormal basis  \eqref{KineticMetric:ONbasis}, the $\eta$ matrix at a generic point of the $\ZR_2$-invariant plane $\Sigma_2$ has a block structure similar to that at the DI saddle point:
\Eq{
(\eta_{ab})=\Mat{M_1 & A \\ A & M_2}\oplus \Mat{M_3 & B \\ B & M_4} \oplus \Mat{M_5 & 0 \\ 0 & M_6},
}
where each of the three blocks corresponds to the 2D directions tangent to  $\Sigma_2$, the $z-\theta_3$ plane and the $\theta_1-\theta_2$ plane, respectively.  The entries of the matrices are
\begin{align}
M_1 &= \frac{g^2}{32V}\insbra{f_1(X)\left(3g_2(X)-s^2-\frac1{s^2}\right)f_1(Y)+2g_1(X)(3f_2(X)-2)},\\
M_2 &= \frac{g^2}{32V}\insbra{g_3(X)+\inpare{\frac6{s^2}-3s^2}X+\inpare{6s^2-\frac3{s^2}}\frac1{X}}f_1(Y),\\
A &= \frac{\sqrt3 g^2}{32V}\inpare{\frac{s^2}{X}-\frac{X}{s^2}}f_1(X)^2 \inpare{Y-\frac1{Y}},\\
M_3 &= \frac{g^2}{32V}\insbra{f_1(X)\inpare{3g_2(X)-s^2-\frac1{s^2}}f_1(Y)-16g_1(X) },\\
M_4 &=\frac{g^2}{32V}\insbra{g_3(X)+\inpare{\frac6{s^2}-3s^2}X+\inpare{6s^2-\frac{3}{s^2}}\frac1X }f_1(Y),\\
B &=-\frac{\sqrt3 g^2}{16V} \inpare{g_3(X)+\frac{s^4+2}{s^2}X + \frac{2s^4+1}{s^2X}},\\
M_5 &= \frac{g^2}{64V}\insbra{\inrbra{5g_3(X)+\frac{6-3s^4}{s^2}X+\frac{6s^4-3}{s^2X}}f_1(Y)
-6 g_1(X)(f_2(X)+6)},\\
M_6 &= \frac{g^2}{64V}\insbra{\inrbra{g_3(X)+\frac{6-3s^4}{s^2}X+\frac{6s^4-3}{s^2X}}f_1(Y)
-2\inrbra{ g_3(X)+\frac{6-3s^4}{s^2}X + \frac{6s^4-3}{s^2X}}}\,,
\end{align}
where
\Eq{
f_n (u) =u^n + \frac{1}{u^n}\,,\qquad
g_n (X) = \frac{X^n}{s^2}+ \frac{s^2}{X^n}\,.
}
%

\begin{figure}
\centerline{
\includegraphics[width=7cm]{\FigDir/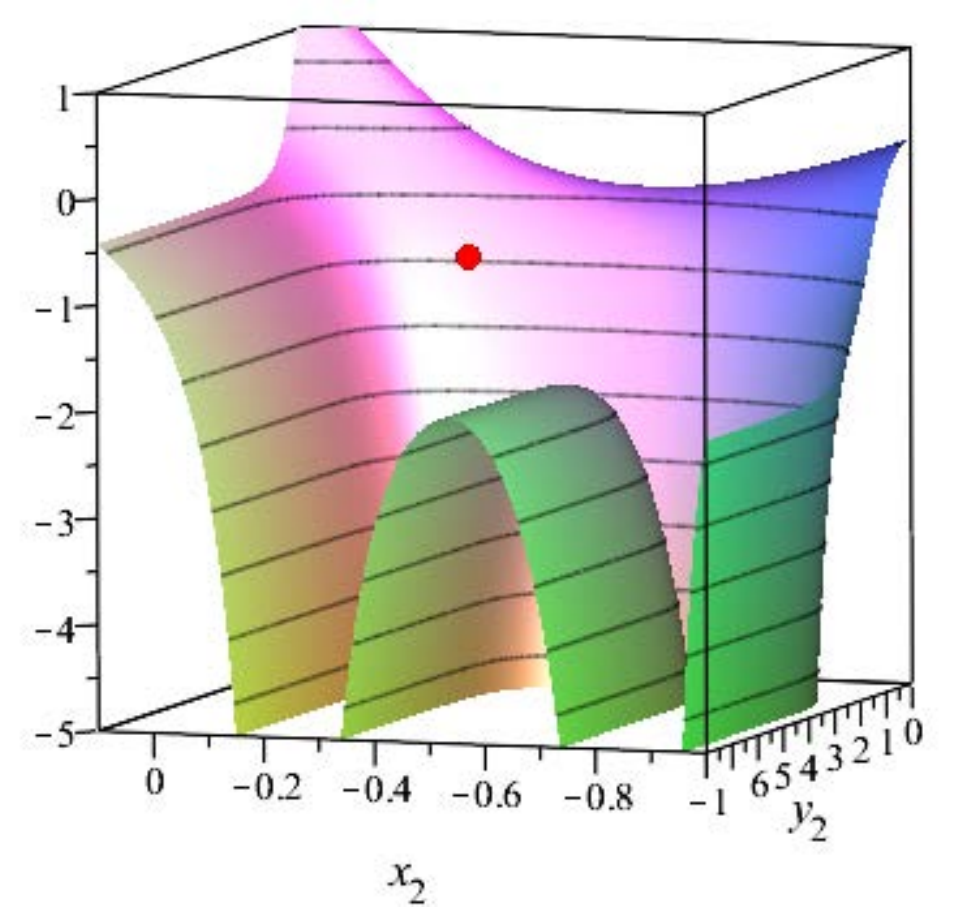}
\hspace{0.5cm}
\includegraphics[width=7cm]{\FigDir/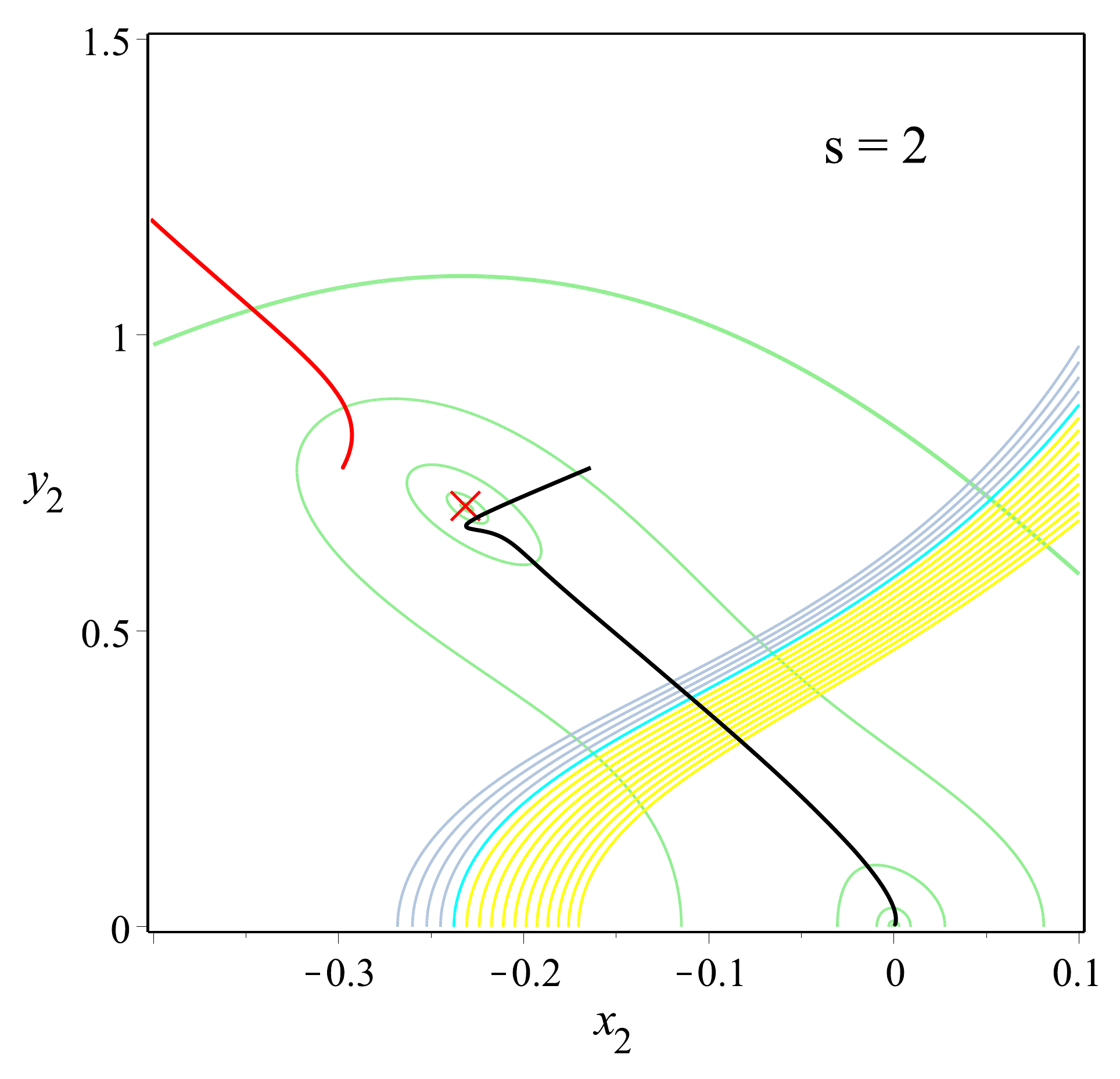}
}
\caption{The left panel is a 3D plot of $\eta_V$ on $\Sigma_2$, and the right panel shows contour plots of $\epsilon_V$ (green curves) and $\eta_V$ ($=-0.04, -0.06, -0.08, -0.1$ for grey curves, $=-0.02$ for the light blue curve, and $=0, 0.02, \cdots$ for yellow curves). Both are for $s=2$. Two examples of trajectories starting near the DI saddle point (red cross) on $\Sigma_2$ are shown by the black and red curves.}
\label{fig:etaplot:Z2:s=2}
\end{figure}

\def\nde#1#2{ #1 \cdot 10^{#2}}
\begin{table}
\begin{center}
{
\small
\begin{tabular}{|l|c|c|c|c|l|}
\hline
$s-s_c$ & $x_*$ & $y_*$ & $V_*$ & $H_*$ & $\eta$ \\
\hline
$\infty$      & $-0.191$ & $0.573$ & $0.294$ & $0.313$ & $[-0.145, 2.309; -1.652, -1.633, 0.8069; 0]$\\
$10^{-1}$ & $-0.304$ & $1.01$ & $0.420$ & $0.618$ & $[-0.7908, -0.9002; -0.5643, 3.37, 1.48; 0]$\\
$10^{-2}$ & $-0.389$ & $1.60$ & $0.536$ & $0.661$ & $[-0.3126, 8.52; -0.2075, 6.42, 3.66; 0]$\\
$10^{-3}$ & $-0.423$ & $2.19$ & $0.592$ & $0.690$ & $[-0.09968, 26.7; -0.05396, 24.7, 17.4; 0]$\\
$10^{-4}$ & $-0.435$ & $2.77$ & $0.611$ & $0.701$ & $[-0.03130, 85.1; -0.01603, 83.1, 61.3; 0]$\\
$4\cdot 10^{-5}$ & $-0.437$ & $3.00$ & $0.615$ & $0.703$ & $[-0.01976, 134; -0.01003, 132, 98.6; 0]$\\
$1.4\cdot 10^{-5}$ & $-0.438$ & $3.26$ & $0.617$ & $0.705$ & $[-0.01168, 228; -0.005891, 226, 168; 0]$\\
$10^{-5}$ & $-0.438$ & $3.34$ & $0.618$ & $0.705$ & $[\nde{-9.868}{-3}, 270; \nde{-4.971}{-3}, 268, 200; 0]$\\
$1.4\cdot 10^{-6}$ & $-0.440$ & $3.84$ & $0.620$ & $0.706$ & $[\nde{-3.688}{-3}, 722; \nde{-1.849}{-3}, 720, 539; 0]$\\
$10^{-6}$ & $-0.440$ & $3.92$ & $0.620$ & $0.706$ & $[\nde{-3.117}{-3}, 850; \nde{-1.562}{-3}, 853, 639; 0]$\\
$10^{-10}$ & $-0.440$ & $6.22$ & $0.621$ & $0.707$ & $[\nde{-3.115}{-5}, \nde{8.55}{4};$\\
           &          &         &         &        & $\qquad \nde{-1.557}{-5}, \nde{8.55}{4}, \nde{6.41}{4}; 0]$\\
$0$          & $-0.440$ & $\infty$ & $0.621$ & $0.707$ & $[-0, +\infty; -0, +\infty, +\infty; 0]$\\
\hline
\end{tabular}
}
\end{center}
\caption{Saddle point data}
\label{tbl:SaddlePointData}
\end{table}

At the $\SO(4)\times\SO(4)$-invariant critical point at the origin, the $\eta$ matrix  is simply
\Eq{
\bm{\eta}=\Mat{2 & 0 \\ 0 & 1}\oplus \Mat{-1 & -\sqrt{3}\\-\sqrt3 & 1}\oplus\Mat{-2 & 0\\ 0 & 0}.
}
Hence, the eigenvalues become negative in two directions orthogonal to $\Sigma_2$ (one in the second block and the other in the third block), and positive in the $\Sigma_2$ direction.  All the eigenvalues are of order unity and independent of the deformation parameter $s$.  Off of the center, negative directions change at each point, but the modulus of the eigenvalues are of order unity for a generic value of the deformation parameter $s$.

Fig. \ref{fig:etaplot:Z2:s=2} shows the behavior of $\eta_V$ on $\Sigma_2$ for $s=2$. As shown in Fig. \ref{fig:Potential:Z2:s=2}, the potential for $s=2$ has two DI saddle points in addition to the central extremum on $\Sigma_2$. The potential rapidly falls off beyond these DI saddle points. Despite this distinguished behavior, nothing special happens for the behavior of $\eta_V$ around these saddle points. There appears a narrow strip on which $\eta_V$ is around $-0.02$, but outside this strip $|\eta_V|$ is large and hence no inflation occurs.

To be precise, there appear two types of trajectories: those falling toward the waterfall valley (the red trajectory in Fig. \ref{fig:Potential:Z2:s=2}), and  those rolling down to the central extremum. In the former case, the spacetime immediately collapses into the anti-de Sitter phase, while, in the latter case,  the expanding  FLRW universe with $\Lambda>0$ is achieved during the roll. If the inflaton starts exactly from the $\ZR_2$-invariant plane $\Sigma_2$ and the initial velocity is tangent to this plane, the inflaton remains inside this surface due to the $\ZR_2$ symmetry of the potential, and the universe asymptotes to the de Sitter universe. In reality, quantum fluctuations render such a fine-tuning of the initial condition impossible and there inevitably exists a deviation from the $\Sigma_2$ plane. This deviation becomes substantial as the inflaton approaches the origin, and in the end falls suddenly to the waterfall directions perpendicular to $\Sigma_2$. Since the potential has no lower bounds in our system, it leads to an immediate collapse of the universe into the anti-de Sitter phase. When the potential is modified to have a realistic physical minimum by the coupling to an appropriate matter sector, one may be tempted to expect that the universe would evolve to a hot big-bang universe.  However, this hope is not satisfactorily realized since the total amount of e-foldings cannot exceed $O(10)$ for $s=2$ (see Fig. \ref{fig:alpha:Z2:s=2}), as seen from the behavior of $\epsilon_V$ (see Fig. \ref{fig:etaplot:Z2:s=2}) together with the standard formula of the e-folding number $N$ for the single-inflaton slow roll inflation,
\Eq{
N =\int \frac{d\phi}{\sqrt{2\epsilon}}\,.
\label{Nbyepsilon}
}
It follows that the inflation is unlikely to occur when $s$ is not close to the critical value $s=s_c$.

The situation changes drastically when we tune the deformation parameter $s$ to be close to the critical value $s=s_c$. As discussed in section~\ref{subsec:SO44:extremal}, two negative eigenvalues of the $\eta$ matrix go to zero as $s\tend s_c$, and the one corresponding to the $\Sigma_2$ direction is the smallest when $s \sim s_c$. Hence, we can identify this negative eigenvalue as $\eta_V$. In the region where $\eta_V<0$, trajectories starting from a point close to the $\Sigma_2$ plane converge to an attractor slow roll trajectory.  Hence, one can expect that the case with small $|\eta_V|$ at the DI saddle point would drive inflation,  provided the inflaton starts from a point near the DI saddle point. Our numerical computations confirm that this expectation is indeed true.  As is clear from Fig. \ref{fig:eta-s} and  Table \ref{tbl:SaddlePointData}, we see that the tuning of the deformation parameter $s$ to the level $s-s_c \lsim 10^{-5}$ is necessary.

\begin{figure}
\centerline{
\includegraphics[width=7cm]{\FigDir/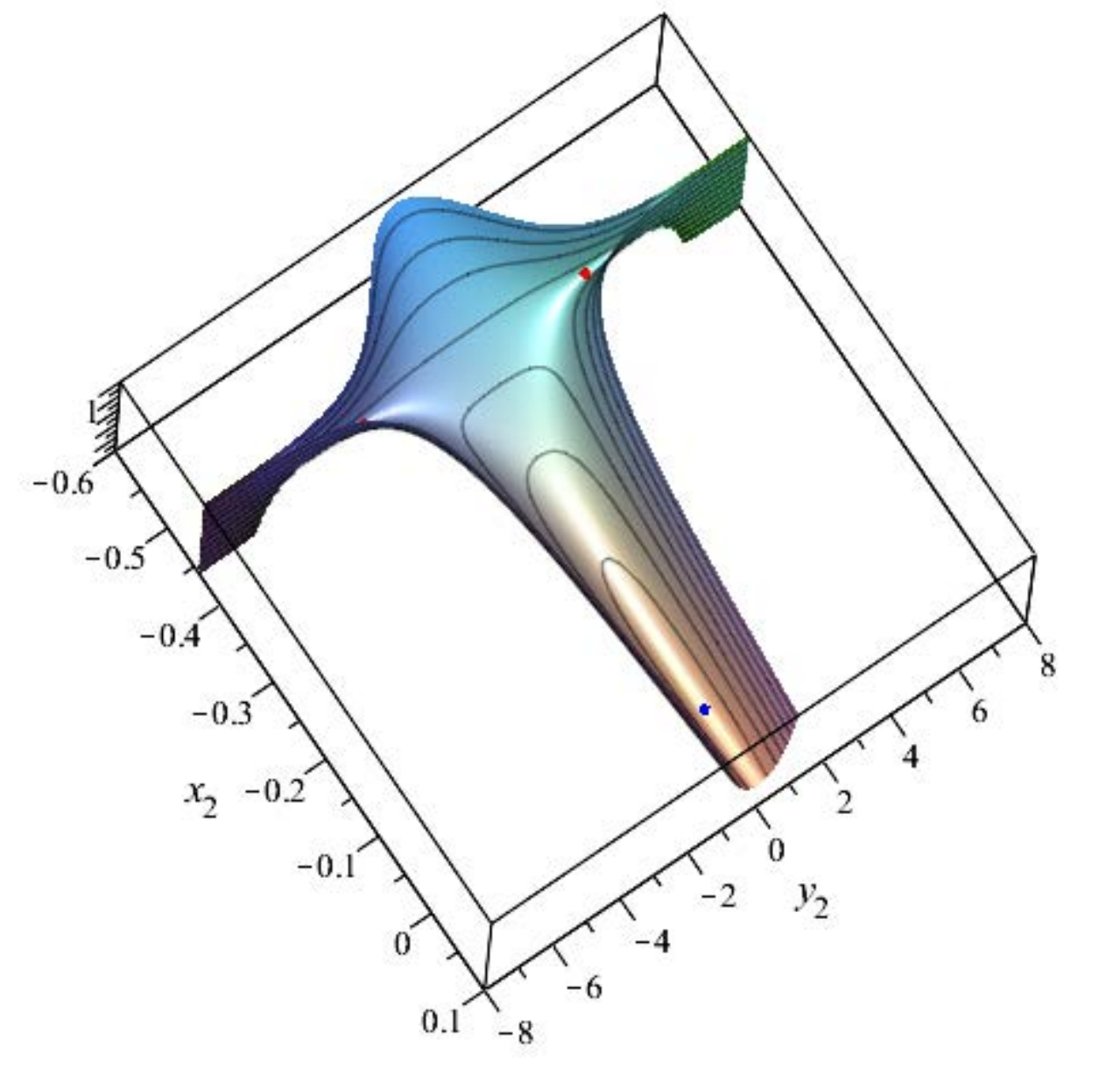}
\hspace{0.5cm}
\includegraphics[width=8cm]{\FigDir/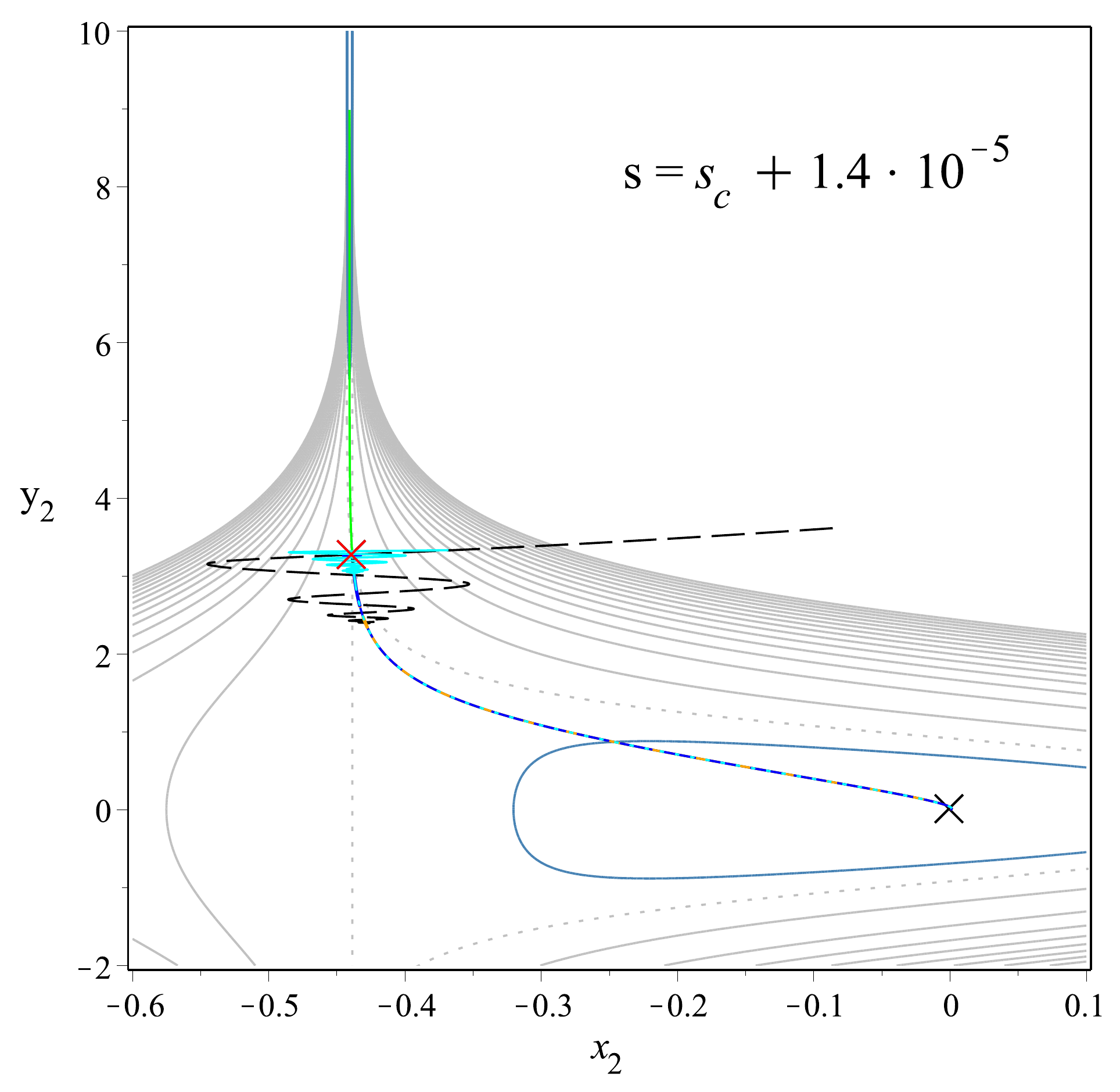}
}
\caption{The left panel is a 3D plot of the potential on $\Sigma_2$ for $s=s_c+1.4\cdot 10^{-5}$, and the right panel is its contour plot. The DI saddle points and the central dS extremum are shown by red and blue dots, respectively, as for $s=2$.  Some examples of trajectories on the $\Sigma_2$ plane are shown on the right panel.}
\label{fig:Potential:Z2:s=scr+1.4e-5}
\end{figure}

\begin{figure}
\centerline{
\includegraphics[width=8cm]{\FigDir/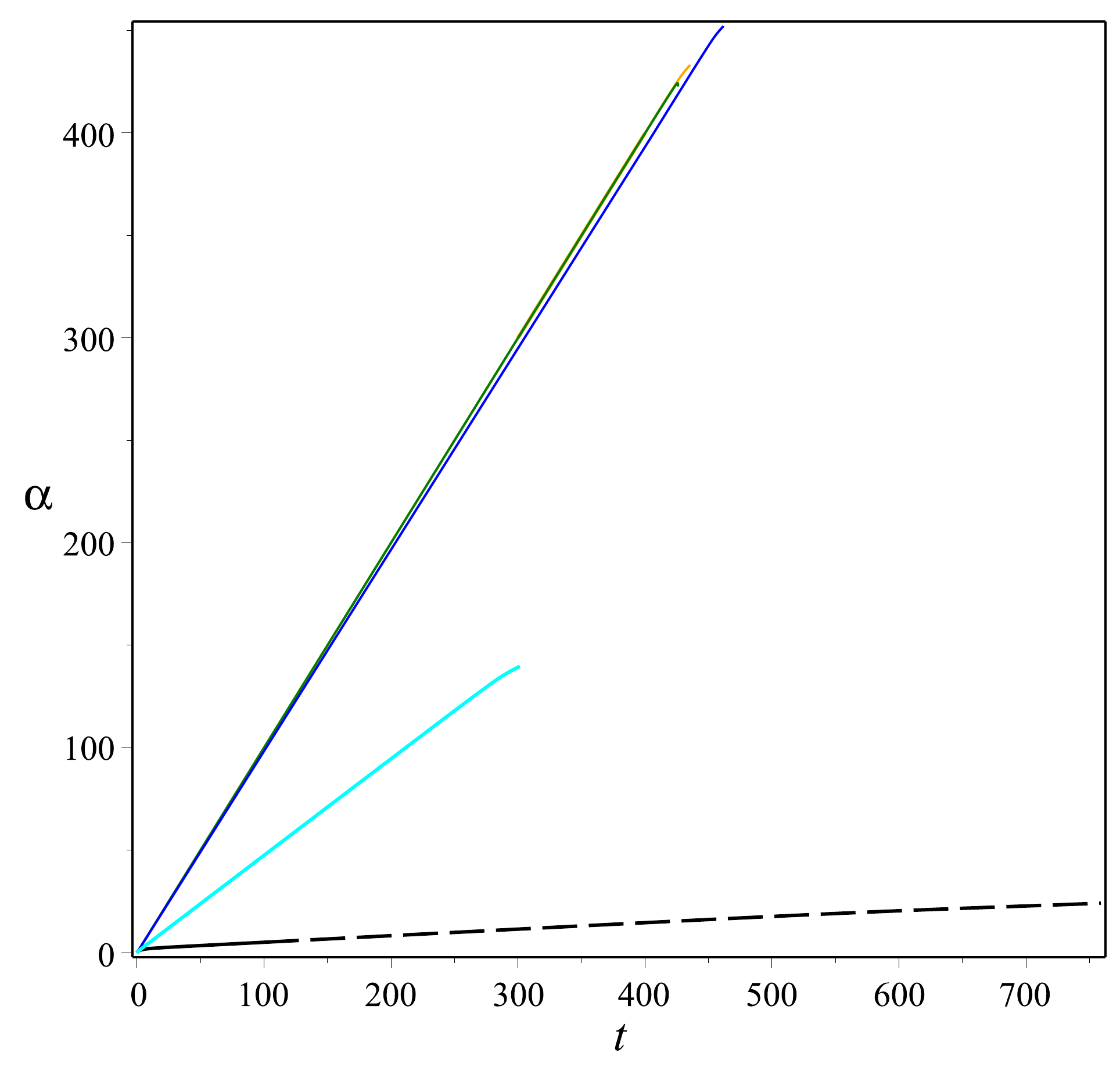}
\hspace{5mm}
\includegraphics[width=8cm]{\FigDir/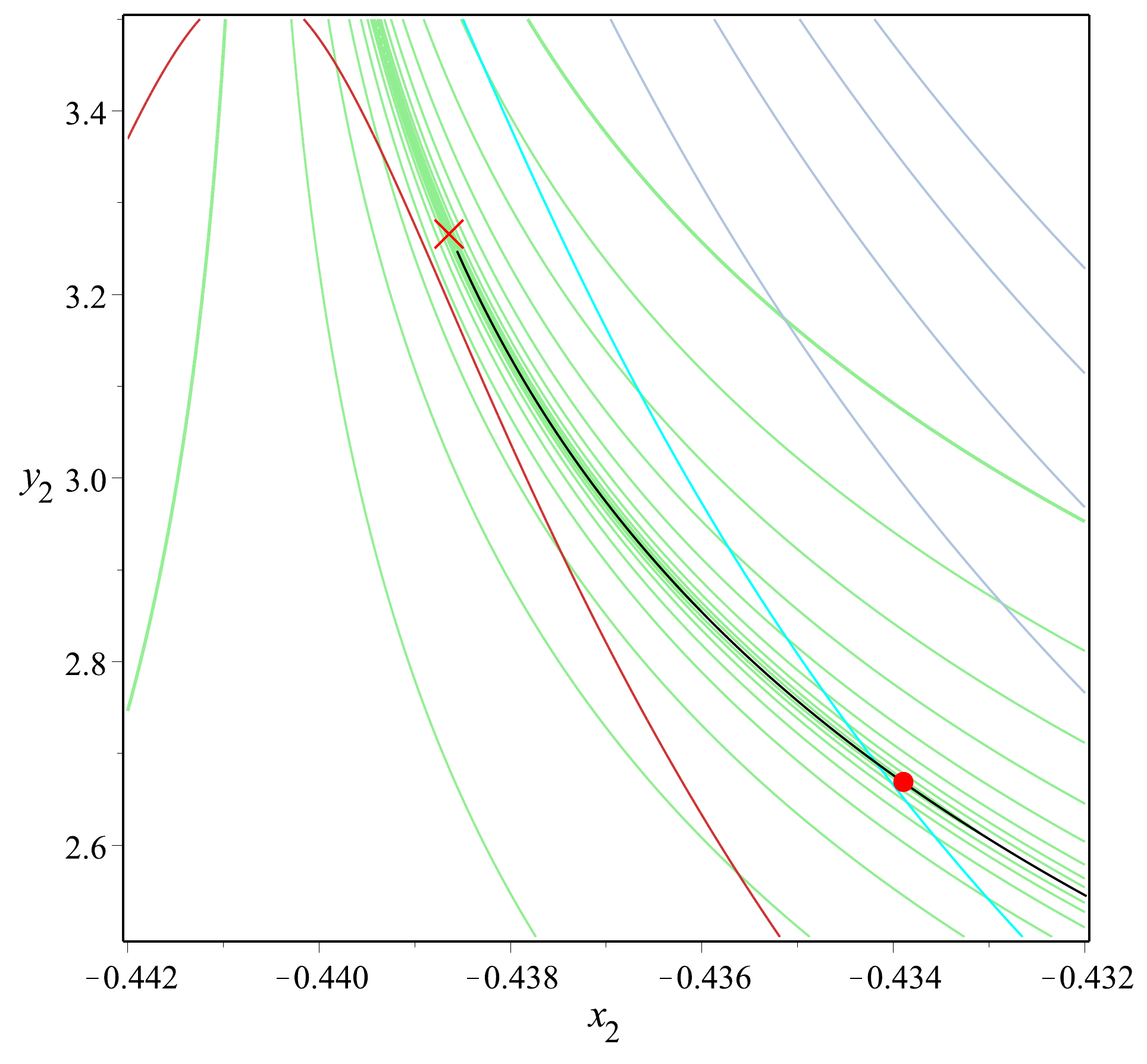}
}
\caption{The left panel shows the time evolution of $\alpha(t)$ for trajectories shown in Fig.\ref{fig:Potential:Z2:s=scr+1.4e-5}. The trajectory and $\alpha(t)$ for the same solution are drawn with the same color. In the right panel,  the blue inflationary trajectory is shown with the contours of $\epsilon_V$ (green curves) and $\eta_V$ (grey, lightblue and yellow curves). $\epsilon_V=1$ for the outermost thick green curve and decreases toward inner curves by the factor $10^{-1/2}$. The orange and lightblue curves correspond to $\eta_V=-0.01$ and $-0.02$, respectively, and the grey curves correspond to $\eta_V=-0.03,-0.04,\cdots$. The red dot on the trajectory indicates the $N=60$ point.}
\label{fig:alpha:Z2:s=scr+1.4e-5}
\end{figure}

\begin{figure}
\centerline{
\includegraphics[width=10cm]{\FigDir/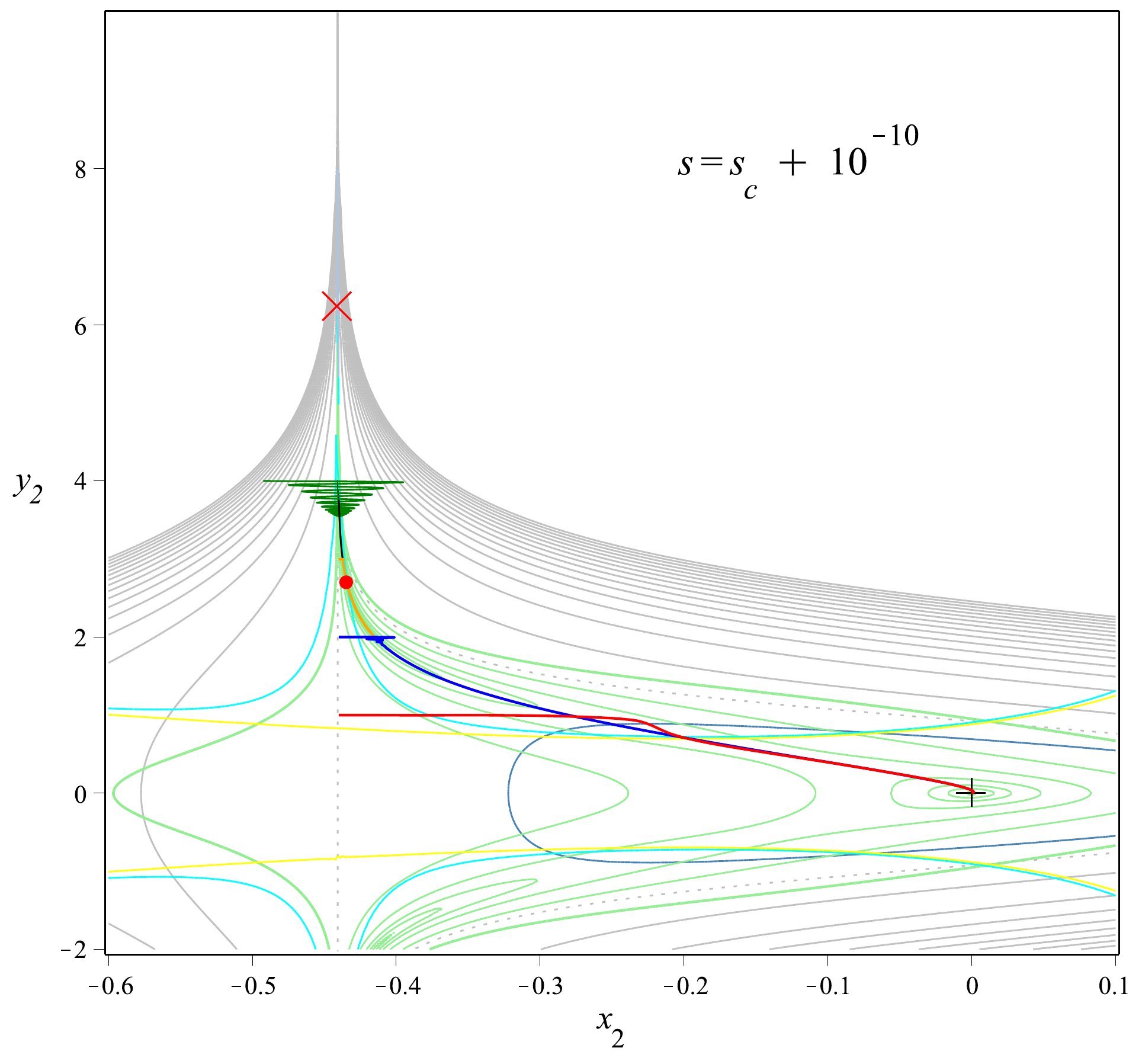}
\hspace{5mm}
\includegraphics[width=5cm]{\FigDir/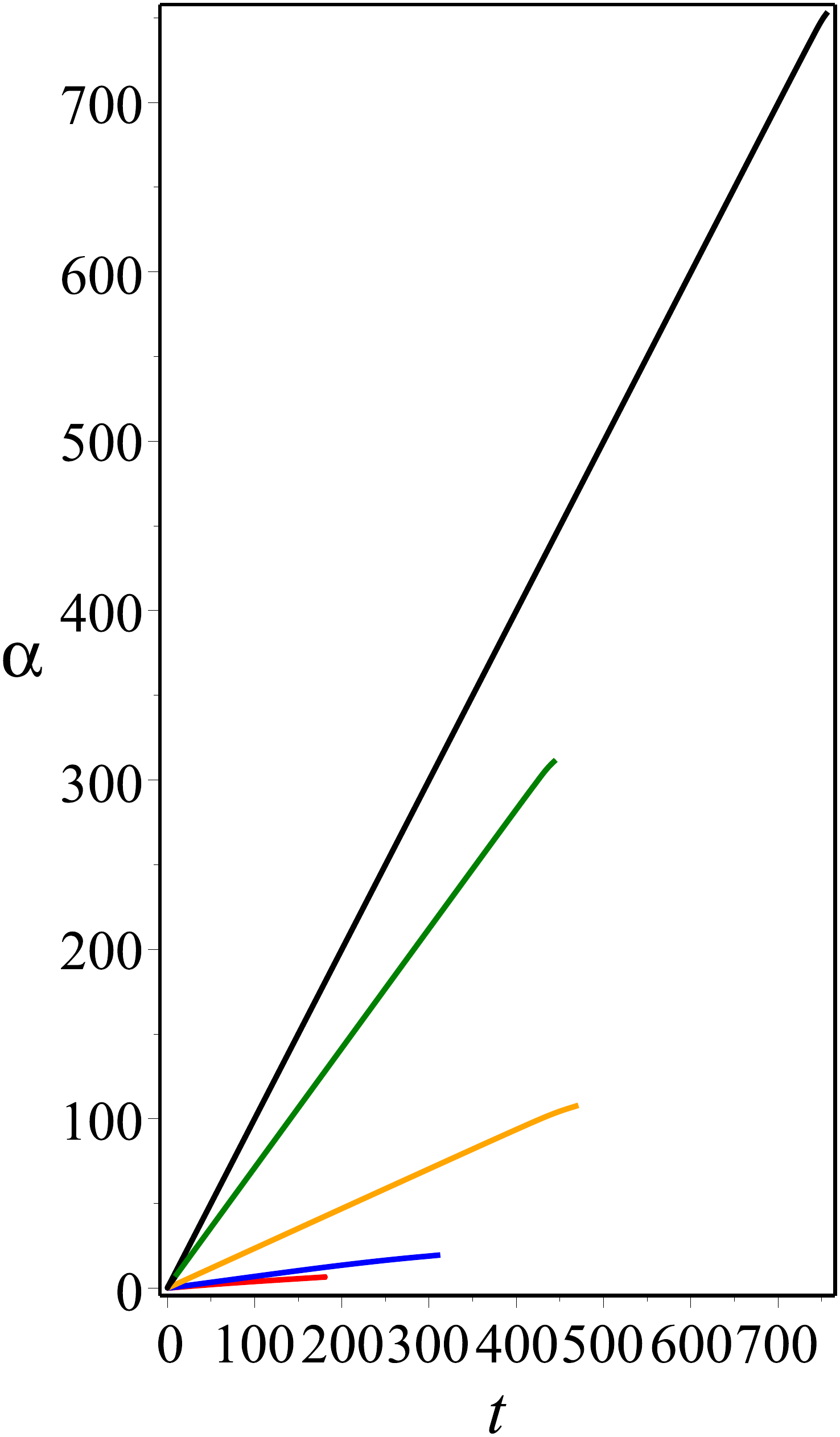}
}
\caption{The left panel shows examples of trajectories with the contours of $V$ (grey curves), $\epsilon_V$ (light green curves) and $\eta_V$ ($=-0.02$ (light blue) and $=0$ (yellow)) for $s=s_c + 10^{-10}$. $\epsilon_V=1$ for the outermost thick light green curve and decreases toward inner curves by the factor $10^{-1}$. The trajectories with black, orange, blue and read colors start from $(x_2,y_2)=(x_*, 4)$, $(x_*, 3)$, $(x_*,2)$ and $(x_*,1)$, respectively, while the green trajectory starts from $(-0.4924, 4)$ where $V=10 V_*$.  The red dot on the inflationary trajectories marks the $N=60$ point again. The right panel shows the time evolution of $\alpha(t)$ for trajectories on the left panel. The time variable $t$ is appropriately scaled for each trajectory to make the figure clearer.}
\label{fig:alpha:Z2:s=scr+1e-10}
\end{figure}

\begin{figure}
\centerline{
\includegraphics[width=4.4cm]{\FigDir/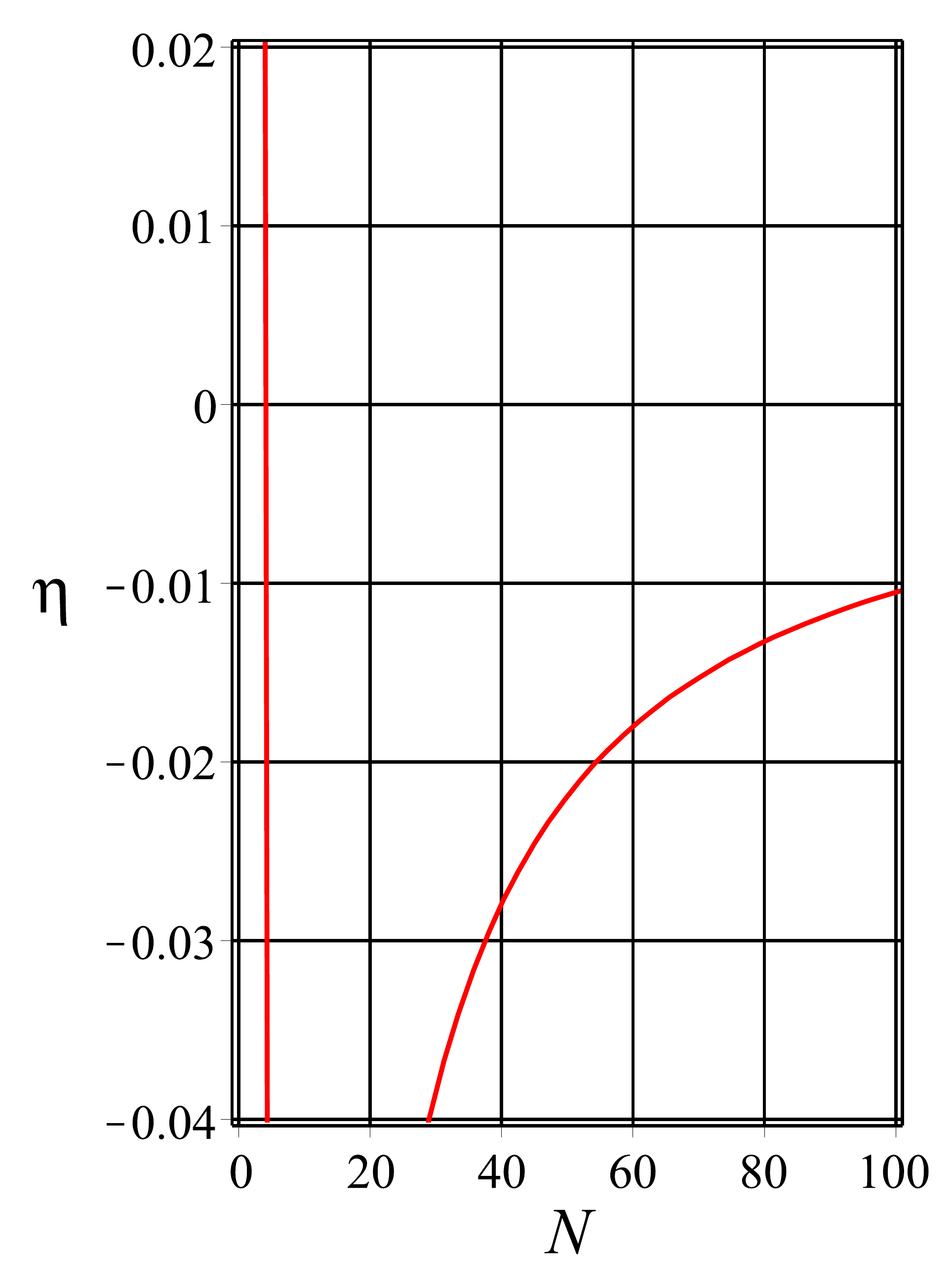}
\hspace{5mm}
\includegraphics[width=4.4cm]{\FigDir/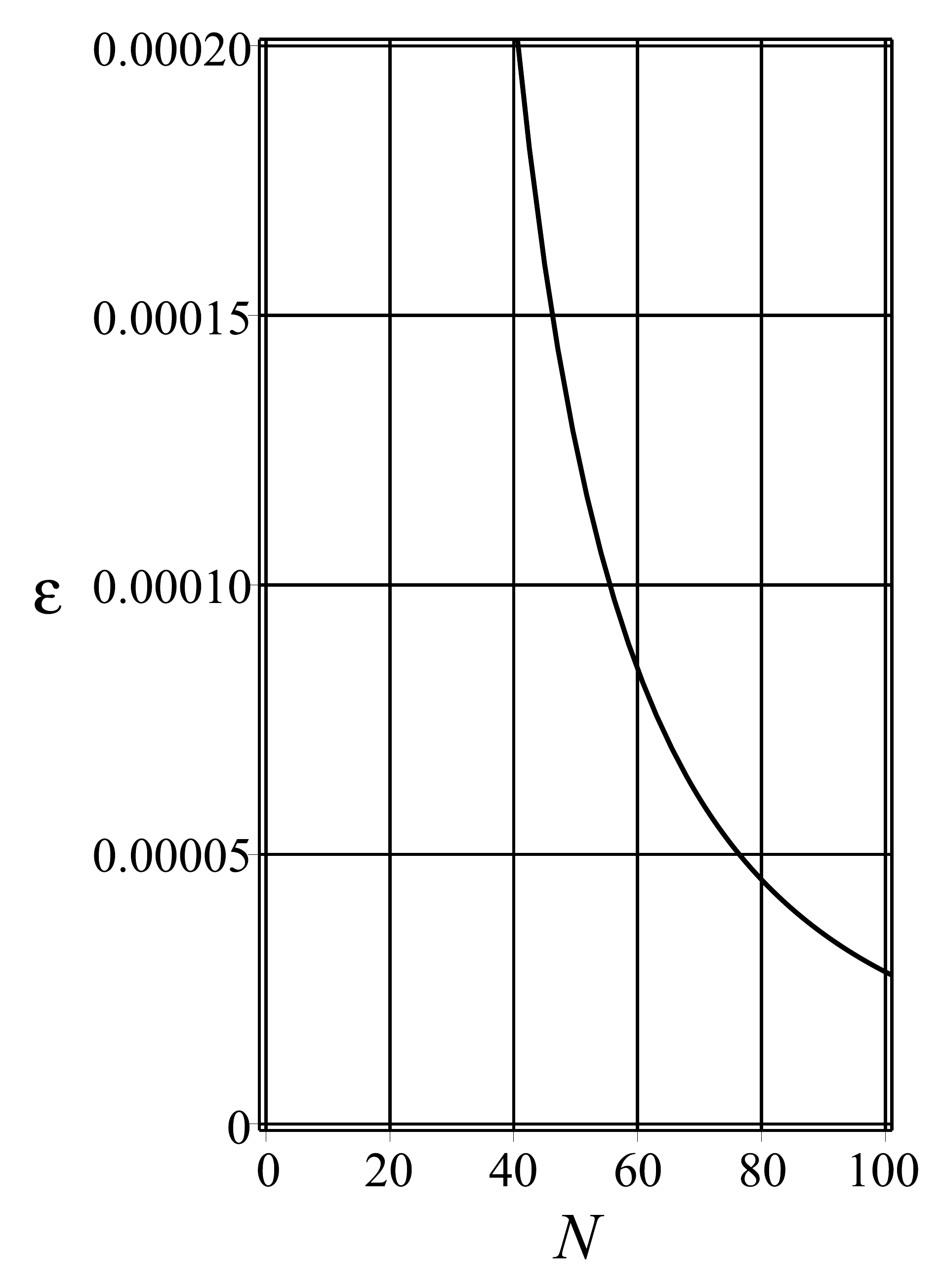}
\hspace{5mm}
\includegraphics[width=4cm]{\FigDir/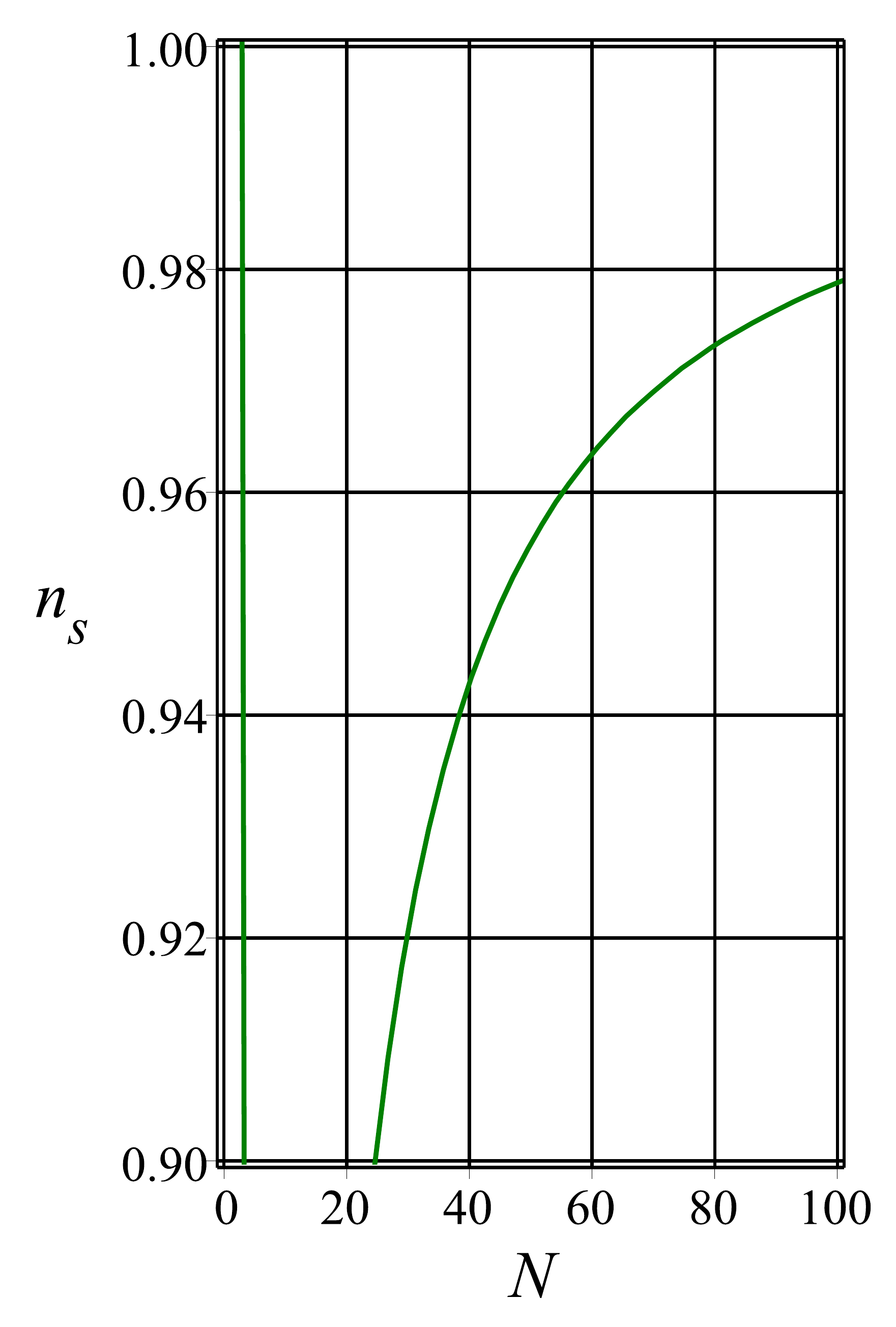}
}
\caption{These three graphs show the values of $\eta_H$, $\epsilon_H$ and $n_s$ as functions of the e-folding number $N$ between the observational point and the end of inflation for the attractor trajectory in the center direction. When $s$ is very close to $s_c$, these functions are independent of $s$. }
\label{fig:SRparam:s=scr+1e-10}
\end{figure}

\subsection{Dynamics in the $\ZR_2$-invariant 2D plane $\Sigma_2$}

Let us describe results of our numerical studies and discuss the condition under which we can realize the inflation consistent with observations  in the six-dimensional $\SO(3)\times\SO(3)$-invariant sector by tuning the deformation parameter $s$. As we have already mentioned, we cannot construct a cosmological model describing both the inflationary phase and the subsequent big-bang phase in the framework of maximal gauged supergravity, since the potential does not have a lower bound and there exists no local minimum at which the reheating occurs. In this paper,  we only discuss the features specific to the inflationary phase, such as the total e-folding number, the spectral index $n_s$ of the scalar curvature perturbation and the tensor-scalar ratio $r$. The issue of reheating is an interesting future work.

Let us first discuss the case in which the inflaton starts from a point on the $\ZR_2$-invariant plane $\Sigma_2$ with a vanishing initial velocity. In this case, its trajectory stays inside $\Sigma_2$ due to $\ZR_2$-invariance. The reflection invariance of the $y$ coordinate allows us to confine to the upper half of the $\Sigma_2$ plane. We mainly consider trajectories starting around the DI saddle point in this upper half plane.

Figures  \ref{fig:Potential:Z2:s=scr+1.4e-5} and \ref{fig:alpha:Z2:s=scr+1e-10} depict the behavior of the potential for $s=s_c+1.4\times 10^{-5}$ and $s=s_c+10^{-10}$, respectively.  One sees that the potential for all $s>s_c$ has the same qualitative structure as that for $s=2$ shown in  Fig. \ref{fig:Potential:Z2:s=2}. As $s$ approaches $s_c$, the $y$-coordinate of the DI saddle point increases, and along with it, the waterfall regions turn into very narrow valleys with steep side walls.  The contours of $V$ passing the DI saddle point consist of two curves: one is the $x=x_*$ line and the other
approaches the one parallel to the former. The separation of these two lines decreases in proportion to $s-s_c$. In particular, the DI saddle point runs away to infinity at the limit $s=s_c$, and the waterfall region disappears. There remains only a narrow infinitely long asymptotically flat aisle whose width goes to zero at infinity.

The right panel of Fig. \ref{fig:Potential:Z2:s=scr+1.4e-5} shows some examples of trajectories for $s=s_c+1.4\times 10^{-5}$.  As this figure shows, there appear two attractor slow-roll trajectories as in the case of $s=2$. Although all trajectories approach either of them eventually, the total inflation rate of each trajectory is sensitive to the moment when the inflaton settles down to a slow roll trajectory. In fact, as the left panel of Fig. \ref{fig:alpha:Z2:s=scr+1.4e-5} shows, if the inflaton starts from a point with a large potential value, it initially oscillates with a large amplitude before it settles down to an attractor trajectory. During this oscillatory phase the cosmic expansion is not inflationary, because the kinetic energy dominates over the potential energy. As a consequence, the total inflation rate becomes small.  This places a constraint upon the initial point of the inflaton. Nevertheless, this constraint is not so tight than expected. In fact,  inflation with a total e-folding number larger than 100 is achieved even if the offset of the initial point of the inflaton from the DI saddle point is of order 0.1.

Let us next look at the values of slow roll parameters at the time $t=t_{\rm O}$ when the comoving scale corresponding to the present horizon scale comes out from the Hubble horizon during inflation. From the requirement that the flatness and horizon problems are resolved,  the e-folding number $N$ from that time to the end of inflation is constrained to be $50\lesssim N \lsim  70$. This wide range of ambiguity comes from the uncertainties of the inflation energy scale and the length of preheating phase. In the present paper, we assume that $N=60$ corresponds to $t=t_{\rm O}$ for definiteness. The results for other values of $N$ can be easily read out from the data given in the present paper.

We give in Table \ref{tbl:ns:nf} the obtained values of the slow roll parameters $\eta_H$ and $\epsilon_H$ at $N=60$  for five solutions with sufficiently long inflation starting from different points around the DI saddle point. The offset of the initial point from the DI saddle point is the order of $H_*$ (the value of $H$ at the saddle point) with $g=0.01$ or $g=0.1$.  Note that the solutions to the equations of motion
($\phi(\tau)$, $\alpha(\tau)$ and $H(\tau)/H_*$) are independent of $g$,  where $\tau =H_* t$
is a dimensionless time variable.  Hence the value of $g$ only affects the initial position through the value of $H_* \propto g$. When we compare inflationary solutions with observations, the value of $H_*$ is determined by the observed amplitude of the scalar perturbation as we explain later. The corresponding value of $g$ is around $g=4\times 10^{-6}$ (see eq. (\ref{gval}) below). It follows that the deviation of $H_*(g=0.1)$ corresponds to $2\times 10^4$ times the actual value of $H$.

A very important feature of our results is that $\eta_H$ at $N=60$ is around $-0.02$ independent of the initial conditions. This is not related to $\eta_*$, the value of $\eta$ at the DI saddle point, because it is around $-0.01$ for $s=s_c+1.4\times 10^{-5}$ (see Table \ref{tbl:SaddlePointData}), hence $\eta_*$ is  half of $\eta$. We can attribute this feature to the fact that all inflationary trajectories immediately settle down to one of the two attractor slow-roll trajectories. On each attractor trajectory, the inflation rate is determined by the formula \eqref{Nbyepsilon}. As a consequence, the $N=60$ point is almost uniquely determined by $\epsilon_V$. Since $\eta_H$ is almost equal to $\eta_V$ for a slow roll trajectory, it turns out that the value of $\eta$ at $N=60$ becomes independent of the initial condition. In fact, the values of $\epsilon_H$ and $\eta_H$ at $N=60$ are almost equal for the first two solutions converging to the slow roll path down the waterfall, and for the subsequent three solutions approaching the center, separately. The value $2\epsilon_H\approx 10^{-4}$ at $N=60$ is also consistent with \eqref{Nbyepsilon} (see the right panel in Fig. \ref{fig:alpha:Z2:s=scr+1.4e-5}). Though, it is rather surprising that the value of $n_s$ at $N=60$ has a much smaller dispersion than $\eta_H$ and is almost the same for the two different attractor trajectories. What happens here is that the variation of $\epsilon$ cancels the variation of $\eta$ so that $n_s=1+2\eta-6\epsilon$ become nearly constant.

These interesting features become clearer when the value of $s$ becomes closer to $s_c$.   In Fig. \ref{fig:alpha:Z2:s=scr+1e-10}, we display trajectories starting from four different points on the line $x_2=x_*$ with $y=y_0=1,2,3,4$ for $s=s_c + 10^{-10}$.  We find that for $y_0=1$ and $2$, the initial point is below the $N=60$ point on the attractor trajectory with $y\simeq 2.69$, and as a result, inflation lasts only for a short period. In the meanwhile, for $y_0\ge 3$, sufficiently long inflation is realized. As is clear from the explanation above, we will obtain the same values for $n_s$ and $\epsilon$ for any trajectories whose initial value of $y$ is between $3$ and $y_*$, provided the initial potential height is not so high as illustrated by the green trajectory in Fig. \ref{fig:alpha:Z2:s=scr+1e-10}. This is  because the stability of the values of slow roll parameters at the $N=60$ point arises due to the uniqueness of the attractor trajectory.  In fact, Table \ref{tbl:ns:nf} shows that the value of $n_s$ at the $N=60$ point is independent of the initial condition with a good accuracy.

Even if $s-s_c$ becomes smaller further, the behavior of trajectories around and after the $N=60$ point does not change significantly because the DI saddle point runs away further and further from that region, and the potential around the $N=60$ point converges to the value for $s=s_c$. Fig. \ref{fig:SRparam:s=scr+1e-10} shows the $N$-dependence of the slow roll parameters $\eta_H$ and $\epsilon_H$ and the spectral index $n_s$ in this limiting region of $s$.

\begin{figure}
\centerline{
\includegraphics[width=7cm]{\FigDir/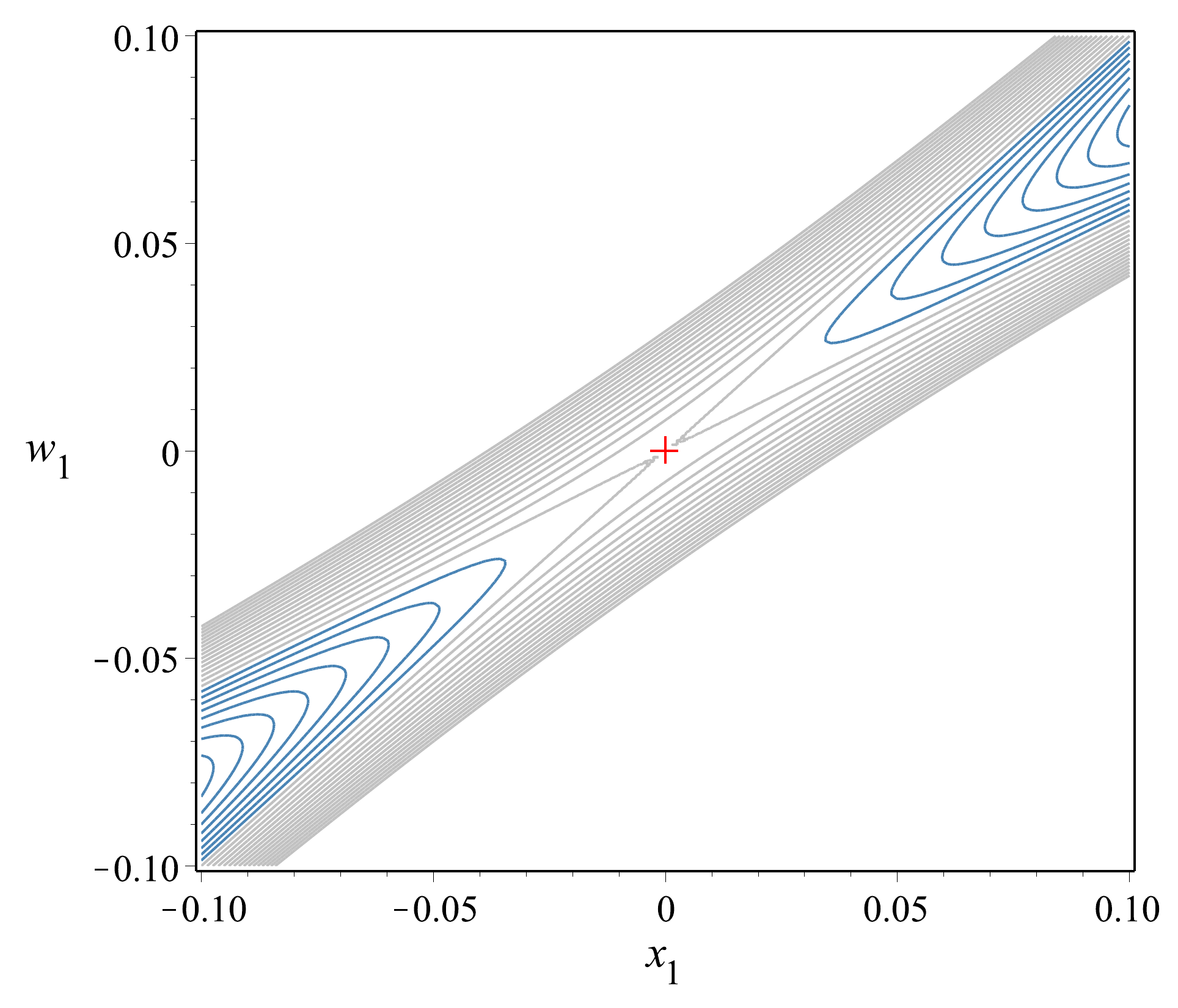}
\hspace{5mm}
\includegraphics[width=7cm]{\FigDir/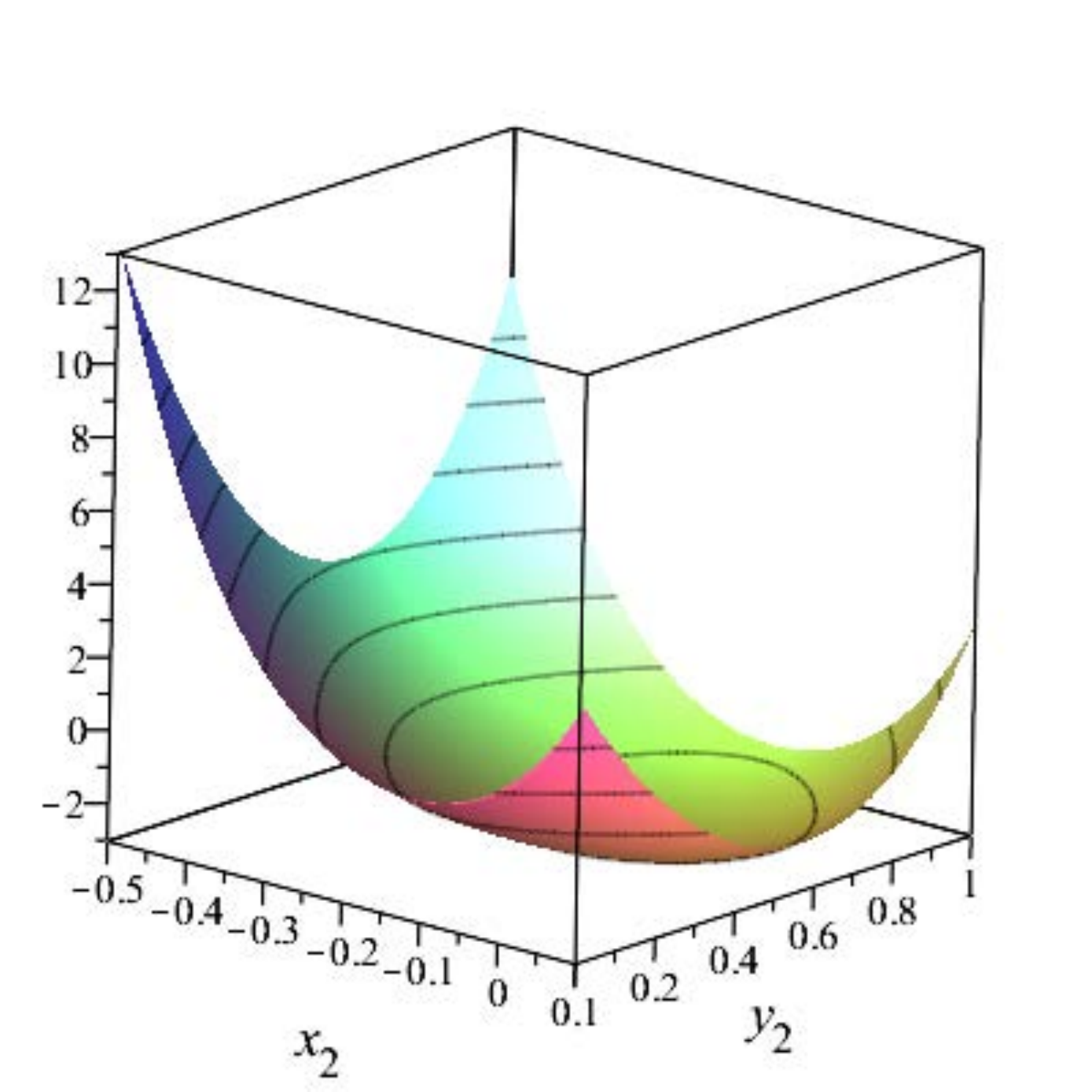}
}
\caption{The left panel is a contour plot of the potential on the $(x_1,w_1)$ passing through the DI saddle point orthogonally to $\Sigma_2$ for $s=s_c+10^{-10}$.  The right panel shows a 3D plot of the potential on the $(x_2,y_2)$ plane with $x_1=-2, w_1=-0.1$. }
\label{fig:Pot:OffZ2:s=scr+1e-10}
\end{figure}

\begin{figure}
\centerline{
\includegraphics[width=7cm]{\FigDir/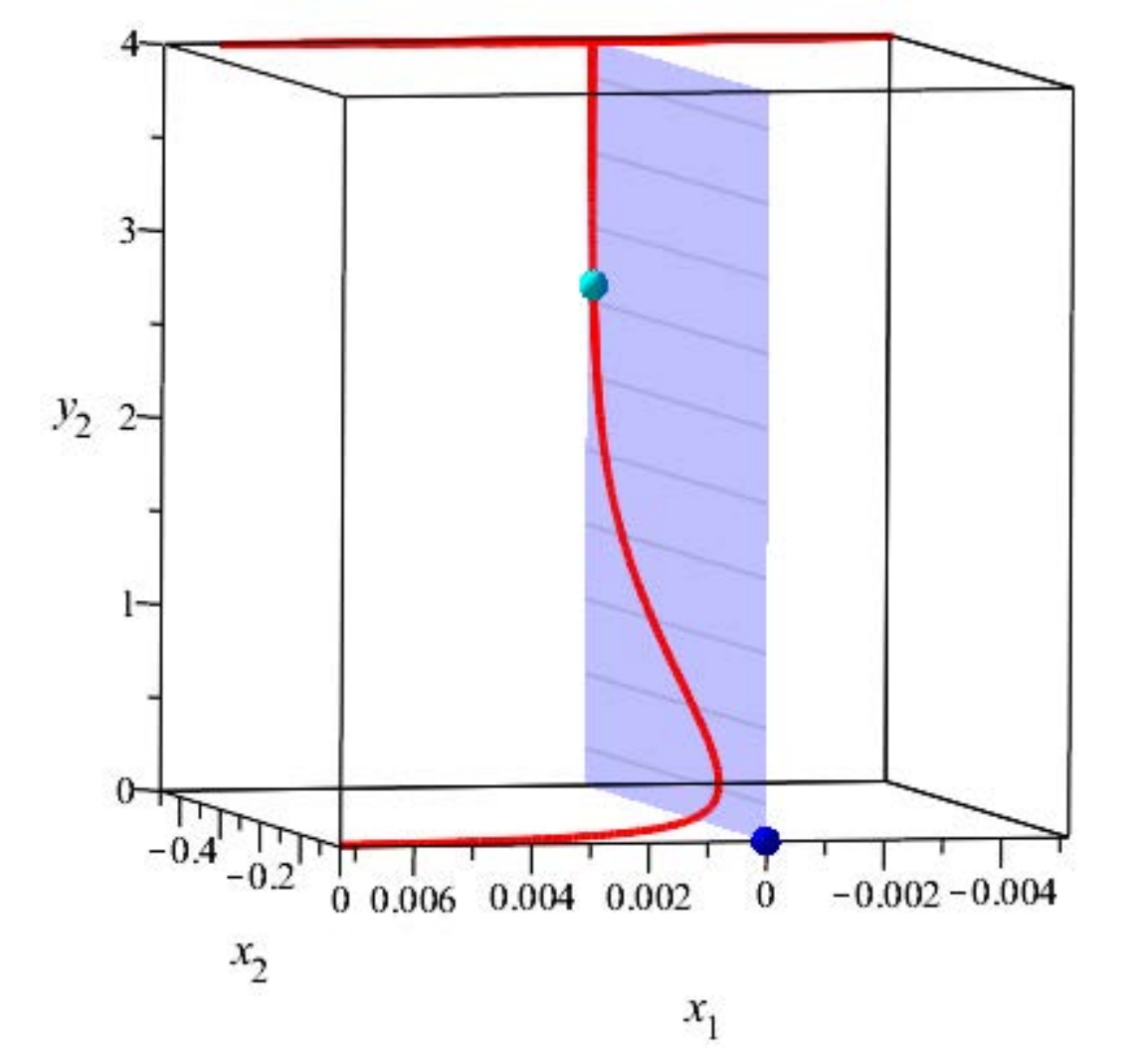}
\hspace{5mm}
\includegraphics[width=7cm]{\FigDir/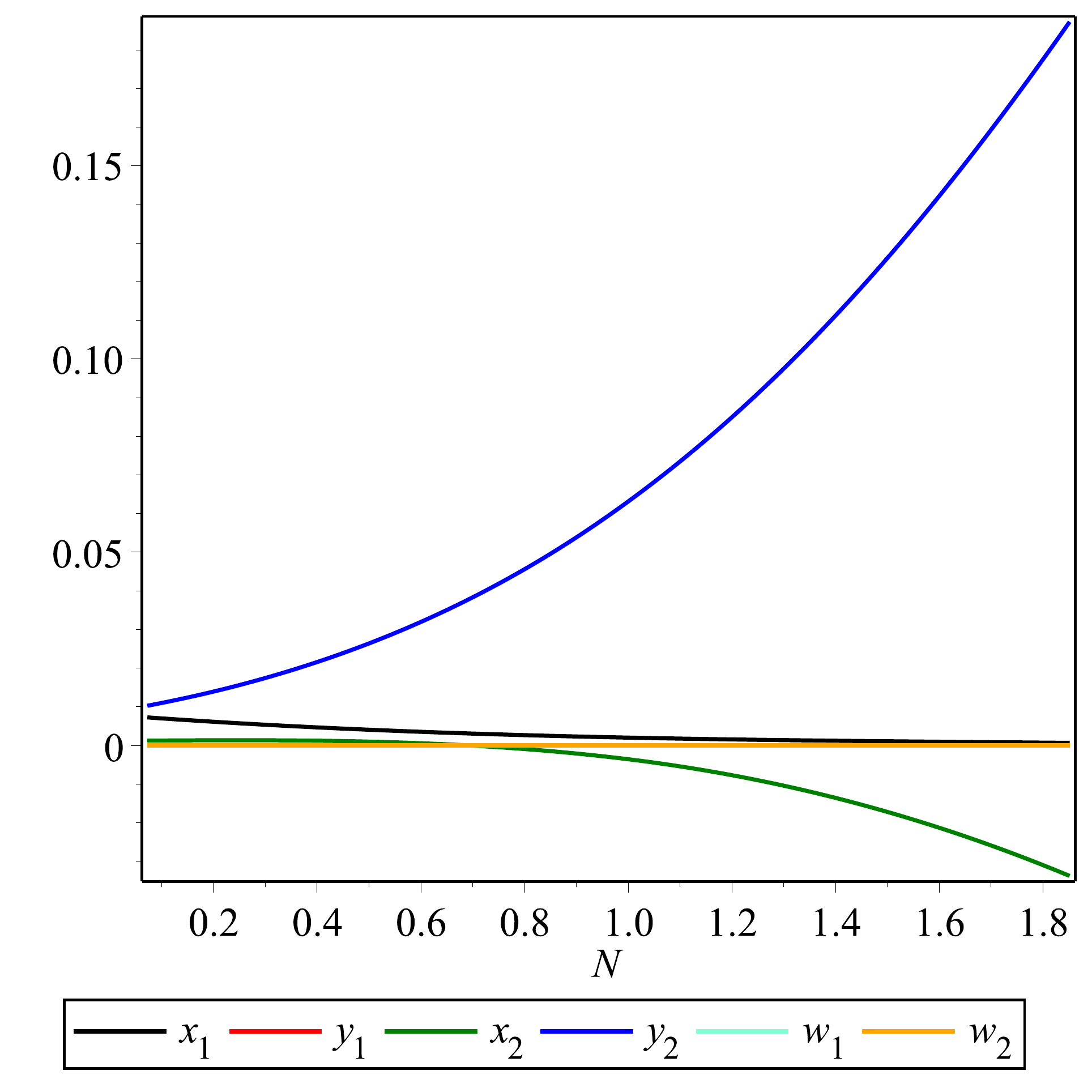}
}
\caption{The left panel shows the 3D trajectory of the solution for $s=s_c+10^{-10}$
with the initial condition: $ x=x_*, y=4, z=H_*(g=0.006), \theta_1=\pi/2, \theta_2=0,\theta_3=3\pi/4$.  The deep blue dot and the lightblue dot on the trajectory show the extremum at the origin and the $N=60$ point, respectively. The right panel shows the behavior of the six coordinates  around the end of inflation. }
\label{fig:3Dtraj:s=scr+1e-10}
\end{figure}

\begin{figure}
\centerline{
\includegraphics[width=7cm]{\FigDir/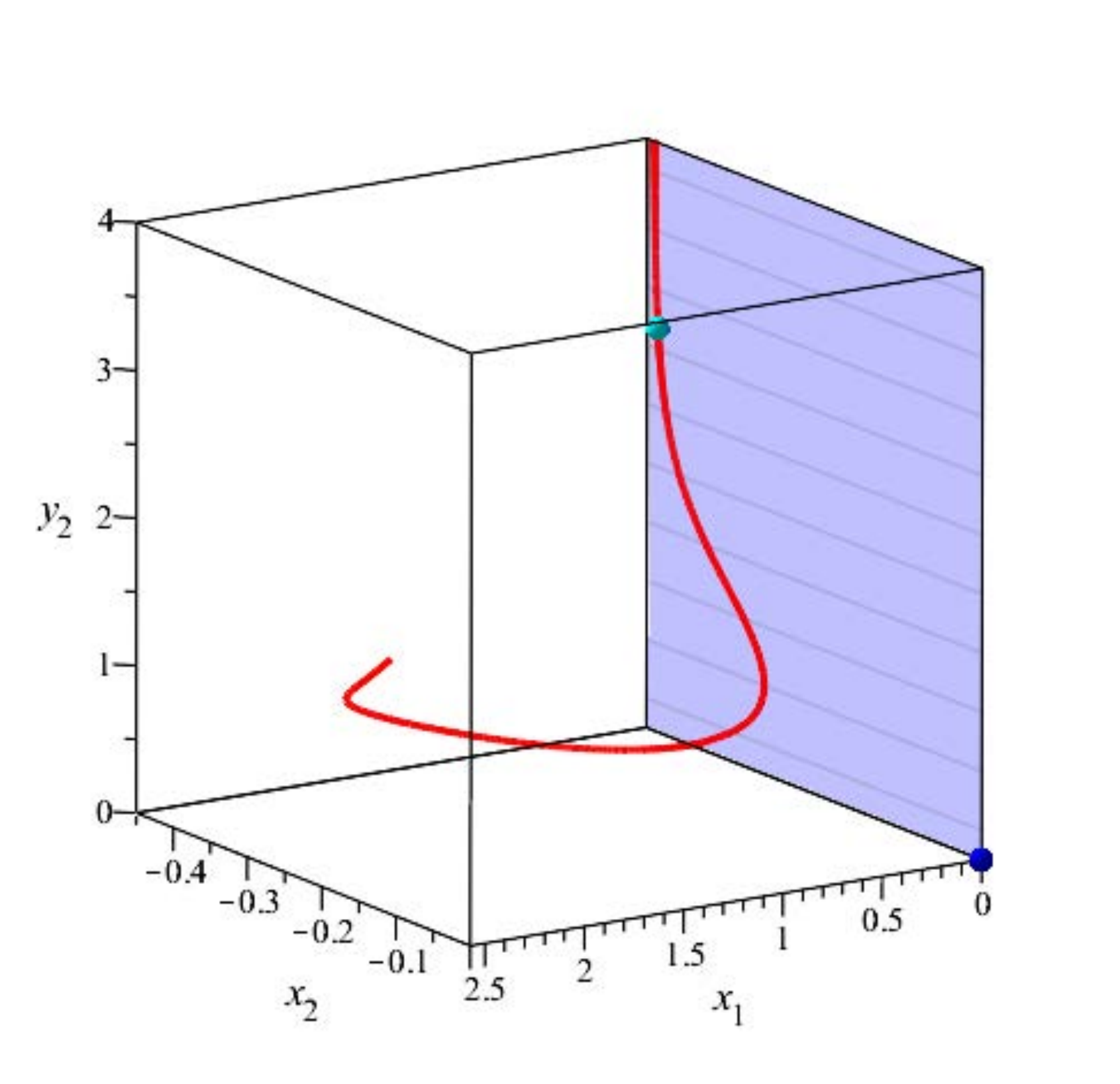}
\hspace{5mm}
\includegraphics[width=7cm]{\FigDir/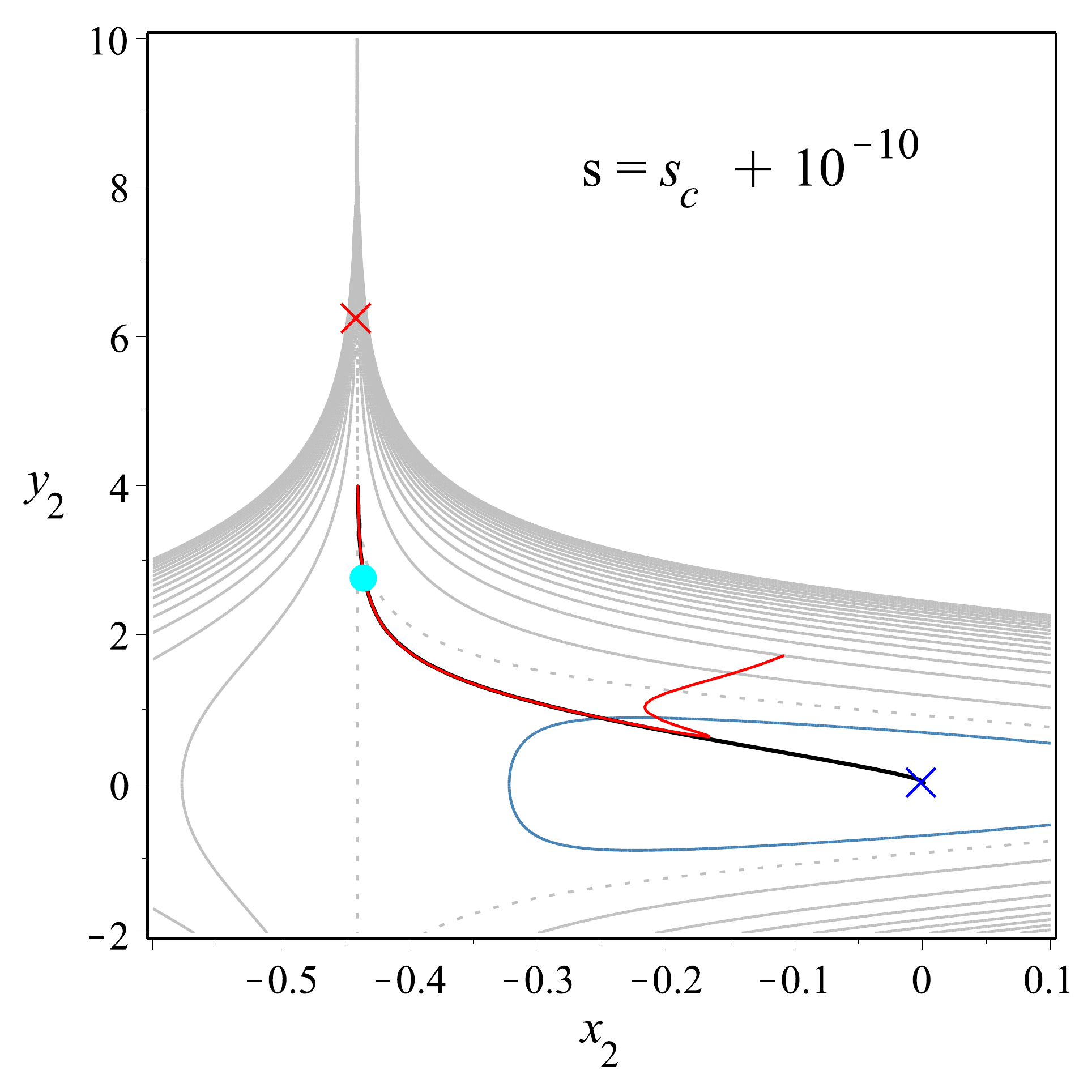}
}
\caption{The left panel shows the 3D trajectory of the solution for $s=s_c+10^{-10}$
with the initial condition: $ x=x_*, y=4, z=0, -y\sinh(2y)\cos(2\theta_3)=H_*(g=0.06)\sqrt2 \pi, \theta_1=\pi/2, \theta_2=0$. The meanings of the colored dots are the same as in Fig. \ref{fig:3Dtraj:s=scr+1e-10}. The right panel shows the projection of this trajectory (red) and that in Fig. \ref{fig:3Dtraj:s=scr+1e-10} on the $\Sigma_2$ plane. }
\label{fig:3Dtraj2:s=scr+1e-10}
\end{figure}

\subsection{Off $\Sigma_2$ trajectories}

Next, we report the results for the trajectories starting from a point outside  the $\ZR_2$-invariant plane $\Sigma_2$. As we saw in \ref{subsec:SO44:extremal}, the $\eta$ matrix has two eigenvectors with negative eigenvalues at the DI saddle point: one is parallel to $\Sigma_2$ and the other is tangent to the
$(x_1,w_1)$ plane (see the left panel of Fig. \ref{fig:Pot:OffZ2:s=scr+1e-10}).  Hence, the inflaton tends to move into the direction off of the $\Sigma_2$ plane when it starts outside this plane.
Nonetheless, the offset from the $\Sigma_2$ plane remains small until the inflaton approaches the origin around which all the negative directions become perpendicular to $\Sigma_2$.
This is because the eigenvalue in this direction is about one half of that in the $\Sigma_2$ direction (see Fig. \ref{fig:eta-s}), hence the inflaton acquires a large velocity in the $\Sigma_2$ direction.

We show illustrative examples of this feature in Fig. \ref{fig:3Dtraj:s=scr+1e-10} and Fig. \ref{fig:3Dtraj2:s=scr+1e-10}.
The solution displayed in Fig. \ref{fig:3Dtraj:s=scr+1e-10} admits an initial offset of the inflaton in the positive eigenvalue direction of the $\eta$ matrix.  In this case, the inflaton initially oscillates with a large amplitude passing through the $\Sigma_2$ plane orthogonally, but quickly settles down to a trajectory very close to the attractor slow-roll trajectory on $\Sigma_2$. In this stage, there remains a small offset from $\Sigma_2$, which grows rapidly at the final stage.
For the solution in Fig. \ref{fig:3Dtraj2:s=scr+1e-10}, the initial offset of the inflaton is of order $10^4H_*(g=4\times 10^{-6})$ with respect to the scalar manifold metric $K_{\alpha\beta}$ and in the negative eigenvalue direction of the $\eta$ matrix. In this case, the offset from the $\Sigma_2$ plane continues to grow from the beginning. As a consequence, the offset becomes of order unity before it comes close to the origin. The inflaton thus feels a potential force in the $(x_2,y_2)$ direction quite different from that on $\Sigma_2$, as illustrated by the right panel of Fig. \ref{fig:Pot:OffZ2:s=scr+1e-10}. It follows that the trajectory abruptly changes its direction when the offset grows to some critical value as shown in Fig. \ref{fig:3Dtraj2:s=scr+1e-10}.  The inflationary stage is thus destroyed by this property.

This violent offset behavior, however, occurs only at the final stage of inflation and does not affect the  evolution of inflaton around and before the $N=60$ point (marked by a lightblue dot in Figs. \ref{fig:3Dtraj:s=scr+1e-10} and \ref{fig:3Dtraj2:s=scr+1e-10}). It follows that the observational parameters take values similar to those for trajectories on the $\Sigma_2$ plane as shown in Table \ref{tbl:ns:nf} (the last two rows).


\subsection{Flux contribution}

Having discussed the case without flux, let us next examine the effects of non-vanishing gauge flux on the inflationary dynamics. The authors of ref.~\cite{Adshead:2012kp,Dimastrogiovanni:2012st} proposed a new mechanism
for slowing down the inflaton motion.  This class of models is referred to as the chromo-natural inflation and caused by the non-Abelian gauge field with a Chern-Simons coupling  of the form
\Eq{
e^{-1}\LL_{\rm CS}= \frac{\lambda}{f_a} \phi \Tr(F* F) \,,
\label{CScoupling}
}
where $\phi$ is an inflaton field, and $f_a$ is analogous to the axion decay constant and specifies the  scale where it comes into play. Fascinating features of this model are that it does not demand a flat potential and gives quite different observational predictions \cite{Adshead:2013nka}.  The only price to pay is that the coefficient $\lambda$ of the Chern-Simons term must be quite large. To be specific, it must be as big as 200 in the Planck units to achieve the 60 e-foldings. In the context of extended supergravities, the Chern-Simons coupling of the type (\ref{CScoupling}) is naturally incorporated, although the effective coupling constant $\Re(N)$ depends on the scalar fields in a more intricate fashion. It is therefore interesting to see whether a large effective Chern-Simons coupling of this sort appears in our $\SO(4,4)$ gauging model, in which there are no controllable parameters except for the gauge coupling constant $g$ and the deformation parameter $s$.

Another motivation to consider non-vanishing gauge flux is its relevance to anisotropic inflation models \cite{Watanabe:2009ct,Yokoyama:2008xw}.  Anisotropic inflation is a mechanism driven by  the vector or higher-rank tensor fields which admit a background configuration with an anisotropic energy-momentum tensor. Anisotropic inflation is interesting also from the general relativistic point of view, since it deserves a
novel counterexample to the cosmic no hair conjecture. In general, such an anisotropy quickly decays dynamically with cosmic expansion. For example, the free magnetic field $B$ decays in proportional to $1/a^2$ due to the flux conservation, where $a$ is the cosmic scale factor.  A possible mechanism to prevent this kind of decay of anisotropy is to allow an inflaton-dependent gauge coupling of the form
\Eq{
e^{-1}\LL_1 = -\frac12 f(\phi) F \cdot F \,.
}
The Maxwell equation reads $\nabla_\mu (f(\phi) F^{\mu\nu})=0$, implying that the combination $a^2 f(\phi) B$ is conserved. Hence a constant magnetic field is maintained if $a^2 f(\phi)$ stays constant.
Since $f$ is proportional to the inverse square of the gauge coupling constant, this means that the inflaton should evolve toward a strong coupling region with time in order for the gauge field to drive inflation.  It is thus interesting to see if the $N=8$ supergravity displays the behavior of this kind.

\begin{figure}[t]
\begin{center}
\includegraphics[width=7cm]{\FigDir/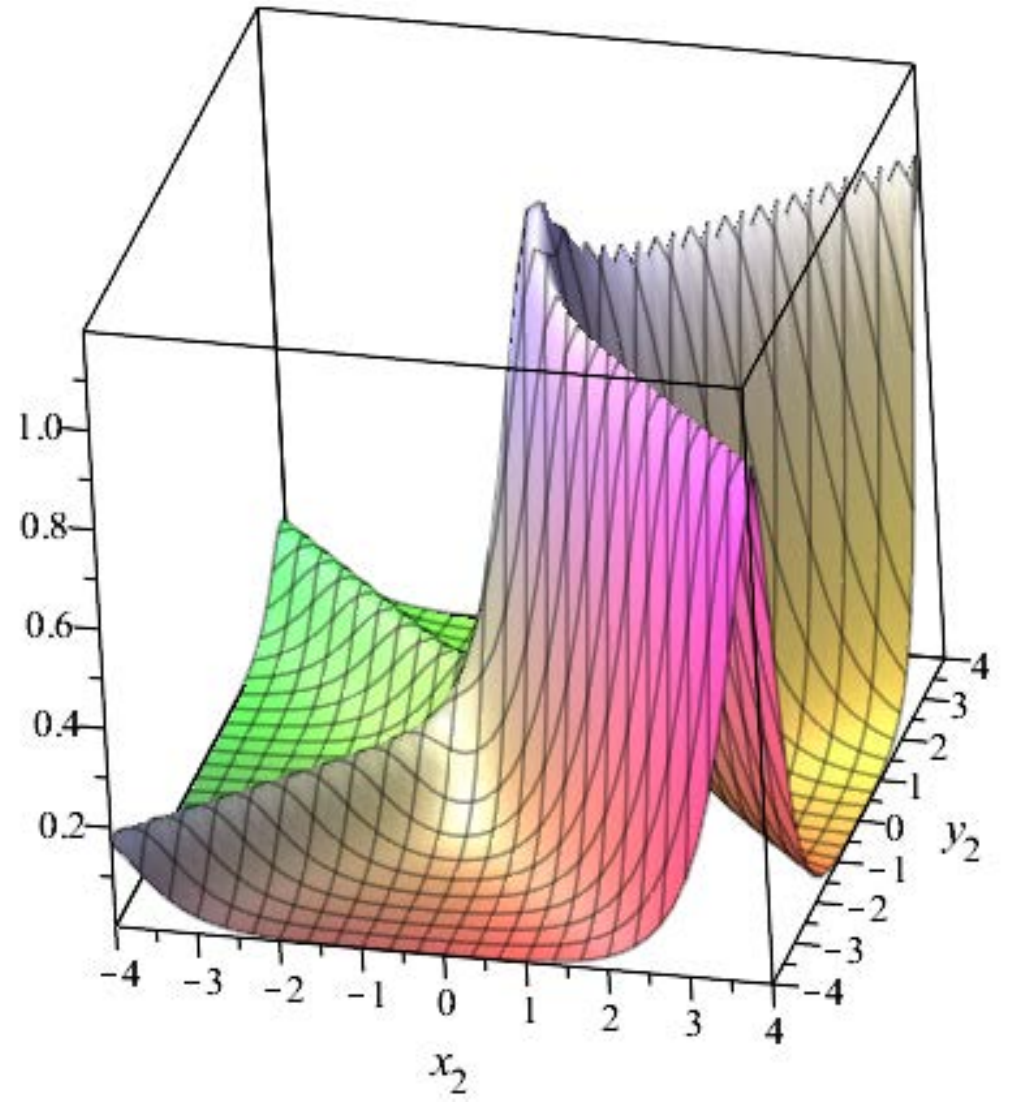}
\hspace{0.5cm}
\includegraphics[width=7cm]{\FigDir/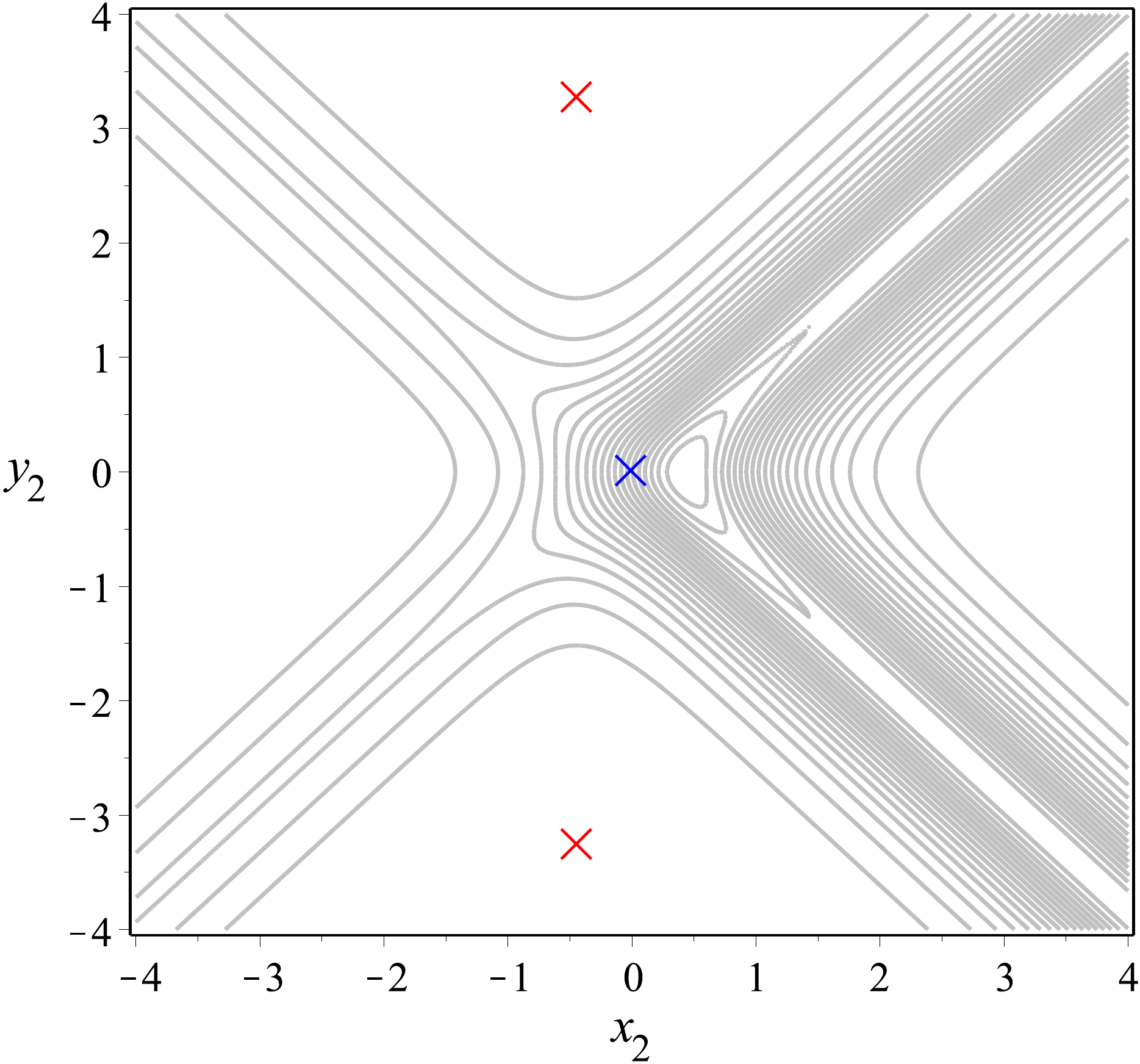}
\end{center}
\caption{The left panel is a 3D plot of $\Im(N_{(+)})$ on $\Sigma_2$ for $s=s_c+1.4\cdot 10^{-5}$, and the right panel is its contour plot. The potential critical points are marked by crosses.}
\label{fig:NI:Z2:s=scr+1.4e-5}
\end{figure}

\begin{figure}[h]
\begin{center}
\includegraphics[width=7cm]{\FigDir/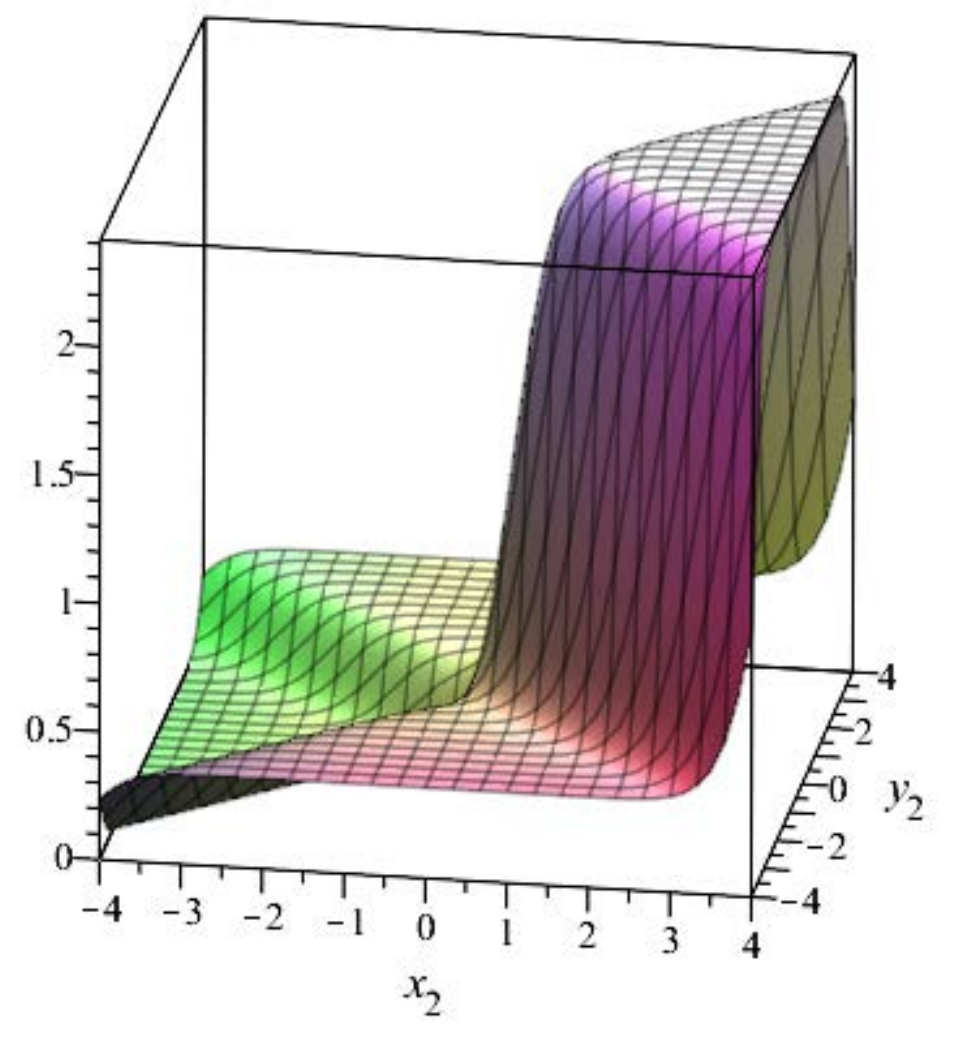}
\hspace{0.5cm}
\includegraphics[width=7cm]{\FigDir/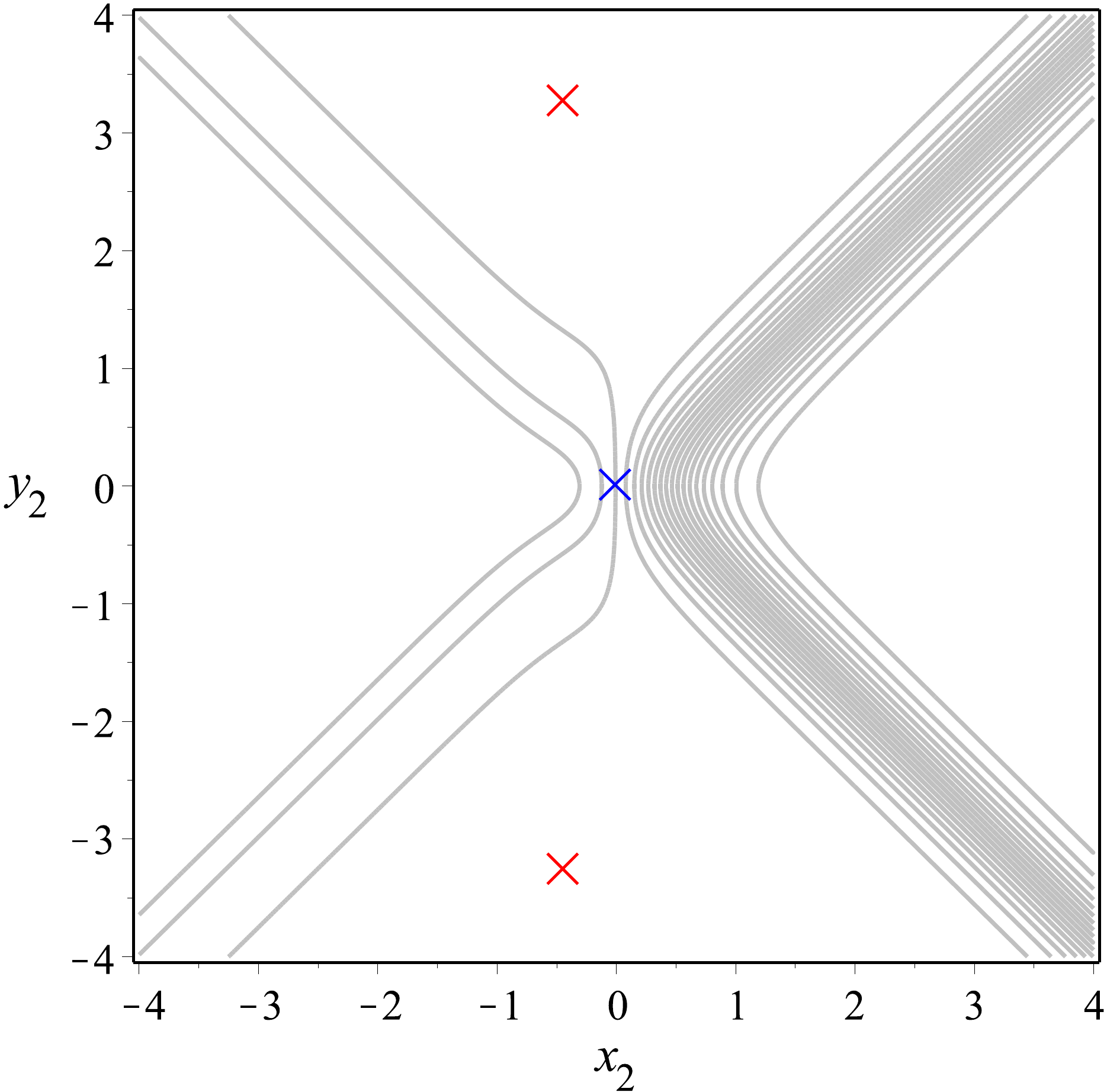}
\end{center}
\caption{The left panel is a 3D plot of $\Re(N_{(+)})$ on $\Sigma_2$ for $s=s_c+1.4\cdot 10^{-5}$, and the right panel is its contour plot.}
\label{fig:NR:Z2:s=scr+1.4e-5}
\end{figure}

Let us now examine these possibilities. The argument discussed in \ref{subsec:gauge_sector} demands us  to truncate the gauge sector to two $\SO(3)$ gauge fields $A^{\pm I}$ ($I=1,2,3$). In order to see the possibility of the chromo-natural type inflation, we require that the gauge flux is spatially homogenous and isotropic. Then, $A^{\pm I}$ should be gauge equivalent to
\Eq{
A^{\pm I}= a(t)\psi_\pm (t) dx^I \,.
}
By inserting this into \eqref{SO44:SO3xSO3:GaugeKineticTerm}, we find that the gauge contribution to the action is given by
\Eq{
S_1 = \int dt
 a^3\sum_{\epsilon=\pm} \insbra{\frac32 \Im(-N_\epsilon)\inrbra{\frac1{\N}\inpare{\dot\psi_\epsilon
 + \dot\alpha \psi_\epsilon}^2-g^2 \N \psi_\epsilon^4} + \frac{g}{2}\N  \Re(N_\epsilon) \psi_\epsilon^3}.
}
The explicit expressions for $\Im(N_\pm)$ and $\Re(N_\pm)$ are given in Appendix \ref{sec:GaugeCouplingFunctions}.  By virtue of the intricacy of these expressions, it seems less obvious to extract physical information. Our numerical computations nevertheless give rise to a simple and surprising result. Figures \ref{fig:NI:Z2:s=scr+1.4e-5} and  \ref{fig:NR:Z2:s=scr+1.4e-5} show 3D plots and contour plots of $\Im(N_+)$ and $\Re(N_+)$ on $\Sigma_2$, respectively.  Fig.\ref{fig:NI:Z2:s=scr+1.4e-5} implies that the coupling function $\Im(N_+)$ is constant with good accuracy except for crossing strips parallel to $y_2=\pm x_2$. Away from these strips, the function $\Im(N_+)$ exponentially damps to zero as $|x_2|$ and $|y_2|$ deviate from these strips. This implies that those flattened regions correspond to the strong coupling regime. The DI saddle point is inside this region. In contrast, $\Im(N_+)$ becomes of order unity on the strips around $y_2=\pm x_2$ ($x_2\gsim0$). These strips contain the critical point at the origin. This implies that for inflationary trajectories that terminate at the origin, the inflaton must undergo the transition from a strong coupling to a weak coupling. It turns out that anisotropic inflation does not work along that kind of trajectory, since the strong coupling is needed so as not to dilute the gauge fields. For inflationary trajectories that end in the waterfall valley, the gauge coupling function decreases, but its total change is not substantial.

\begin{figure}[t]
\begin{center}
\includegraphics[width=7cm]{\FigDir/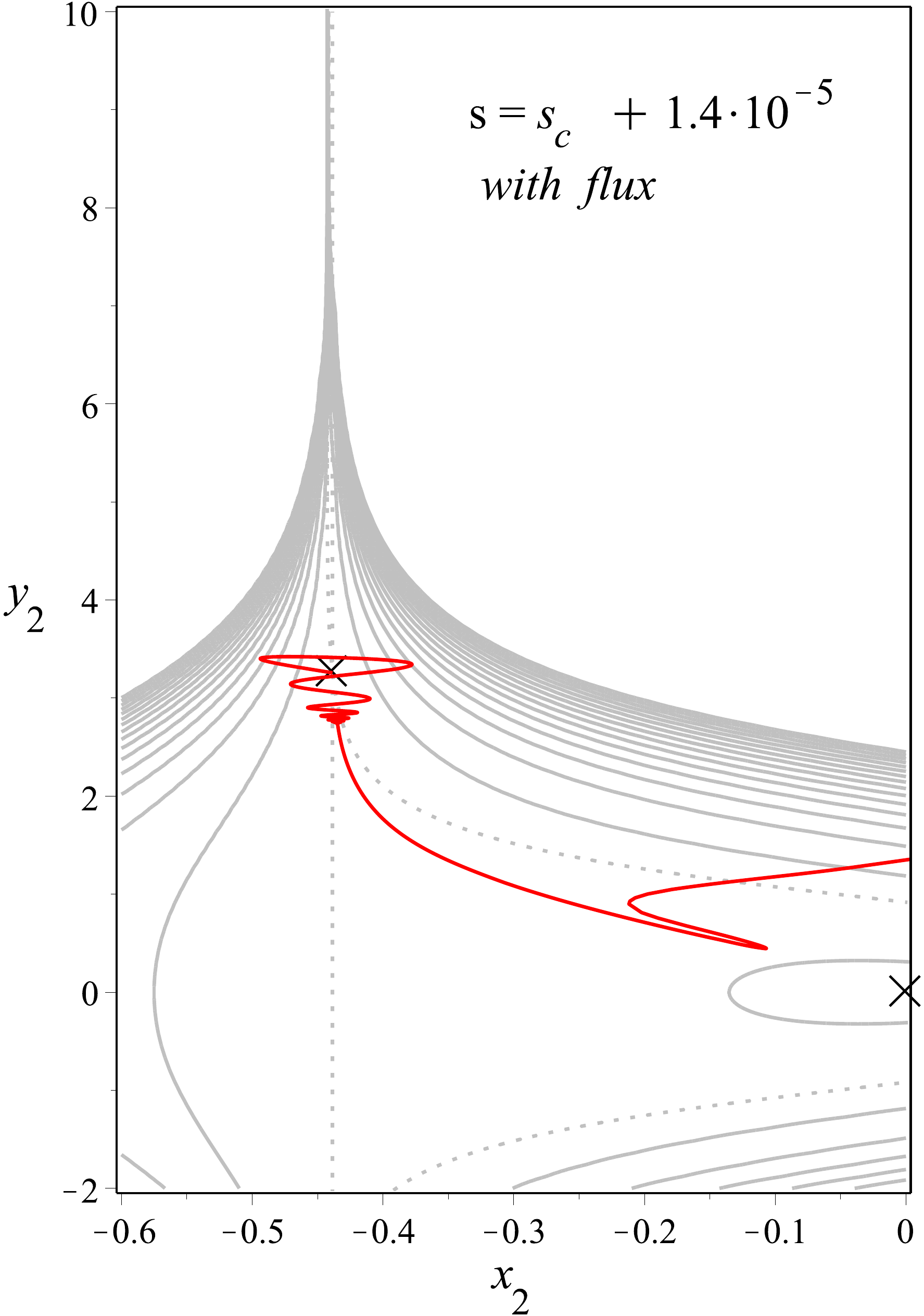}
\hspace{5mm}
\includegraphics[width=7cm]{\FigDir/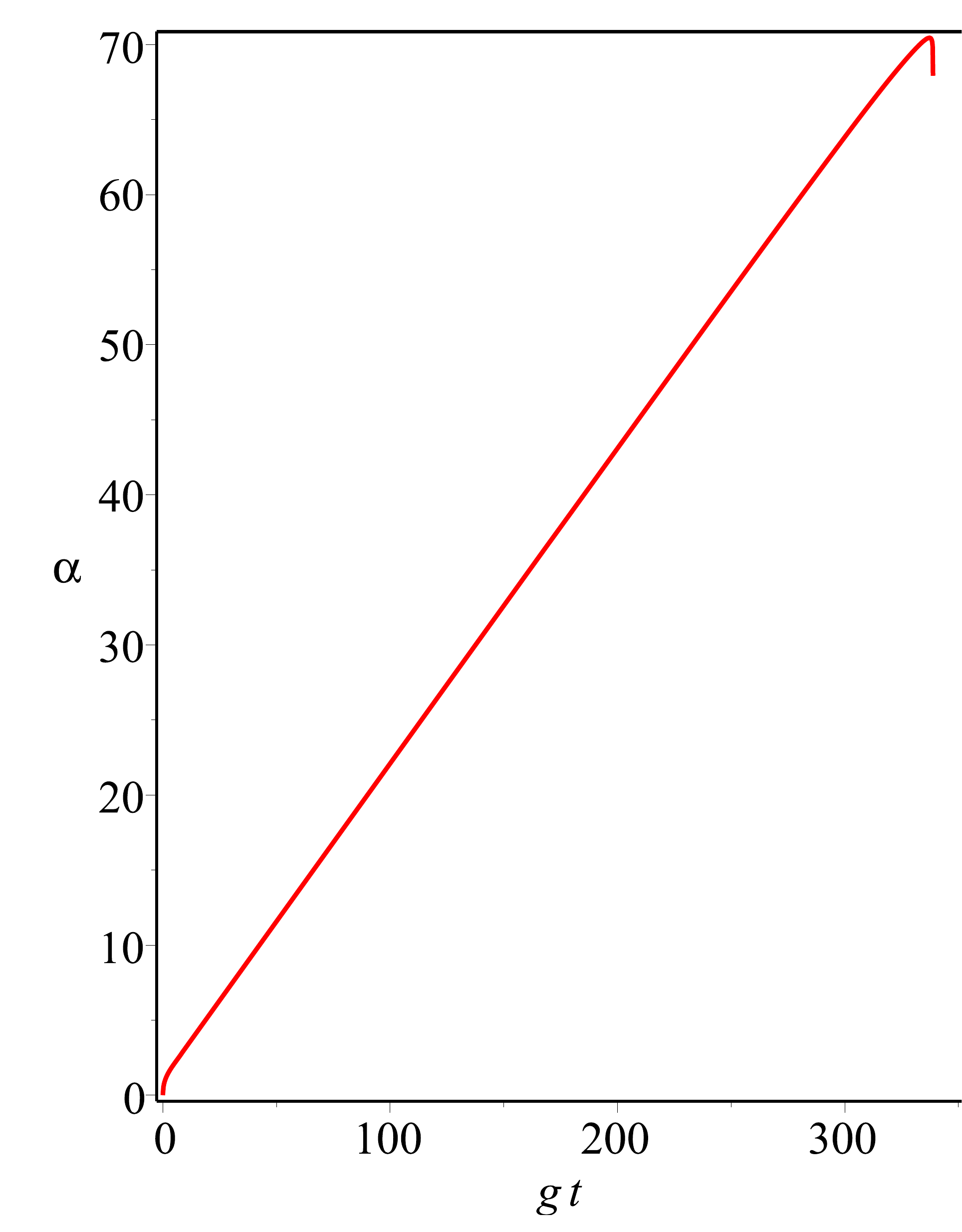}
\end{center}
\caption{The left panel shows the $\Sigma_2$ projection of the trajectory of the solution
with the initial conditions: $ x=x_*, y=y_*-H_*(g=0.01)$ on $\Sigma_2$ with flux $\psi=10, \dot\psi/H_*=-10$.  The right panel shows $\alpha(t)$ of this solution. }
\label{fig:2Dtraj_wf:s=scr+1.4e-5}
\end{figure}

\begin{figure}
\begin{center}
\includegraphics[width=7cm]{\FigDir/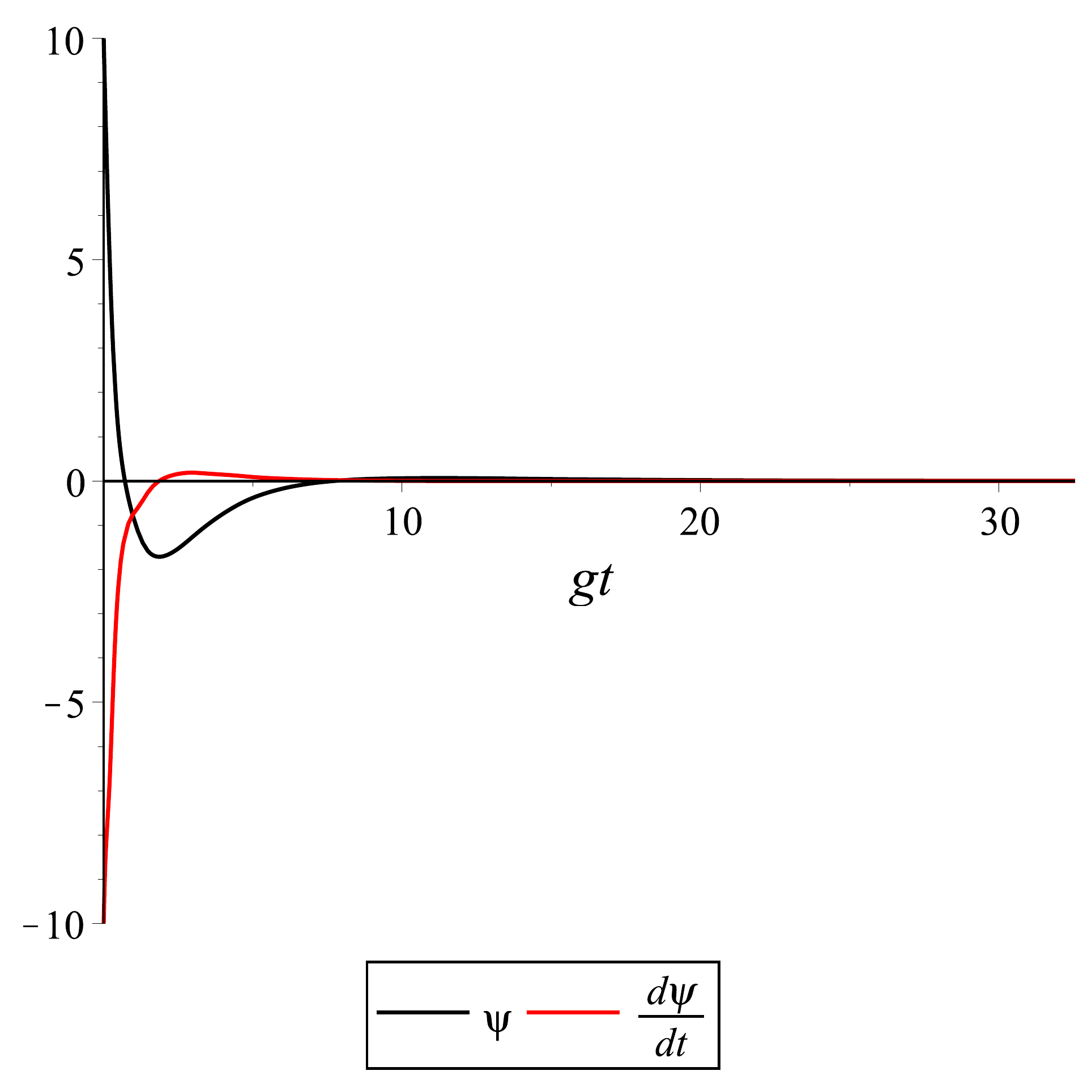}
\end{center}
\caption{Behavior of the magnetic flux for the solution in Fig. \ref{fig:2Dtraj_wf:s=scr+1.4e-5}.}
\label{fig:flux_wf:s=scr+1.4e-5}
\end{figure}

\begin{figure}
\centerline{
\includegraphics[width=7cm]{\FigDir/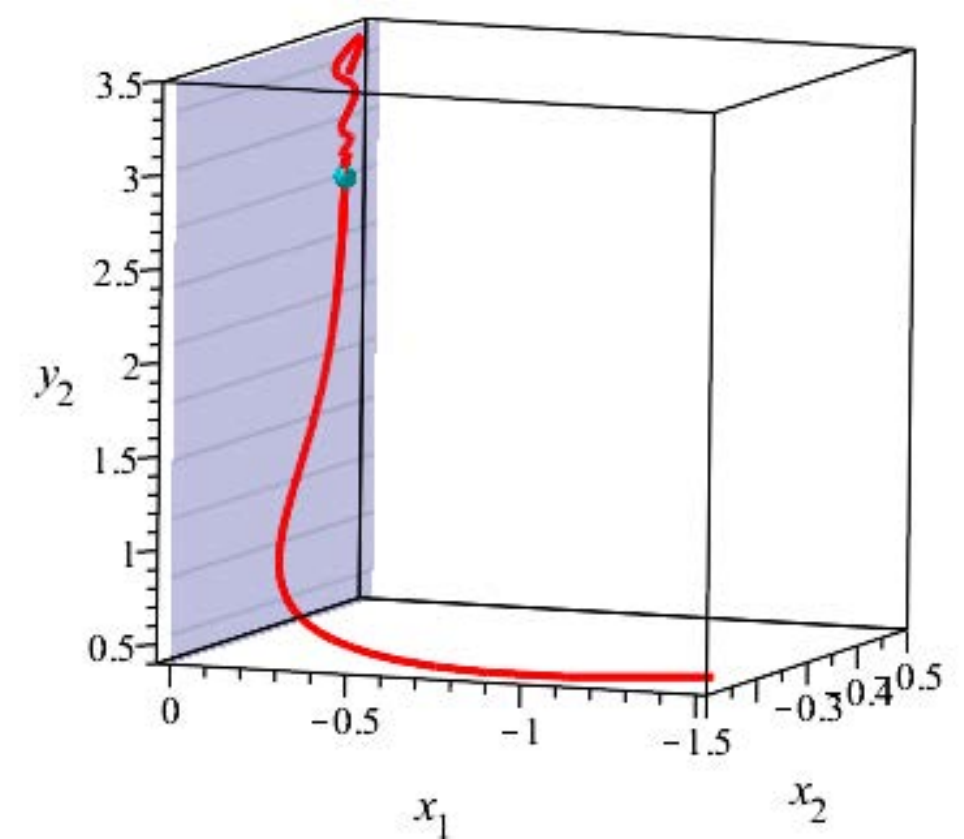}
\hspace{5mm}
\includegraphics[width=7cm]{\FigDir/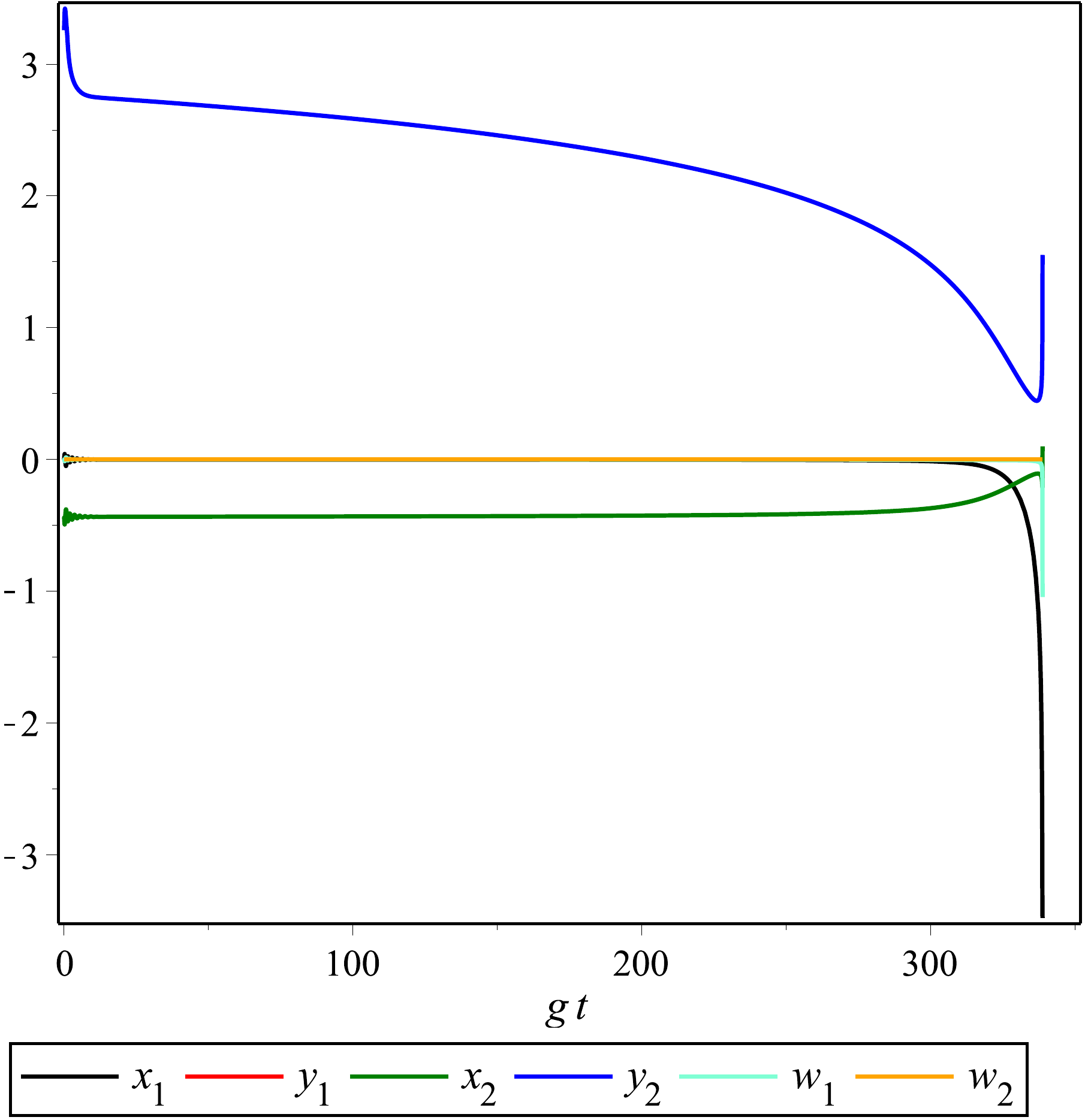}
}
\caption{The left panel shows the 3D trajectory of the solution in Fig. \ref{fig:2Dtraj_wf:s=scr+1.4e-5}.  The right panel shows the behavior of the six coordinate $\bm{\phi}_{(0)}$ of this solution.}
\label{fig:3Dtraj_wf:s=scr+1.4e-5}
\end{figure}

Let us next discuss the behavior of the coefficient of the Chern-Simons term.
Fig.\ref{fig:NR:Z2:s=scr+1.4e-5} indicates that there exist several different regions where
$\Re(N_+)$ remains approximately constant in each domain. We see
that the higher value of $\Re(N_+)$ is at most of order unity. This makes it difficult to drive chromo-natural inflation. We confirmed this statement more rigidly by solving the equations of motion of the 6-dimensional scalar system coupled with gravity and a single $\SO(3)$ gauge field. The inflaton trajectory and $\alpha(t)$ for $s=s_c+1.4\times 10^{-5}$ are depicted in Fig. \ref{fig:2Dtraj_wf:s=scr+1.4e-5}. This solution starts from the $\Sigma_2$ plane with zero velocity. We take the gauge fields to have a vanishing initial electric flux ($E\propto \dot\psi + H \psi$) but turn on a large amplitude of magnetic flux ($\psi=\psi_+=10,\psi_-\equiv0$). As we can see from the left panel of Fig. \ref{fig:2Dtraj_wf:s=scr+1.4e-5}, the inflaton initially oscillates with a large amplitude due to the magnetic flux, even though the initial point is close to the attractor slow-roll trajectory. The oscillation quickly damps and the inflaton settles down to the inflationary attractor trajectory since the magnetic flux decays in a short time (see  Fig. \ref{fig:flux_wf:s=scr+1.4e-5}). A notable difference from the case without flux is that  the trajectory exhibits behavior similar to the ones with initial conditions outside $\Sigma_2$ plane and the total e-folding number is smaller than the ones on $\Sigma_2$ for a similar starting point, as shown in Fig. \ref{fig:2Dtraj_wf:s=scr+1.4e-5}.  As Fig. \ref{fig:3Dtraj_wf:s=scr+1.4e-5} reveals, this behavior arises because the oscillatory motion at the beginning is not confined in the $\Sigma_2$ plane. The magnetic flux effectively produces a large offset of the inflationary trajectory from the $\Sigma_2$ plane. Thus, the initial strong magnetic flux affects the final stage of inflation significantly. As far as this solution, however, the $N=60$ point (marked by a lightblue dot in Fig. \ref{fig:3Dtraj_wf:s=scr+1.4e-5}) is on the attractor slow-roll trajectory below the oscillatory path. As a result, the slow roll parameters at the $N=60$ point take similar values to those for solutions without flux. Refer to Table \ref{tbl:ns:wf} for details.

\begin{table}[t]
{\small
\begin{tabular}{c|c|c|c|c|c|c|c|c|c|c}
\hline
$s-s_c$ & \multicolumn{4}{c|}{Initial Condition} & $N_t$ & \multicolumn{5}{c}{$N=60$ point}  \\
\hline
        & $x_1$ & $x_2$ & $y_2$ & $w_1$ &   & $x_2$ & $y_2$ & $\epsilon_H$[$\epsilon_V$]& $\eta_H$ [$\eta_V$] & $n_s$ \\
\hline
\raisebox{-10pt}{$1.4\cdot\e{-5}$}
                & 0     & $x_*$     & $y_*+H_*$ & 0 & 417 & -.4399 & 3.747 & .429{\rm e-4} & -.0175 & .964 \\
                &       &           & $(g=0.01)$&   &     &        &       & [.434{\rm e-4}]&[-.0176]& \\
\cline{2-11}
                 & 0     & $x_*+H_*$ & $y_*+H_*$ & 0 & 564 & -.4399 & 3.791 & .541{\rm e-4} & -.0187 & .962 \\
                 &       &$(g=0.01)$& $(g=0.01)$ &   &     &        &       & [.448{\rm e-4}]&[-.0189]& \\
\cline{2-11}
                 & 0      & $x_*     $ & $y_*-H_*$ & 0 & 433 & -.4338 & 2.667 & .750{\rm e-4} & -.0205 & .958 \\
                 &        &            & $(g=0.01)$ &   &     &        &       & [.760{\rm e-4}]&[-.0205]& \\
\cline{2-11}
                 & 0       & $x_*-H_*$  & $y_*$      & 0 & 452 & -.4338 & 2.666 & .755{\rm e-4} & -.0205 & .958 \\
                 &         &  $(g=0.01)$ &           &   &     &        &       & [.766{\rm e-4}]&[-.0207]& \\
\cline{2-11}
                 & 0       & $x_* +H_*$  & $y_*+H_*$  & 0 & 139 & -.4339 & 2.669 & .744{\rm e-4} & -.0204 & .958 \\
                 &         &  $(g=0.1)$  &  $(g=0.1)$  &   &     &        &       & [.756{\rm e-4}]&[-.0206]& \\
\cline{2-11}
                 &1.05  & $x_* $      & $y_*-H_*$  & -.352e-2 & 381 & -.4340 & 2.679 & .709e-4 & -.0202 & .961 \\
                 &         &            &$(g=0.1)$    &          &     &        &       & [.718e-4]&[-.0204]& \\
\hline
\raisebox{-10pt}{$1.4\cdot\e{-6}$}
                 &  0      & $x_* $     & $y_*-H_*$  &  0       & 727 & -.4341 & 2.685 & .832e-4 & -.0182 & .962 \\
                 &         &            &$(g=0.1)$    &          &     &        &       & [.842e-4]&[-.0184]& \\
\hline
\raisebox{-10pt}{$\e{-10}$}
                 & 0       & $x_* $      &  3         & 0        & 108 & -.4340 & 2.683 & .855e-4 & -.0181 & .963 \\
                 &         &            &             &          &     &        &       & [.866e-4]&[-.0183]& \\
\cline{2-11}
                 & 0       & $x_* $      &  4         & 0        & 753 & -.4341 & 2.690 & .833e-4 & -.0179 & .963 \\
                 &         &            &             &          &     &        &       & [.843e-4]&[-.0181]& \\
\cline{2-11}
                 & 0       & $-.4924$      &  4         & 0        & 511 & -.4341 & 2.685 & .848e-4 & -.0180 & .963 \\
                 &         &            &             &          &     &        &       & [.858e-4]&[-.0182]& \\
\cline{2-11}
                &6.36e-3  & $x_* $      &  4         & -2.12e-3 & 740 & -.4341 & 2.690 & .830e-4 & -.0178 & .963 \\
                 &         &            &             &          &     &        &       & [.840e-4]&[-.0180]& \\
\cline{2-11}
                 &1.26e-3  & $x_* $      &  4         &  1.26e-3 & 752 & -.4348 & 2.743 & .673e-4 & -.0160 & .967 \\
                 &         &            &             &          &     &        &       & [.680e-4]&[-.0163]& \\
\hline
\end{tabular}
}
\caption{Examples of numerical results for models without gauge flux. For all models, the initial velocity $\dot \phi^\alpha$ and the initial values of $y_1$ and $w_2$ are zero. The value of $g$ for the initial value of $x_2$ and $y_2$ indicates that $H_* (\propto g)$ is estimated for that value of $g$.}
\label{tbl:ns:nf}
\end{table}

\begin{table}[t]
\begin{center}
{\small
\begin{tabular}{c|c|c|c|c|c|c|c|c|c|c}
\hline
$s-s_c$ & \multicolumn{4}{c|}{Initial Condition} & $N_t$ & \multicolumn{5}{c}{$N=60$ point}  \\
\hline
        & $x_1$ & $x_2$ & $y_2$ & $w_1$ &   & $x_2$ & $y_2$ & $\epsilon_H$[$\epsilon_V$]& $\eta_H$ [$\eta_V$] & $n_s$ \\
\hline
\raisebox{-10pt}{$1.4\cdot\e{-5}$}
        & 0     & $x_*$     & $y_*+H_*$ & 0 & 71 & -.4344 & 2.713 & .600{\rm e-4} & -.0189 & .961 \\
        &       &           & $(g=0.01)$&   &     &        &       & [.608{\rm e-4}]&[-.0192]& \\
\hline
\end{tabular}
}
\end{center}
\caption{An exampel of numerical results for the model with gauge flux. }
\label{tbl:ns:wf}
\end{table}

\subsection{Comparison with observations}

Finally, let us discuss the observational consistency of inflationary solutions we obtained numerically in the six-dimensional sector of the maximal gauged supergravity with $\SO(4,4)$ gauging.  First of all, as is clear from the explanations above and Tables \ref{tbl:ns:nf} and \ref{tbl:ns:wf}, the spectral index $n_s$ at the $N=60$ point takes values around $0.96$, irrespective of the initial conditions.  This is within the $1\sigma$ range $n_s=0.9639\pm0.0047$ obtained by Planck for a wide range of initial conditions, provided that the deformation parameter $s$ is very close to or equal to the critical value $s_c$. At this critical value, we have only one tunable parameter $g$ which determines the height of the potential. It follows that this consistency with  observations is rather miraculous. The situation is similar to the Starobinsky model~\cite{Starobinsky:1980te} that admits also a single free parameter in the absolute units and predicts  a unique value of $n_s$. The similarity has a clear physical ground. In our model, there exists only one slow-roll attractor trajectory on the $\Sigma_2$ plane in the limit $s=s_c$. This trajectory extends out to infinity and the potential $V$ on that trajectory approaches a non-vanishing constant $V_*$. The difference $V-V_*$ decreases exponentially with the coordinate $y_2$. As a result, the effective one-dimensional potential has the same structure as that for the Starobinsky model in the scalar-tensor theory representation. The only difference is that our model has waterfall fields and inflation ends as in the hybrid inflation model. In some aspects, our model provides a variant of multi-dimensional version of the Starobinsky model.

The values of $n_s$ and its running are fixed by the slow-roll attractor trajectory on the $\Sigma_2$ plane for a given $N$, if the deformation parameter is close to the critical value $s_c$. We can thus read out the value of the running $-d n_s /d N$ from the right graph of Fig.~\ref{fig:SRparam:s=scr+1e-10} as 
\begin{align}
\label{}
-\frac{d n_s}{d N} \simeq 
 \left\{
 \begin{array}{ll}
 -9.3\times 10^{-4}\quad  & (N=50) \,,\\
  -6.7\times 10^{-4} & (N=60) \,, \\
-4.6\times 10^{-4} & (N=70) \,. 
\end{array}
 \right.
\end{align}
These values are consistent with the observational constraint $-0.0033\pm 0.0074$ obtained by Planck temperature data and the CMB lensing~\cite{Ade:2015lrj}. Thus, the scale-dependence of the spectral index is very small in our model.  

The spectral index is independent of the height of a potential. In contrast, the amplitudes of
the curvature perturbations $\P_\zeta(k)$ (representing the dimensionless power spectra) and tensor perturbations $\P_h(k)$ (representing the gravitational waves) depend on the potential height through the cosmic expansion rate $H$ as
\Eq{
\P_\zeta(k) = \frac{1}{2\epsilon(t_k)}\pfrac{H(t_k)}{2\pi \mpl}^2,\qquad
\P_h(k) = 8 \pfrac{H(t_k)}{2\pi \mpl}^2,
}
where we have restored the Planck mass and $t_k$ denotes the cosmic time when the comoving scale corresponding to the comoving wavenumber $k$ comes out of the Hubble horizon. From these equations, the famous consistency relation for the tensor-scalar ratio $r$ follows:
\Eq{
r\equiv\frac{\P_\zeta}{\P_h} = 16\epsilon \,.
}
We have to use this formula with care because this formula holds only under a set of assumptions that the slow-roll inflation by a single inflaton with the standard kinetic term and negligible curvaton contribution.

In our numerical solutions, the values of $\epsilon$ at the $N=60$ point has a larger dispersion than that of $n_s$, but still remains of the same order. In particular, for inflationary trajectories ending around the origin, our model predicts
\Eq{
0.96\times 10^{-3} \lsim r \lsim 1.37 \times 10^{-3}.
}
This is about one third of the value in the Starobinsky model and around the border of the sensitivity of the next generation B-mode satellite experiment LiteBIRD~\cite{Matsumura:2013aja}.

Assuming that the major part of scalar curvature perturbations is produced by inflaton,  the CMB normalization by Planck and WMAP~\cite{Planck:2015xua},
\Eq{
\P_\zeta(k_0) \simeq 2\times 10^{-9}\,,\quad (k_0=0.002{\rm Mpc}^{-1} )
}
fixes the value of the gauge coupling constant $g$ as
\Eq{
g\simeq 4 \times 10^{-6} \,.
\label{gval}
}

\section{Concluding remarks}
\label{sec:summary}

In the present paper, we have addressed the problem whether the $N=8$ gauged supergravity with a one-parameter deformation realizes a realistic inflationary cosmology. Our work is motivated in part by the theoretical restriction and the ultraviolet control of $N=8$ supergravity. If cosmological predictions of such a theory that is constrained by general physical principles are consistent with observations, it will provide us with a hint for deep understanding of the fundamental laws of Nature in contrast to the phenomenological framework provided by the $N=1$ supergravity.  From this perspective, we have made  elaborated studies of the dynamical structure in the ${\rm SO}(3)\times {\rm SO}(3)$ invariant sector of the ${\rm SO}(4,4)$ and ${\rm SO}(5,3)$ gaugings of $N=8$ supergravity, extending the work of~\cite{Dall'Agata:2012sx}.

In the ${\rm SL}(8)$ gauge-field frame, we have outlined a general recipe which enables us to compute the potential analytically. This strategy is useful for searching  ``off-center'' critical points. Although the final expression is quite messy for the full ${\rm SO}(3)\times {\rm SO}(3)$-invariant sector where six nonvanishing scalars survive, our work may serve as a cornerstone to the analytic scan of the critical points  on the whole scalar manifolds in extended supergravities.  Our formulation indeed works also for the potential of the ${\rm SO}(8)$ gaugings obtained in the literature~\cite{Dall'Agata:2012bb,Borghese:2012zs,Borghese:2013dja}. The present analytic way of obtaining the potential is applicable as far as the scalar fields take values in the coset space, e.g, $N\ge 3$ supergravities. The application of the present study for these cases is a plausible future work.

Unfortunately, our numerical computations show that there appear no critical points except for the known ones for the ${\rm SO}(3)\times {\rm SO}(3)$ invariant sector of ${\rm SO}(4,4)$  or ${\rm SO}(5,3)$  gaugings.  In order to find out the other critical points, we need to consider a larger scalar sector that does not respect the ${\rm SO}(3)\times {\rm SO}(3)$ invariance, or other gauge groups. At present, no de Sitter critical points have been found in a gauge group other than
${\rm SO}(4,4)$  and ${\rm SO}(5,3)$. As far as the authors know, however, there seems no no-go theorem which prohibits the de Sitter solution in other gaugings.

Although metastable de Sitter critical points are of importance in the context of dark energy, the nonexistence of these extrema is not a serious problem in the context of inflationary model-building, because otherwise the inflation never ends.  In light of this, we have extensively investigated numerically the dynamics of six scalar fields coupled with gravity  and gauge fields in the ${\rm SO}(4,4) $ gaugings. We concluded that a sufficient duration of inflation can be easily achieved, provided the deformation parameter is very close to or equal to the critical value. The initial point of the inflaton should be around the saddle point found by Dall'Agata-Inverso~\cite{Dall'Agata:2012sx}, but the offset can be order unity in the Planck units.  We have further determined the values of the slow roll parameters at the $N=60$ point. We have found that the spectral index $n_s$ for the scalar curvature perturbation is miraculously insensitive to the initial condition and the value of initial flux, and in the $1\sigma$ range reported by Planck satellite~\cite{Ade:2013zuv}. We have explained why such a miraculous situation arises on the basis of the existence of special attractor slow roll trajectories. We have pointed out that the situation is analogous to the Starobinsky model and that our model is a multi-dimensional extension of the Starobinsky model. In fact, our model predicts a small tensor-scalar ratio around $10^{-3}$, which is close to that of the Starobinsky model.

We have also explored the possibility of the chromo-natural inflation due to the Chern-Simons terms and the anisotropic inflation due to the non-trivial inflaton dependence on the gauge coupling function.  We have denied these possibilities based on the inspection of the behavior of the gauge coupling functions and the numerical studies of the evolution of the system with non-vanishing flux.  Because it is a common structure of gauged supergravities with dyonic gauging that the real part of the gauge coupling function appearing in the Chern-Simons term is bounded in the electric frame, we can conclude that the chromo-natural inflation fails to occur in the framework of $N=8$ gauged supergravity.

\section*{Acknowledgements}

MN benefited from valuable discussions with  Gianguido Dall'Agata in the early stages of this work.
MN is supported in part by a grant for research abroad by JSPS and INFN. HK is supported by the Grant-in-Aid for Scientific Research (A) (26247042) from Japan Society for the Promotion of Science (JSPS).

\appendix

\section{Autonne-Takagi factorization}
\label{sec:ATF}

In this appendix, we briefly outline how to implement the
Autonne-Takagi factorization for a $3\times 3$ complex symmetric matrix:
\begin{align}
 \Phi=U\Delta {}^{T\!}U \,, \qquad
 \Delta={\rm diag}(\mu_1,\mu_2,\mu_3) \,, \qquad
 U\in {\rm U}(3)\,,
\label{N8_ATfac_3x3}
\end{align}
where $\mu_i$ are real eigenvalues.
Eq. (\ref{N8_ATfac_3x3}) is tantamount to the following eigenvalue problem
\begin{align}
 \Phi u_{(i)}=\mu_i \bar u_{(i)} \,, \qquad
U_{ij}=(\bar u_{(j)})_i\,.
\end{align}
The $3\times 3$ complex symmetric matrix $\Phi$ that
we encountered in the body of text takes the universal form
\begin{align}
 \Phi=\left(
\begin{array}{ccc}
a &iy_1 & i y_2 \\
i y_1 & b & w \\
i y_2 & w & c
\end{array}
\right) \,,
\label{Phi_univ}
\end{align}
where $a, b, c, w, y_1, y_2$ are real quantities.
Decomposing the eigenvector into real and imaginary parts as
$u_i=X_i+i Y_i$, we have
 \begin{align}
 \left(
\begin{array}{ccc}
 \mu_i-a& i y_1& i y_2\\
i y_1 & \mu_i+b & w \\
i y_2 & w & \mu_i+c
\end{array}
\right)\left(
\begin{array}{c}
 X_1\\
iY_2 \\
i Y_3
\end{array}
\right)+
\left(
\begin{array}{ccc}
\mu_i+a & i y_1&i  y_2 \\
iy_1 & b-\mu _i & -w \\
i y_2 & -w & c-\mu_i
\end{array}
\right)\left(
\begin{array}{c}
iY_1 \\
X_2 \\
X_3
\end{array}
\right)=0 \,.\label{N8_SO44_ATdec_EVP}
\end{align}
Let us define $Q(\mu)$ as the determinant of the
first matrix in (\ref{N8_SO44_ATdec_EVP}), i.e.,
\begin{align}
Q(\mu )=&\mu^3+(b+c-a)\mu^2 +\{bc-a(b+c)-w^2-y_1^2-y_2^2\} \mu
\nonumber \\
&+a (w^2-bc)-c y_1^2-b y_2^2+2 w y_1 y_2 \,.
\label{Qmueq}
\end{align}
In this case,  one can easily verify that the determinant of the second matrix in  (\ref{N8_SO44_ATdec_EVP}) yields $-Q(-\mu )$. Together with
$\Phi \Phi^\dagger =\mu^2 \mathbb I_3$, we have
the property
\begin{align}
0={\rm det}(\mu^2 \mathbb I_3-\Phi\Phi^\dagger )=-Q(\mu )Q(-\mu ) \,.
\end{align}
Setting  $Q(\mu )=0$ loses no generality and  this choice  requires $Y_1=X_2=X_3=0$.
It therefore turns out that the complex eigenvalue problem (\ref{N8_ATfac_3x3}) now
reduces to the following real eigenvalue problem:
\begin{align}
A \vec v = \mu \vec v
\,, \qquad A=
\left(
\begin{array}{ccc}
a& -y_1& -y_2\\
-y_1 & -b & -w \\
-y_2 & -w & -c
\end{array}
\right) \,, \qquad \vec v=\left(
\begin{array}{c}
X_1       \\
Y_2    \\
Y_3
\end{array}
\right)\,.
\label{ATfac_real}
\end{align}
 A straightforward computation shows that
\begin{align}
\vec v_i = \left(
\begin{array}{c}
 X_{1(i)}\\
 Y_{2(i)}\\
 Y_{3(i)}
\end{array}
\right)=
\left(
\begin{array}{c}
\mu_i^2+(b+c)\mu_i+bc-w^2 \\
-y_1(\mu_i+c)+wy_2 \\
-y_2 (\mu_i+b)+w y_1
\end{array}
\right)\in \mathbb{RP}^3 \,. \label{N8_SO44_AT_vec}
\end{align}
This gives
\begin{align}
U= \left(
\begin{array}{ccc}
C_1X_{1(1)} &C_2 X_{1(2)}  & C_3 X_{1(3)} \\
iC_1Y_{2(1)} & i C_2 Y_{2(2)} & i C_3 Y_{2(3)}\\
i C_1Y_{3(1)} & i C_2 Y_{3(2)}& i C_3 Y_{3(3)}
\end{array}
\right)  \,, \qquad
C_i\equiv (X_{1(i)}^2+Y_{2(i)}^2+Y_{3(i)}^2)^{1/2}\,.
\label{N8_SO44_AT_Umat}
\end{align}
This form of matrix $U$ fulfills
\begin{align}
 U{}^{T\!}U={\rm diag}(1,-1,-1)\,, \qquad
U_1 {}^{T\!}U_2=0 \,, \qquad U=U_1+i U_2 \,,
\end{align}
leading to
\begin{align}
O\equiv U_1+U_2 \,, \qquad O{}^{T\!}O=\mathbb I_3\,.
\label{orth_mat_U12}
\end{align}
Note that $O$ is an orthogonal matrix
appearing in the diagonalization of $3\times 3$ real eigenvalue
problem (\ref{ATfac_real}).
A suitable choice of the phase allows us to set $O\in {\rm SO}(3)$,
so that one can employ a parametrization in terms of
Euler angles ($\theta_1 ,\theta_2 , \theta_3 $),
\begin{align}
O&=\left(
\begin{array}{ccc}
\cos\theta_1  & -\sin\theta_1\cos \theta_3   & \sin\theta_1 \sin \theta_3  \\
\sin\theta_1 \cos\theta_2 & ~\cos\theta_1 \cos\theta_2 \cos\theta_3 -\sin\theta_2 \sin\theta_3
~&~ -\cos\theta_1 \cos\theta_2 \sin\theta_3-\sin\theta_2 \cos\theta_3  \\
\sin\theta_1 \sin\theta_2 & \cos\theta_1 \sin\theta_2 \cos\theta_3 +\cos\theta_2 \sin\theta_3 &
-\cos\theta_1 \sin\theta_2 \sin\theta_3  +\cos\theta_2 \cos\theta_3
\end{array}
\right)\,.
\label{orth_mat}
\end{align}

\section{Coordinate Systems of the six-dimensional subspace $\H_6$}
\label{sec:CoordTrf}

In this appendix we discuss the coordinate transformations for the target space in the
${\rm SO}(4,4)$ gaugings.
The original coordinate system $\bm{\phi}_{(0)}=(x_1,x_2, y_1,y_2,w_1,w_2)$ defined in (\ref{so44_scalar_full})  is regular on the whole space $\Sigma_6=\exp(\H_6)$. From eqs.~(\ref{SO44_Phi0}),  (\ref{N8_SO44_AT_Umat}), (\ref{orth_mat_U12}) and (\ref{orth_mat}), one finds that
these coordinates can be written in terms of the polar coordinates $\bm{\phi}_{(1)}=(x_2,\mu_2,\mu_3,\theta_1,\theta_2,\theta_3)$ as
\Eqrsubl{CartesianByPolar}{
x_1 &=& \inrbra{ -\frac12 c_1^2 (1+c_2^2) (1+c_3^2)-\frac12+c_3^2+\frac12 c_2^2(1+ s_3^2) +c_1 s_2 c_2 s_3 c_3} \mu_2 \notag\\
     && + \inrbra{-\frac12 c_1^2 (1+c_2^2) (1+s_3^2)-\frac12+s_3^2+\frac12 c_2^2(1+ c_3^2) -c_1 s_2 c_2 s_3 c_3} \mu_3,\\
y_1 &=& s_1 \insbra{\inrbra{-c_1 c_2 (1+c_3^2)+s_2 s_3 c_3} \mu_2+\inrbra{-c_1 c_2 (1+s_3^2)-s_2 s_3 c_3} \mu_3}, \\
y_2 &=& s_1 \insbra{-\inrbra{c_1 s_2 (1+c_3^2)+c_2 s_3 c_3} \mu_2 -\inrbra{c_1 s_2 (1+s_3^2)-c_2 s_3 c_3} \mu_3},\\
w_1 &=&  \inrbra{-\frac12 c_1^2 s_2^2 (c_3^2+1)+\frac12-c_2^2+\frac12 c_2^2 c_3^2-s_2 s_3 c_1 c_2 c_3  } \mu_2 \notag\\
     && +\inrbra{-\frac12 c_1^2 s_2^2 (s_3^2+1)+\frac12-c_2^2+\frac12 c_2^2 s_3^2+s_2 s_3 c_1 c_2 c_3 } \mu_3,\\
w_2 &=&  \inrbra{ s_2 c_2 (-c_1^2 (c_3^2+1)+s_3^2+1)-c_1 s_3 c_3 (2c_2^2-1)} \mu_2  \notag\\
    &&  +\inrbra{ s_2 c_2 (-c_1^2 (s_3^2+1)+c_3^2+1)+c_1 c_3 s_3 (2c_2^2-1)} \mu_3,
}
where
\Eq{
c_i=\cos(\theta_i),\quad s_i=\sin(\theta_i),\ (i=1,2,3).
}
The right-hand side of this coordinate relation is invariant under the following six transformations:
\begin{subequations}
\Eqr{
T_1&:& \theta_3 \tend \theta_3+\pi,\\
T_2&:& \theta_1 \tend \theta_1 +\pi,\quad \theta_3\tend -\theta_3,\\
T_3&:& \theta_1 \tend -\theta_1,\quad \theta_2\tend \theta_2+\pi,
}
\Eqr{
T_4&:& (\mu_2,\mu_3)\tend (\mu_3,\mu_2),\quad \theta_3\tend \theta_3+\frac{\pi}{2} \\
T_5&:&  \mu_2\tend -(\mu_2+\mu_3),\notag\\
&&  (s_1,c_1)\tend (\pm\sqrt{1-s_1^2 c_3^2},-s_1c_3),\notag\\
&&  (s_2, c_2) \tend N_1 (c_1s_2c_3+s_2s_3,c_1c_2c_3-s_2s_3),\notag\\
&&  (s_3,c_3) \tend N_2 (s_1s_3,c_1),\\
T_6&:&  \mu_3\tend -(\mu_2+\mu_3),\notag\\
&&  (s_1,c_1)\tend (\pm\sqrt{1-s_1^2 s_3^2},s_1s_3),\notag\\
&&  (s_2, c_2) \tend N_1 (-c_1s_2s_3+c_2c_3,-(c_1c_2s_3+s_2c_3)),\notag\\
&&  (s_3,c_3) \tend N_2 (-c_1,s_1c_3)
}
\end{subequations}
Here, note that the subspaces $\mu_2=\mu_3$, $\mu_1=\mu_2$ and $\mu_3=\mu_1$  with $\mu_1+\mu_2+\mu_3=0$ are invariant under the transformations $T_4$, $T_5$ and $T_6$, respectively. Further, each of these transformations exchanges two of these three subspaces. Hence, six wedges cut out from the $\mu_2-\mu_3$ plane by these three lines are all equivalent if we neglect the angular coordinates $\theta_i$ ($i=1,2,3$). If we select one of the wedges as the fundamental region in the $\mu_2-\mu_3$ plane with the help of the transformations $T_1$, $T_2$ and $T_3$, we can restrict the ranges of the Euler angles as
\Eq{
0\le \theta_1 < \pi,\quad
0\le \theta_2 < \pi,\quad
0\le \theta_3 < \pi.
}
Instead, if we select two of the wedge regions as the fundamental region, say, the region cut out by two lines $2\mu_2+\mu_3=0$ and $\mu_2 + 2\mu_3=0$,
\Eq{
(\mu_2,\mu_3)=p(2,-1)+q(1,-2)=(2p+q,-p-2q),\quad pq\ge0,
\label{FundamentalRegion:mu2-mu3}
}
we can reduce the range of $\theta_3$ to $\pi/2\le \theta_3<\pi$ with the help of $T_4$. In this choice of the fundamental region,  the $\SO(3)\times\SO(3)$-invariant plane $\Sigma_2$ can be represented as $\mu_2+\mu_3=0, \theta_1=\pi/2, \theta_2=0,\theta_3=3\pi/4$ as used in the present paper.

Because these three subspaces $\mu_i=\mu_j$ ($i>j=1,2,3$) in the coordinate system $\bm{\phi}_{(1)}$ correspond to a single subspace in $\H_6$ and play the role of an axis for the Euler angles, this coordinate system is singular at this axis. This can be easily confirmed by calculating the Jacobian of the transformation:
\Eq{
\left| \frac{D\bm{\phi}_{(0)}}{D\bm{\phi}_{(1)}}\right| = \frac12\sin(\theta_1)
(\mu_2-\mu_3)(2\mu_2+\mu_3)(\mu_2+2\mu_3).
}
In particular, the $x$-axis on $\Sigma_2$ is contained in the  singular subspace $\mu_2=\mu_3$.

Although the coordinate singularity at the intersection of these three subspaces are quite nasty and can be removed only by going back to the original coordinate system $\bm{\phi}_{(0)}$, singularity outside this intersection  can be removed by introducing Cartesian-like coordinates. As an example, let us consider the fundamental region cut out by two lines $\mu_2=\mu_3$ and $\mu_2+2\mu_3=0$, and introduce $z$ and $r$ by
\Eq{
\mu_2=z+r\,, \qquad \mu_3=z-r\,, \qquad
r\ge0\,, \qquad z\ge \frac{r}{3} \,.
}
In terms of the variable $r$,  we can introduce the two-dimensional Cartesian coordinates $(u,v)$ by
\Eq{
u=r\cos(2\theta_3)\,, \qquad v=r\sin(2\theta_3).
}
The original coordinate system $\bm{\phi}_{(0)}$ can thus be expressed in terms of the new coordinate system $\bm{\phi}_{(2)}=(u,v,z,\theta_1,\theta_2)$ as
\begin{subequations}
\label{phi2}
\begin{align}
w_1+x_1 &= s_1^2 u + (1-3c_1^2)z,\\
w_1-x_1 &= \inrbra{-2c_1c_2s_2 v +(c_1^2c_2^2-s_2^2)u}+(1-3s_1^2c_2^2)z,\\
w_2 &= c_1(1-2c_2^2)v - (1+c_1^2)c_2s_2 u+ 3s_1^2c_2s_2 z,\\
y_1 &= s_1\inpare{s_2 v - c_1c_2 (u+3z)},\\
y_2 &= s_1\inpare{-c_2 v - c_1s_2(u+3z)}.
\end{align}
\end{subequations}
The Jacobian of this transformation is given by
\Eq{
\left| \frac{D\bm{\phi}_{(0)}}{D\bm{\phi}_{(2)}}\right| = \sin(\theta_1) (9z^2-r^2).
}
Hence, the coordinate system $\bm{\phi}_{(2)}$ is singular only at $r=3z$, i.e., $\mu_2+2\mu_3=0$ apart from the singularity of the polar coordinate at $\theta_1=0$.

When we use this coordinate system in the study of trajectories around $\Sigma_2$, we have to map the fundamental region we adopted in the present paper to the above one by the transformation $T_5$ or $T_6$.  We should be also careful about the fact that this coordinate system is still singular on the $x$-axis on $\Sigma_2$.  When we consider a trajectory that pass through this axis, we have to adopt the original coordinate $\bm{\phi}_{(0)}$ near the axis where an approximate expression for the action can be obtained by the Taylor expansion.

The kinetic term of the effective Lagrangian for the scalar fields can be written in terms of the coordinate system $\bm{\phi}_{(2)}$ as
\Eqr{
ds_T^2 &=& dx_2^2 + du^2 + dv^2 + 3dz^2 \notag\\
&& + \frac{{\rm jh}(2r)^2-1}{r^2} (udv - v du)^2 \notag\\
&& + 4{\rm jh}(2r)^2 ( udv-vdu + r^2 \cos(\theta_1) d\theta_2)\cos(\theta_1) d\theta_2 \notag\\
&& +\frac12 \inpare{\cosh(6z)\cosh(2r)-1}(d\theta_1^2+\sin^2\theta_1 d\theta_2^2) \notag\\
&& + {\rm jh}(2r)\sinh(6z)\inrbra{ u(d\theta_1^2-\sin^2\theta_1 d\theta_2^2)-2v\sin\theta_1 d\theta_1 d\theta_2}.
}
The potential for the scalar fields reads
\Eqr{
g^{-2}V &=& \inpare{\frac{X^3}{s^2}+\frac{s^2}{X^3}}\Big[D_{31}\cosh(6z)\cosh(2r)
 + \sinh(6z){\rm jh}(2r)(D_{32}u+D_{33}v) \notag\\
&&\quad
  + {\rm jh}(2r)^2 (D_{34}u^2+D_{35}uv+D_{36}v^2) + D_{37} \Big]
  \notag\\
&& + \frac3{32}\inpare{s^2X + \frac1{s^2X}}\Big[ -c_1^2\cosh(4z)-s_1^2\cosh(2z)\cosh(2r)
\notag\\
&& \qquad -s_1^2{\rm jh}(2r)\sinh(2z)u+ 6 \Big]
 \notag\\
&& +\inpare{\frac{X}{s^2}+\frac{s^2}{X}} \Big[\frac3{16}\inpare{-\cosh(4z)+\cosh(2z)\cosh(2r)}s_1^2\cos(2\theta_2)
\notag\\
&& +  \frac38{\rm jh}(2r)\sinh(2z)\inrbra{-(1+c_1^2)\cos(2\theta_2)u+2c_1\sin(2\theta_2)v } \Big],
}
where
\Eqrsub{
D_{31} &=& \frac{s_1^2}{64}\inrbra{s_1^2\cos(4\theta_2)-1-c_1^2},\\
D_{32} &=& -\frac{s_1^2}{16}\inrbra{1+(1+c_1^2)\sin^2(2\theta_2)} ,\\
D_{33} &=& \frac{s_1^2c_1}{16}\sin(4\theta_2),\\
D_{34} &=& \frac1{256}\{\cos(4\theta_1)+28\cos(2\theta_1)+35)\cos^2(2\theta_2) \notag\\
       &&\quad -\cos(4\theta_1)+4\cos(2\theta_1)-3 \},\\
D_{35} &=& -\frac1{32}(7\cos(\theta_1)+\cos(3\theta_1))\sin(4\theta_2),\\
D_{36} &=& \frac1{256}\inrbra{(\cos(4\theta_1)+28\cos(2\theta_1)+35)\sin^2(2\theta_2)},\\
D_{37} &=& \frac1{512}\inpare{-\cos(4\theta_1)+4\cos(2\theta_1)-3}\cos(4\theta_2) \notag\\
       && +\frac1{512}\inpare{\cos(4\theta_1)+4\cos(2\theta_1) +27 }.
}

\newpage
\section{Gauge Coupling Functions for the $\SO(3)\times\SO(3)$ gauge fields in the $\SO(4,4)$ gauging}
\label{sec:GaugeCouplingFunctions}

The gauge coupling functions $N_\pm$ in \eqref{SO44:SO3xSO3:GaugeKineticTerm} are related to the general gauge coupling matrix $\N'{}^{[ab][cd]}$ in the electric frame as

\Eq{
N_+ = \N'{}^{[12][12]},\quad
N_- = \N'{}^{[56][56]}.
}
In terms of
\Eq{
X=e^{-2x_2},\quad Y=e^{2\mu_2},\quad Z=e^{2\mu_3},
}
$N^+$ can be written as
\Eq{
\Im(N_+)= \frac{B}{A},\quad \Re(N_+)=\frac{C}{A},
}
with
\Eqrsub{
A &=& s^4 X^6 Y^2 Z^2 + X^4 (a_{442} Y^4Z^2+a_{433}Y^3Z^3 + a_{424} Y^2 Z^4 +a_{421}Y^2Z  \notag\\
&& + a_{412} YZ^2 + a_{400})
 + X^2(a_{244} Y^4Z^4 + a_{232}Y^3Z^2+a_{223}Y^2Z^3 \notag\\
 && + a_{220}Y^2 + a_{211}YZ + a_{202}Z^2)
 + s^8 Y^2 Z^2,
 \\
B &=& -s^4X \Big[X^4 (b_{423}Y^2Z^3+b_{432}Y^3Z^2+b_{411}YZ)
\notag\\
&&\quad
  +  X^2 (b_{243}Y^4Z^3+b_{234}Y^3Z^4+ b_{231} Y^3Z+  b_{222}Y^2Z^2
  \notag\\
&&\quad
  + b_{213}YZ^3 + b_{210}Y + b_{201}Z )
  \notag\\
&&\quad
 + b_{033}Y^3Z^3+ b_{012}YZ^2 + b_{021}Y^2Z  \Big],
 \\
C &=& s^2\Big[ X^4 (c_{442} Y^4Z^2+c_{433}Y^3Z^3 + c_{424}Y^2Z^4
\notag\\
&&\quad
 +c_{421}Y^2Z + c_{412}YZ^2 + c_{400} )
 \notag\\
&&\quad
 +X^2(c_{244}Y^4Z^4+c_{232}Y^3Z^2+c_{223}Y^2Z^3
 \notag\\
 &&\quad
 + c_{220}Y^2 + c_{211}YZ + c_{202}Z^2) + s^8 Y^2Z^2 \Big].
}
Here, $a_{ijk}$,  $b_{ijk}$ and  $c_{ijk}$ are functions of the Euler angle $\theta_i$ ($i=1,2,3$). $N_-$ is easily obtained from $N_+$ by
\Eq{
N_- = N_+( Y\tend 1/Y, Z \tend 1/Z).
}

The coefficient functions $a_{ijk}$, $b_{ijk}$ and $c_{ijk}$ are expressed in terms of
\Eqrsub{
c_i=\cos(\theta_i),\quad s_i=\sin(\theta_i)\ (i=1,2,3),
}
as follows:
\begin{subequations}
\begin{align}
a_{442} =& \{(-4 c_1^4 s_2^4+4 c_1^4 s_2^2-24 c_1^2 s_2^4-c_1^4+24 c_1^2 s_2^2-4 s_2^4-2 c_1^2+4 s_2^2-1) c_3^4 \notag\\
        &  +8 c_1 c_2 s_2 s_3 (2 c_2^2-1) (c_1^2+1) c_3^3  \notag\\
        &   +(24 c_1^2 s_2^4-24 c_1^2 s_2^2+8 s_2^4+2 c_1^2-8 s_2^2+2) c_3^2  \notag\\
        &   -8 c_1 c_2 s_2 s_3 (2 c_2^2-1) c_3-(2 s_2^2-1)^2 \} s^8  \notag\\
        &  +\{ -2 (2 s_2^2-1)  s_1^2  (c_1^2+1) c_3^4
           -8 c_1 s_1^2 s_2  c_2 s_3 c_3^3+2 c_3^2 (2 s_2^2-1) s_1^2\} s^4  \notag\\
        &   -c_3^4 s_1^4
\end{align}
\Eqr{
a_{433} &=& \{(8 c_1^4 s_2^4-8 c_1^4 s_2^2+48 c_1^2 s_2^4+2 c_1^4-48 c_1^2 s_2^2+8 s_2^4+4 c_1^2-8 s_2^2+2) c_3^4  \notag\\
        && -16 c_1 c_2 s_2 s_3 (2 c_2^2-1) (c_1^2+1) c_3^3  \notag\\
        && +(-8 c_1^4 s_2^4+8 c_1^4 s_2^2-48 c_1^2 s_2^4-2 c_1^4+48 c_1^2 s_2^2-8 s_2^4-4 c_1^2+8 s_2^2-2) c_3^2  \notag\\
        && +8 c_1 c_2 s_2 s_3 (2 c_2^2-1) (c_1^2+1) c_3-8 c_1^2 s_2^2 c_2^2 \} s^8  \notag\\
        &&  +\{ 4 (2 s_2^2-1) s_1^2  (c_1^2+1) c_3^4+16 c_1 s_1^2 s_2 c_2 s_3 c_3^3  \notag\\
        && +(-4 s_1^2 (2 s_2^2-1)) (c_1^2+1) c_3^2
         -8 c_1 s_1^2 s_2 c_2 s_3 c_3  \} s^4  \notag\\
        && +2 c_3^4 s_1^4-2 s_1^4 c_3^2
              \\
a_{424} &=& \{(-4 c_1^4 s_2^4+4 c_1^4 s_2^2-24 c_1^2 s_2^4-c_1^4+24 c_1^2 s_2^2-4 s_2^4-2 c_1^2+4 s_2^2-1) c_3^4  \notag\\
         && +8 c_1 c_2 s_2 s_3 (2 c_2^2-1) (c_1^2+1) c_3^3  \notag\\
         &&  +2 c_1^2 (4 c_1^2 s_2^4-4 c_1^2 s_2^2+12 s_2^4+c_1^2-12 s_2^2+1) c_3^2  \notag\\
         &&  -8 s_2 c_2 c_1^3 s_3 (2 c_2^2-1) c_3-c_1^4 (2 s_2^2-1)^2\} s^8  \notag\\
         &&  +\{ -2 (2 s_2^2-1) s_1^2  (c_1^2+1) c_3^4 - 8 s_1^2 c_1 s_2 c_2 s_3 c_3^3  \notag\\
         &&  + 2 (2 s_2^2-1) s_1^2 (2 c_1^2+1) c_3^2+ 8 c_1 s_1^2 s_2 c_2 s_3 c_3
           -2 c_1^2 s_1^2 (2 s_2^2-1)\} s^4  \notag\\
         &&  -s_1^4 s_3^4,
         \\
a_{421} &=& s_1^2 \{ (-8 c_1^2 s_2^4+8 c_1^2 s_2^2-8 s_2^4-2 c_1^2+8 s_2^2) c_3^2
            +8 c_1 c_2 s_2 s_3 (2 c_2^2-1) c_3 - 8 s_2^2 c_2^2\} s^8  \notag\\
        &&     +s_1^2 \{4 c_1^2 (2 s_2^2-1) c_3^2+8 s_2 c_2 c_3 c_1 s_3\} s^4
             -2 s_1^2 c_1^2 c_3^2,  \\
a_{412} &=& s_1^2 \{(8 c_1^2 s_2^4-8 c_1^2 s_2^2+8 s_2^4+2 c_1^2-8 s_2^2) c_3^2
      -8 c_1 c_2 s_2 s_3 (2 c_2^2-1) c_3-2 c_1^2 (2 s_2^2-1)^2\} s^8  \notag\\
        &&     +s_1^2\{-4 c_1^2 (2 s_2^2-1) c_3^2-8 s_2 c_2 c_3 c_1 s_3+4 c_1^2 (2 s_2^2-1)\} s^4
             -2 c_1^2 s_1^2 s_3^2,
        \\
a_{400} &=& -(2 s_2^2-1)^2 s_1^4 s^8-2 s_1^2 c_1^2 (2 s_2^2-1) s^4-c_1^4,
\\
a_{244} &=& -c_1^4 s^{12}-2 s_1^2 c_1^2 (2 s_2^2-1) s^8-(2 s_2^2-1)^2 s_1^4 s^4,
\\
a_{232} &=&  -2 s_1^2 c_1^2 s_3^2 s^{12}+s_1^2\{-8 s_2 c_2 c_3 c_1 s_3+4 c_1^2 s_3^2 (2 s_2^2-1)\} s^8  \notag\\
        &&     +s_1^2\{(-8 c_1^2 s_2^4+8 c_1^2 s_2^2-8 s_2^4-2 c_1^2+8 s_2^2) s_3^2  \notag\\
        &&     -8 c_1 c_2 s_2 s_3 (2 c_2^2-1) c_3-8 s_2^2 c_2^2 \} s^4,
\\
a_{223} &=& -2 s_1^2 c_1^2c_3^2 s^{12}
+s_1^2\{ -4 c_1^2 s_3^2 (2 s_2^2-1)+8 s_2 c_2 c_3 c_1 s_3+4 c_1^2 (2 s_2^2-1)\} s^8  \notag\\
        &&     +s_1^2 \{(8 c_1^2 s_2^4-8 c_1^2 s_2^2+8 s_2^4+2 c_1^2-8 s_2^2) s_3^2  \notag\\
        &&     +8 c_1 c_2 s_2 s_3 (2 c_2^2-1) c_3-2 c_1^2 (2 s_2^2-1)^2\} s^4,
\\
a_{220} &=& -s_1^4 s_3^4 s^{12}
            -s_1^2\{(2 (2 s_2^2-1)) (c_1^2+1) s_3^4-(2 (2 s_2^2-1)) s_3^2
            +8 s_1^2 c_1 c_3 s_2 s_3^3  c_2\} s^8  \notag\\
        &&    +\{(-4 c_1^4 s_2^4+4 c_1^4 s_2^2-24 c_1^2 s_2^4-c_1^4+24 c_1^2 s_2^2-4 s_2^4-2 c_1^2+4 s_2^2-1) s_3^4  \notag\\
        &&    +(24 c_1^2 s_2^4-24 c_1^2 s_2^2+8 s_2^4+2 c_1^2-8 s_2^2+2) s_3^2  \notag\\
        &&    +8 s_2 c_2 c_3 c_1 (2 c_2^2-1) (c_1^2 c_3^2-c_1^2+c_3^2) s_3-(2 s_2^2-1)^2\} s^4,
}
\Eqr{
a_{211} &=& -2 s_1^4 c_3^2 s_3^2 s^{12}
            +s_1^2 \{ -4 (2 c_2^2-1)  (c_1^2+1) c_3^4 + 16 c_1 s_2  c_2 s_3 c_3^3  \notag\\
         &&     + 4 (2 c_2^2-1)  (c_1^2+1) c_3^2 -8 c_1 s_2 c_2 s_3 c_3\} s^8  \notag\\
         &&    +\{(8 c_1^4 c_2^4-8 c_1^4 c_2^2+48 c_1^2 c_2^4+2 c_1^4-48 c_1^2 c_2^2+8 c_2^4+4 c_1^2-8 c_2^2+2) c_3^4  \notag\\
         &&    -16 c_1 c_2 s_2 s_3 (2 c_2^2-1) (c_1^2+1) c_3^3  \notag\\
        &&     +(-8 c_1^4 c_2^4+8 c_1^4 c_2^2-48 c_1^2 c_2^4-2 c_1^4+48 c_1^2 c_2^2-8 c_2^4-4 c_1^2+8 c_2^2-2) c_3^2  \notag\\
        &&     +8 c_1 c_2 s_2 s_3 (2 c_2^2-1) (c_1^2+1) c_3-8 c_2^2 c_1^2 s_2^2\} s^4,
\\
a_{202} &=& -s_1^4 c_3^4 s^{12}
              +s_1^2\{-2 (2 s_2^2-1) (c_1^2+1) c_3^4-8 c_1 s_2 c_2 s_3 c_3^3+2 c_3^2 (2 s_2^2-1)\} s^8  \notag\\
         &&    +\{(-4 c_1^4 s_2^4+4 c_1^4 s_2^2-24 c_1^2 s_2^4-c_1^4+24 c_1^2 s_2^2-4 s_2^4-2 c_1^2+4 s_2^2-1) c_3^4  \notag\\
        &&     +8 c_1 c_2 s_2 s_3 (2 c_2^2-1) (c_1^2+1) c_3^3
             +(24 c_1^2 s_2^4-24 c_1^2 s_2^2+8 s_2^4+2 c_1^2-8 s_2^2+2) c_3^2  \notag\\
        &&     -8 c_1 c_2 s_2 s_3 (2 c_2^2-1) c_3-(2 s_2^2-1)^2\} s^4.
}
\end{subequations}

\begin{subequations}
\Eqr{
b_{423} &=& -s_1^2 s_3^2,\quad
b_{432} = -s_1^2 c_3^2,\quad
b_{411} = -c_1^2,\\
b_{243} &=& s_1^2\{(-4 c_1^2 s_2^2 c_2^2-4 s_2^2c_2^2+1) c_3^2-4 c_1 c_2 s_2 s_3 (2 c_2^2-1) c_3-(2 s_2^2-1)^2\}s^4,
\\
b_{234} &=& s_1^2\{(+4c_1^2s_2^2c_2^2+4s_2^2c_2^2-1)c_3^2 +4 c_1 c_2 s_2 s_3 (2 c_2^2-1) c_3-4 c_1^2 s_2^2 c_2^2 \}s^4,\\
b_{231} &=& \{(4 c_1^4 s_2^4-4 c_1^4 s_2^2+24 c_1^2 s_2^4-24 c_1^2 s_2^2+4 s_2^4+4 c_1^2-4 s_2^2) c_3^2s_3^2 \notag\\
        &&     -4 c_1 c_2 s_2 s_3 (2 c_2^2-1) (c_1^2+1) c_3 (1-2c_3^2) -c_1^2 (2 s_2^2-1)^2\}s^4, \\
b_{222} &=& \{(-8 c_1^4 c_2^4+8 c_1^4 c_2^2-48 c_1^2 c_2^4+48 c_1^2 c_2^2-8 c_2^4-8 c_1^2+8 c_2^2) c_3^2s_3^2 \notag\\
        &&     +8 c_1 c_2 s_2 s_3 (2 c_2^2-1) (c_1^2+1) c_3 (1-2c_3^2) \notag\\
        && +8 c_2^2 (c_2-1) (c_2+1) (c_1^4-c_1^2+1)\}s^4, \\
b_{213} &=& \{(4 c_1^4 s_2^4-4 c_1^4 s_2^2+24 c_1^2 s_2^4-24 c_1^2 s_2^2+4 s_2^4+4 c_1^2-4 s_2^2) c_3^2s_3^2  \notag\\
        &&    -4 c_1 c_2 s_2 s_3 (2 c_2^2-1) (c_1^2+1) c_3 (1-2c_3^2) -c_1^2 (2 s_2^2-1)^2\}s^4.\\
b_{210} &=& s_1^2\{ (4c_1^2c_2^2s_2^2+4c_2^2s_2^2-1)c_3^2 +4 c_1 c_2 s_2 s_3 (2 c_2^2-1) c_3-4 c_2^2 c_1^2 s_2^2\}s^4,\\
b_{201} &=& s_1^2\{ (-4c_1^2c_2^2s_2^2-4c_2^2s_2^2+1)c_3^2 -4 c_1 c_2 s_2 s_3 (2 c_2^2-1) c_3-(2 c_2^2-1)^2 \}s^4,\\
b_{033} &=& -c_1^2 s^8,\quad
b_{012} = -s_1^2c_3^2 s^8,\quad
b_{021} = -s_1^2s_3^2 s^8.
}
\end{subequations}
\vspace*{-1cm}
\begin{subequations}
\Eqr{
c_{442} &=& s_1^2\{(2 c_2^2-1) (c_1^2+1) c_3^4-4 c_1 c_2 s_2 s_3 c_3^3+(-2 c_2^2+1) c_3^2\} s^4-s_1^4 c_3^4,\\
c_{433} &=&  s_1^2\{8 c_1 c_2 s_2 s_3 c_3^3+(2 (2 c_2^2-1)) (c_1^2+1) c_3^2 s_3^2-4 s_2 c_2 c_3 c_1 s_3\}s^4
          -2c_3^2 s_3^2s_1^4,\\
c_{424} &=& s_1^2\{ -4 c_1 c_2 s_2 s_3 c_3^3-(2 c_2^2-1) (2 c_1^2+1) c_3^2s_3^2+4 s_2 c_2 c_3 c_1 s_3+c_1^2 (2 c_2^2-1)\}s^4\notag\\
        &&  -s_1^4s_3^4,\\
c_{421} &=& s_1^2\{-2 c_1^2 (2 c_2^2-1) c_3^2+4 s_2 c_2 c_3 c_1 s_3\} s^4-2s_1^2 c_1^2 c_3^2,\\
c_{412} &=& s_1^2\{-4 s_2 c_2 c_3 c_1 s_3+2 c_1^2 s_3^2 (2 s_2^2-1)\} s^4-2s_1^2 c_1^2 s_3^2,\\
c_{400} &=& -s_1^2c_1^2(2s_2^2-1)s^4 -c_1^4,\\
c_{244} &=& -s_1^2c_1^2(2s_2^2-1)s^8-s_1^4(2s_2^2-1)^2 s^4,
}
\Eqr{
c_{232} &=&  s_1^2\{-4 s_2 c_2 c_3 c_1 s_3+2 c_1^2 (2 s_2^2-1)s_3^2\} s^8
           +s_1^2\{(8 c_1^2 s_2^4-8 c_1^2 s_2^2+8 s_2^4+2 c_1^2-8 s_2^2) c_3^2 \notag\\
         &&  -8 c_1 c_2 s_2 s_3 (2 c_2^2-1) c_3-2 c_1^2 (2 s_2^2-1)^2\} s^4,\\
c_{223} &=& s_1^2\{2 c_1^2 (2 s_2^2-1) c_3^2+4 s_2 c_2 c_3 c_1 s_3\} s^8
            +s_1^2\{(-8 c_1^2 s_2^4+8 c_1^2 s_2^2-8 s_2^4-2 c_1^2+8 s_2^2) c_3^2\notag\\
        &&    +8 c_1 c_2 s_2 s_3 (2 c_2^2-1) c_3+8 s_2^2 (s_2-1) (s_2+1)\} s^4,
\\
c_{220} &=&  s_1^2\{(2 c_2^2-1) (c_1^2+1) c_3^4-4 c_1 s_2 c_2 s_3 c_3^3\notag\\
        &&   \quad -(2 c_2^2-1)(2 c_1^2+1) c_3^2+4 c_1 s_2 c_2 s_3 c_3+c_1^2 (2 c_2^2-1)\} s^8\notag\\
         &&   +\{(-4 c_1^4 c_2^4+4 c_1^4 c_2^2-24 c_1^2 c_2^4-c_1^4+24 c_1^2 c_2^2-4 c_2^4-2 c_1^2+4 c_2^2-1) c_3^4 \notag\\
         &&    +8 c_1 c_2 s_2 s_3 (2 c_2^2-1) (c_1^2+1) c_3^3+2 c_1^2 (4 c_1^2 c_2^4-4 c_1^2 c_2^2+12 c_2^4+c_1^2-12 c_2^2+1) c_3^2 \notag\\
        &&     -8 s_2 c_2 c_1^3 s_3 (2 c_2^2-1) c_3-c_1^4 (2 c_2^2-1)^2\} s^4,\\
c_{211} &=& s_1^2\{8 c_1 c_2 s_2 s_3 c_3^3-2 (2 s_2^2-1) (c_1^2+1) c_3^2s_3^2-4 s_2 c_2 c_3 c_1 s_3\}s^8  \notag\\
        && +\{-16 c_1 c_2 s_2 s_3 (2 c_2^2-1) (c_1^2+1) c_3^3 \notag\\
        && +(-8 c_1^4 s_2^4+8 c_1^4 s_2^2-48 c_1^2 s_2^4-2 c_1^4+48 c_1^2 s_2^2-8 s_2^4-4 c_1^2+8 s_2^2-2) c_3^2s_3^2 \notag\\
        && +8 c_1 c_2 s_2 s_3 (2 c_2^2-1) (c_1^2+1) c_3-8 c_1^2 s_2^2 c_2^2\} s^4,\\
c_{202} &=& s_1^2\{-(2 s_2^2-1) (c_1^2+1) c_3^4-4 c_1 c_2 s_2 s_3 c_3^3+(2 s_2^2-1) c_3^2\} s^8\notag\\
        && +\{(-4 c_1^4 s_2^4+4 c_1^4 s_2^2-24 c_1^2 s_2^4-c_1^4+24 c_1^2 s_2^2-4 s_2^4-2 c_1^2+4 s_2^2-1) c_3^4  \notag\\
        && +8 c_1 c_2 s_2 s_3 (2 c_2^2-1) (c_1^2+1) c_3^3+(24 c_1^2 s_2^4-24 c_1^2 s_2^2+8 s_2^4+2 c_1^2-8 s_2^2+2) c_3^2\notag\\
        && -8 c_1 c_2 s_2 s_3 (2 c_2^2-1) c_3-(2 s_2^2-1)^2\} s^4.
}
\end{subequations}

\end{document}